\documentclass[fleqn,usenatbib]{mnras}

\usepackage{newtxtext,newtxmath}

\usepackage[T1]{fontenc}

\DeclareRobustCommand{\VAN}[3]{#2}
\let\VANthebibliography\thebibliography
\def\thebibliography{\DeclareRobustCommand{\VAN}[3]{##3}\VANthebibliography}

\usepackage{graphicx}	
\usepackage{amsmath}	
\usepackage[dvipsnames]{xcolor} 

\newcommand{\Msol}{M_{\odot}}
\newcommand{\ang}{\textup{\AA}}

\title[KN Post Peak Opacities]{NLTE Effects on Kilonova Expansion Opacities}

\author[Q. Pognan et al.]{
Quentin Pognan,$^{1}$\thanks{E-mail: quentin.pognan@astro.su.se}
Anders Jerkstrand,$^{1}$
Jon Grumer$^{2}$
\\
$^{1}$The Oskar Klein Centre, Department of Astronomy, Stockholm University, AlbaNova, SE-10691 Stockholm, Sweden\\
$^{2}$Theoretical Astrophysics, Department of Physics and Astronomy, Uppsala University, Box 516, SE-751 20 Uppsala, Sweden \
}

\date{Accepted XXX. Received YYY; in original form ZZZ}

\pubyear{2022}

\begin{document}
\label{firstpage}
\pagerange{\pageref{firstpage}--\pageref{lastpage}}
\maketitle

\begin{abstract}
A binary neutron star merger produces a rapidly evolving transient known as a kilonova (KN), which peaks a few days after merger. Modelling of KNe has often been approached assuming local thermodynamic equilibrium (LTE) conditions in the ejecta. We present the first analysis of non-local thermodynamic equilibrium (NLTE) level populations, using the spectral synthesis code \textsc{sumo}, and compare these to LTE values. We investigate the importance of the radiation field by conducting NLTE excitation calculations with and without radiative transfer. Level populations, in particular higher lying ones, start to show deviations from LTE a few days after merger. Excitation is lower in NLTE for the majority of ions and states, and this tends to give lower expansion opacities. While the difference is small for the first few days, it grows to factors 2-10 after this. Our results are important both for demonstrating validity of LTE expansion opacities for an initial phase (few days), while highlighting the need for NLTE modelling during later phases. Considering also NLTE ionisation, our results indicate that NLTE can give both higher or lower opacities, depending on composition and wavelength, sometimes by orders of magnitudes.

\end{abstract}




\section{Introduction}

Merging neutron stars are an ideal site for r-process nucleosynthesis, with radioactive decay of unstable isotopes powering a supernova (SN) like transient known as a kilonova (KN) \citep[][]{Symbalisty.Schramm:82,Eichler.etal:89,Li.Paczynski:98,Freiburghaus.etal:99,Rosswog.etal:99,Metzger.etal:10}. There has been much work on predicting ejecta properties such as mass, velocity and composition using hydrodynamic merger simulations and nuclear network calculations \citep[e.g.][]{Rosswog.etal:99,Martinez.etal:07,Metzger.etal:10,Wanajo.etal:14,Rosswog.etal:18,Nedora.etal:21,Zhu.etal:21}. The observation of the first KN AT2017gfo provided the first chance to empirically determine these quantities \citep[see e.g][]{Arcavi.etal:17,Coulter.etal:17,Cowperthwaite.etal:17,Drout.etal:17,Kasen.etal:17,Smartt.etal:17,Tanvir.etal:17}. 

Light curve (LC) models for AT2017gfo make use of local thermodynamic equilibrium (LTE) expansion opacities \citep[see e.g.][]{Barnes.Kasen:13,Tanvir.etal:17,Kasen.etal:17,Hotokezaka.etal:20}. Uncertainties from both the atomic data used for opacity calculations, and the validity of LTE itself, propagate to uncertainties in the ejecta mass and composition and are therefore important to address and understand. While experimental atomic data for r-process elements remain sparse, much work has recently been conducted on the theoretical side \citep[see e.g.][]{Kasen.etal:13,Tanaka.Hotokezaka:13,Gaigalas.etal:19,Gamrath.etal:19,Banerjee.etal:20,Fontes.etal:20,Radziute.etal:20,tanaka:opacities:2020}. As such, there are now LTE expansion opacities available for all r-process elements. However, the LTE assumption remains to be thoroughly investigated. Due to the relatively low ejecta densities in the KN, stemming from low ejecta masses and high ejecta velocities, the ejecta will rapidly transition into a regime where electron densities are below the critical densities needed to maintain detailed collisional balance. 

It has also argued on qualitative grounds that the radiation field may be able to push populations close to LTE even when collisions cannot \citep[][]{Pinto.Eastman:00,Kasen.etal:17}. Whether this occurs has, however, never been investigated, and defines one of the goals in this study. More broadly, we aim to determine under what conditions the LTE assumption is valid, and to gauge when the ejecta transitions into a regime requiring non-local thermodynamic equilibrium (NLTE) calculations. 

Thus far, NLTE studies of KN ejecta are limited to neodymium (Nd, Z = 60) \citep{Hotokezaka.etal:21}, while \citet{Pognan.etal:22} have applied an NLTE approach to a set of models containing Te (Z = 52), Ce (Z = 58), Pt (Z = 78) and Th (Z = 90). Modelling emergent lightcurves and spectra with NLTE requires the calculation of temperature, ionisation and excitation structure, as well as the radiation field, all self-consistently, and therefore represents significant effort both in terms of physical complexity and computational power. It is therefore important to determine from which point in the evolution of the KN it becomes necessary to employ this more complex approach.

Quantities derived from excitation structure, such as expansion opacities used in LC modelling, may be affected by the deviation of level populations from LTE values. Should LTE expansion opacities be significantly different from full NLTE values, it may be worth attempting to update LTE expansion opacity libraries to include NLTE opacities. While replacing these with full NLTE ones is highly complicated, an approach using a simplified NLTE calculation may still be relevant. For example, if one ignores the radiation field and non-thermal excitations, the opacity formally depends on just one more parameter, the electron density $n_e$, than the LTE one. Therefore, it is of interest to investigate how such simplified NLTE quantities relate to the full NLTE ones. This defines an additional goal of the study, tied to the goal of investigating the importance of the radiation field in setting the level populations.

In \citet{Pognan.etal:22}, we presented a grid of simple KN models, and focused the analysis on the significance of time-dependent terms in the energy and ionisation structure equations. We now analyse parts of this model grid with respect to the NLTE excitation and ionisation structure. We define three main goals with this work: 1) Establish the size and significance of LTE deviations in excitation and ionisation at different epochs. 2) Assess the role the radiation field has in maintaining LTE excitation structure for longer than collisions allow for. 3) Assess the accuracy of LTE expansion opacities, widely used so far for virtually all KN modelling. These goals are determined with the purpose of aiding in understanding how accurately LTE models infer properties of AT2017gfo, from ejecta mass estimates, to the interpretation of the two ejecta component model based on spectral energy distribution (SED) changes \citep[see e.g.][]{Kasen.etal:17,Tanvir.etal:17,Villar.etal:17}. This will also provide information on which methods to use for future modelling efforts of KNe. 

The models used here were calculated using the NLTE spectral synthesis code \textsc{sumo} \citep{Jerkstrand.etal:11,Jerkstrand.etal:12} adapted for KNe. We consider outputs at 3, 5, 10 and 20 days after the merger, corresponding roughly to the earliest onset of NLTE effects, until when the NLTE regime should unambiguously be applied. We briefly cover the key parts of the models in Section \ref{sec:models}, and go on to present our findings in Sections \ref{sec:levelpops} and \ref{sec:opacity}. We then discuss the implications for LC and SED modelling in Section \ref{sec:discussion}. Finally, we summarise our findings and the implications for current and future KN modelling in Section \ref{sec:conclusions}.

\section{Ejecta Models and Expansion Opacity}
\label{sec:models}

The models used in this study are taken from those of \citet{Pognan.etal:22}, with the parameters and physics described therein. Our standard model has constant density ejecta with parameters $M_{\rm{ej}} = 0.05 \Msol$ and maximum expansion velocity $v_{\rm{ej}} = 0.1$c. These values correspond roughly to the estimates made for the observed KN AT2017gfo \citep[see e.g.][]{Abbott.etal:17,Kasen.etal:17,Smartt.etal:17,Metzger:19}. We also consider a low density ejecta model with $M_{\rm{ej}} = 0.01 \Msol$ and $v_{\rm{ej}} = 0.2$c, which is expected to fall out of LTE conditions earlier, and may be closer to the parameters for a dynamic, lanthanide rich ejecta.

The elements in the ejecta are Te (Z=52), Ce (Z=58), Pt (Z=78) and Th (Z=90) scaled to their approximate relative r-process solar abundances, such that their their mass fractions here 0.65, 0.08, 0.25 and 0.02 respectively \citep[][]{Sneden.etal:08,Lodders.etal:09,Pritychenko:19,Prantzos.etal:20,Nedora.etal:21}. In other words, isotopes of these elements created solely by the s-process are not included in the determination of mass fraction. We note that this remains an approximate scaling, as the relative contributions of s-process vs r-process for a given isotope varies between works. This composition implies a relatively low electron fraction ($Y_e \lesssim 0.2$) ejecta due to the presence of 2nd and 3rd r-process peak elements (Te and Pt), as well as a lanthanide (Ce) and an actinide (Th) \citep[][]{Rosswog.etal:18,Nedora.etal:21,Zhu.etal:21}. Since we evolve our models only up to 20 days after merger, we consider that a steady-state approach is sufficient \citep[][]{Pognan.etal:22}.

The temperature, ionisation and excitation structures in \textsc{sumo} are calculated self-consistently and coupled to the radiation field by radiative transfer. The level populations are calculated considering internal transitions, recombinations into and out of excited states, and ionisation from excited states. Ionisation into excited states is currently not implemented in \textsc{sumo}, and so all ionisations are allocated to the ground state. Internal collisional processes include thermal collisions, the treatment of which is described in \citet{Pognan.etal:22}. Non-thermal collisional excitation and thermal collisional ionisation are not yet included for r-process elements. However, neither are expected to be particularly important - the relatively high ionisation state of KN ejecta means thermal collisional populations dominate non-thermal ones \citep[][]{Kozma.Fransson:92}, and the relatively low densities combined with moderate temperatures make collisional ionisation inefficient (see Section \ref{subsec:analytic}).

Since a focus of this study is the effect of the radiation field on the excitation structure, it is useful to explicitly write out the radiative internal transitions considered by \textsc{sumo}, where the upper level is denoted as $u$, and lower level by $l$:

\begin{align}
    \label{eq:R_spont}
    &R^{\rm{spont}}_{u,l} = A_{u,l} \beta^{\rm{S}}_{u,l} \\
    \label{eq:R_abs}
    &R^{\rm{abs}}_{l,u} = \rm{From~MC-sim.} \\
    \label{eq:R_stim}
    &R^{\rm{stim}}_{u,l} = \frac{g_l}{g_u} R^{\rm{abs}}_{l,u}
\end{align}

\noindent The transition rates correspond to spontaneous emission, absorption, and stimulated emission respectively. In the above, $A_{u,l}$ is the Einstein A-coefficient for the transition, $g_u$ and $g_l$ are the statistical weights of the upper and lower levels respectively, and $\beta^{\rm{S}}_{u,l}$ is the Sobolev escape probability, given by:

\begin{equation}
    \beta^{\rm{S}}_{u,l} = \frac{1-e^{-\tau^S_{u,l}}}{\tau^S_{u,l}}
    \label{eq:beta_sob}
\end{equation}

\noindent where $\tau^{\rm{S}}_{u,l}$ is the Sobolev optical depth:

\begin{equation}
    \tau^S_{\rm{u,l}} = \frac{\pi e^2}{m_e c} f n_l \left(1 -  \frac{n_u g_l}{n_l g_u}\right) t \lambda_0
    \label{eq:tau_sob}
\end{equation}

\noindent Here $f$ is the oscillator strength of the transition, $n_l$, $n_u$ are the lower and upper level number densities, with associated statistical weights $g_l$ and $g_u$ respectively, $\lambda_0$ is the line center rest wavelength, and $t$ is the time after merger. The term in the brackets accounts for stimulated emission. The radiative absorption rate (equation (\ref{eq:R_abs})) is formally is formally a product of the Einstein B-coefficient, the radiation field in the blue wing of the line, and the Sobolev escape probability. In practice, in \textsc{sumo} the rates are obtained by accumulators in the Monte Carlo simulation. Stimulated emission is related simply to radiative absorption by the levels' statistical weights. 

Though not employed by \textsc{sumo}, which treats radiative transfer line by line, many LTE codes make use of expansion opacities in order to predict the LC and SED evolution. As such it is interesting to test if and how expansion opacity is affected by differences in excitation structure. The bound-bound opacity is calculated from the Sobolev optical depth for a wavelength bin $\Delta \lambda$ following the expansion opacity formalism, which has been largely applied in the context of SNe and KNe \citep[][]{Karp.etal:77,Eastman.Pinto:93,Kasen.etal:06,Kasen.etal:13}: 

\begin{equation}
    \kappa_{\rm{exp}}\left(\lambda\right) = \frac{1}{c t \rho} \sum_i \frac{\lambda_i}{\Delta \lambda} \left(1 - e^{-\tau^S_i} \right)
    \label{eq:opacity}
\end{equation}

\noindent where the sum is conducted over all lines within the wavelength bin $\Delta \lambda$. The sum in Equation \ref{eq:opacity} is somewhat dependent on the size of the wavelength bin, i.e. the ratio $\Delta \lambda / \lambda_i$, which we choose following several considerations. Firstly, the maximum ejecta velocity, $v_{\rm{ej}}/c = 0.1$ in this model, sets a maximum scale, as photons cannot propagate through a Doppler shift larger than this value. A lower limit is set by the thermal broadening, which for a temperature of $T \sim 10^4$K (our highest value), and a mean atomic weight of $A \sim 140$ yields $\Delta^{thermal} \sim 10^{-6}$. Since expansion opacity aims to capture the effect of several well-separated lines, counting $\gtrsim 10$ lines in the wavelength bin each separated by $\gtrsim$10 thermal widths yields a lower limit of $10^{-4}$. We thus test ratios in the range of $\Delta \lambda/ \lambda_i \sim 10^{-4} - 10^{-2}$, finding that higher resolution does not qualitatively change the shape of the opacity curves, but simply makes them noisier. As such, we choose wavelength bins such that we have a ratio of $\Delta \lambda/ \lambda_i = 0.01$, which is consistent with previous studies' choices \citep[][]{Kasen.etal:13,tanaka:opacities:2020}.

The radiative transfer calculations in \textsc{sumo} are conducted using the steady-state assumption, also known as the stationarity approximation. Formally, this is only valid when the photon diffusion time $t_{\rm{diff}}$ is small compared to the evolutionary time, which may not necessarily be the case at early times for models with high opacity. However, we believe the effect of this on our temperature and ionisation solutions will not be too severe, as non-thermal processes are expected to dominate the determination of these quantities \citep[][]{Kozma.Fransson:92}. In general, steady-state radiative transport is well applied at peak LC times and later; this has been tested in the context of 1-D SN simulations \citep[][]{Kasen.etal:06}.

We designate four models in this study: full NLTE, limited NLTE, limited LTE and full LTE, presented in Table \ref{tab:models}. The first three models only vary the method used to calculate excitation structure, while the final model is designed to test the effects of using LTE ionisation. 

We begin by comparing the excitation structure yielded from the full NLTE model to that of the limited LTE solution, i.e. the level populations are from the Boltzmann equation and scaled by the respective ion's abundance in the model. We also consider the excitation structure from the limited NLTE model which omits the radiation field, i.e. $R^{\rm{abs}} = R^{\rm{stim}} = 0$ in equations (\ref{eq:R_abs}) and (\ref{eq:R_stim}) above, and also photoionisation terms. Corresponding expansion opacities are then calculated and compared for each of these models. Thus, the differences in this first comparison lie solely with the level population calculation; all other quantities are taken from the NLTE solution. 

We construct an additional model in "full LTE", where the ionisation and excitation come from the Saha and Boltzmann equations. The expansion opacity from this model is compared to that of the full NLTE model, in an attempt to demonstrate that deviations from LTE occur not only for excitation, but also ionisation, which may have significant impact on opacity.

\subsection{Analytic Estimates of LTE Deviation}
\label{subsec:analytic}

The LTE approximation is expected to breakdown relatively early on after the initial merger due to rapidly dropping densities, $n \sim 6 \times 10^9 t_d^{-3} M_{0.05} v_{0.1c}^{-3}$ cm$^{-3}$, where $M_{0.05} = M_{\rm{ej}}/0.05 \Msol$ and $v_{0.1c} = v_{\rm{ej}}/0.1$c.  It is useful to make some analytic estimates as to when the transition to NLTE is expected.

\subsubsection{LTE Excitation Structure}

Maintaining LTE level populations by collisions requires thermal collisions to dominate spontaneous radiative decays. A 'critical density' may be defined, below which detailed collisional balance cannot be maintained. The critical density will depend on the transition, notably whether it is optically thin or thick. From the equations in \citet{Jerkstrand:17} we derive:

\begin{align}
    \label{eq:ncrit_thick}
    n_e^{\rm{crit}} &\sim 5 \times 10^8 \left(\frac{\lambda}{5000~ \ang}\right)^3 t_d^{-2}M_{0.05}^{-1} v_{0.1c}^{3} \; \mathrm{cm}^{-3} \;  ;\tau^S \gtrsim 1 \\
    \label{eq:ncrit_thin}
    n_e^{\rm{crit}} &\sim 10^7 \mathrm{A \; cm}^{-3} \; ; \tau^S << 1
\end{align}

\noindent While the critical density for an optically thick line depends on ejecta density and transition wavelength, but not the Einstein A-value, the optically thin critical density solely depends on the A-value for that transition. Considering our ejecta model at 3 days after merger, the earliest epoch and thus highest density in this study, we have $n \sim 10^{8} \; \mathrm{cm^{-3}}$, which means levels whose main radiative deexcitation channels are optically thick with $\lambda \lesssim 4000 \ang$, may have their level populations in LTE, though lines emitting at longer wavelengths may not. Similarly, the condition of level populations for the optically thin lines will depend on the specific transition's Einstein A-coefficient. A level that has an allowed radiative transition as a deexcitation channel, i.e. $A \gtrsim 10^5 \; \mathrm{s^{-1}}$, will most likely not be able to maintain an LTE level population by thermal collisional processes alone.

\begin{table}
    \caption{Model types and properties. NLTE ionisation and excitation refer to the solutions yielded from the full \textsc{sumo} simulations. The limited NLTE model calculates excitation structure in a post-processing step and excludes (diffuse) radiative field terms. Boltzmann and Saha equations are LTE excitation and ionisation respectively. In every model, density and temperature are taken from the full NLTE solution.}
    \centering
    \begin{tabular}{c|c|c}
    Name & Ionisation Structure & Excitation Structure \\
    \hline \\
    Full NLTE & NLTE & NLTE \\
    Limited NLTE & NLTE & NLTE without radiative terms \\
    Limited LTE & NLTE & Boltzmann \\
    Full LTE & Saha & Boltzmann 
    \end{tabular}
    \label{tab:models}
\end{table}

A breakdown of LTE for high-lying states can be expected as collisional deexcitation struggles to compete against allowed radiative transitions. Some of these allowed transitions will be to other high-lying, sparsely populated states, and will be optically thin, implying critical densities of $n_e^{\rm{crit}} \gtrsim 10^{12}~\mbox{cm}^{-3}$ according to equation \ref{eq:ncrit_thin}. As this is already higher than the model's density at 1 day after merger, we expect such high-lying states to already be out of LTE conditions at our first epoch, 3 days after merger. It is possible that a strong enough thermal radiation field may push these level populations towards LTE, which is one of the main ideas investigated in this study. We will return to this in Section \ref{subsec:radfield_pop}.

\subsubsection{LTE Ionisation Balance}

A quantitative assessment of the validity of the Saha equation can be gauged by comparing non-thermal ionisation rates to an estimation of the thermal collisional ionisation rates. The treatment of ionisation and recombination in \textsc{sumo} for KNe is presented in detail in \citet{Pognan.etal:22}. Non-thermal collisional cross sections estimated following the formalism in \citet{Lotz:67}. The non-thermal collisional rate for an ion is calculated as:

\begin{equation}
     \Gamma_{\mathrm{nt}}(t) = 10^{10} \; t_d^{-1.3} \; f(t) \; SF_{\mathrm{ion}}(t) \; \frac{A \; m_p}{\chi} s^{-1} 
     \label{eq:gamm_NT}
\end{equation}

\noindent where $f(t)$ is a radioactive decay thermalisation efficiency of order unity at early times, and $SF_{\rm{ion}}(t)$ is the fraction of non-thermal energy allocated to ionisation from solving the Spencer-Fano equation \citep[][]{Spencer.Fano:54,Fransson.Kozma:93}, typically on the order of $0.01 - 0.05$. $\chi$ is the ionisation potential of the ion, while $A$ is the atomic mass of the ion, and $m_p$ the mass of a proton.

In order to consistently compare our non-thermal ionisation rates to thermal collisional ionisation, we consider Equations 1 and 2 from \citet{Shull.Steenberg:82}, which also make use of \citet{Lotz:67}: 

\begin{equation}
    C_{\rm{therm}} = A_{\rm{col}} T^{1/2} (1 + a k_{\rm{B}}T/\chi)^{-1} e^{-\chi/k_{\rm{B}}T} \; \mathrm{cm^{3} \; s^{-1}}
    \label{eq:Ctherm}
\end{equation}

\noindent where $a \approx 0.1$, $\chi$ is the ionisation potential of the ion in erg, and $A_{\rm{col}}$ is given by:

\begin{equation}
    A_{\rm{col}} = 1.3 \times 10^{-8} F N_{\rm{val}} \ \chi_{\rm{eV}}^{-2} \; \mathrm{cm^{3} \; s^{-1}}
    \label{eq:Acol}
\end{equation}

\noindent Here, F is a 'focussing factor' of order unity, and $N_{\mathrm{val}}$ is the number of valence electrons of the ion, taken here to be one in order to be consistent with the non-thermal ionisation treatment. The thermal collisional ionisation rate is then given by $\Gamma_{\rm{them}} = C_{\rm{therm}}n_e$.

Considering the temperature and density solution of our model at 3 days (see Section \ref{sec:opacity}), i.e. $T \sim 4000$K, $n_e = 10^{8} \; \mathrm{cm^{-3}}$, and taking an average ionisation potential of 10 eV, roughly that of TeI, Ce II, Pt I, and Th II, we find a thermal collisional ionisation rate of $\Gamma_{\rm{thermal}} \approx 10^{-13} \; \mathrm{s^{-1}}$. Compared to this, we have non-thermal collisional ionisation rates on the order of $\Gamma_{\rm{NT}} \approx 10^{-4} \; \mathrm{s^{-1}}$, implying non-thermal collisions completely dominate relative to thermal collisions at this early time. 

Although both the non-thermal and thermal collisional rates are dependent on density and thus may decrease with time, temperature tends to increase with time, so it is possible that the relative importance of thermal collisional ionisation may be greater later on. Taking our model at 20 days with $T\sim 10000$K and $n_e \sim 10^6 \; \rm{cm^{-3}}$, we find thermal collisional rates of $\Gamma_{\rm{thermal}} \approx 10^{-7} \; \mathrm{s^{-1}}$, while the non-thermal rates are on the order of $\Gamma_{\rm{NT}} \approx 10^{-5} \; \mathrm{s^{-1}}$. While the rates are now closer, the non-thermal ones are still dominant for most ions. Naturally, the exact importance of thermal collisional ionisation is model dependent, as these in general will not have identical densities, ionisation structure and temperature solutions.

In the same way that a thermal radiation field may push excitation structure towards LTE, ionisation structure could potentially be pushed towards LTE by a quasi-thermal radiation field through photoionisation (PI). Considering the average energy of a thermal photon at $T = 10^4$K, we have $E = k_b T \sim 1 eV$. It is thus possible that photons from the high energy tail of the radiation field may contribute to photoionisation for ions with low ionisation potentials (e.g. Ce I and Th I at $\chi \sim 5$eV), especially when considering PI from excited states which requires less energetic photons. Looking ahead at our models at any epoch, we find that PI rates actually dominate for neutral ions, by an order of magnitude compared to the non-thermal rates for Te I and Pt I ($\Gamma_{\rm{PI}} \approx 10^{-3} \; \mathrm{s^{-1}}$ at 3 days), and by several orders of magnitude for Ce I and Th I. However, the other ions remain dominated by non-thermal collisional ionisation, implying that it remains unlikely that an LTE ionisation structure as given by the Saha equation will be present in our models for the epochs considered in this study. Thus, it seems that for a full application of LTE in both ionisation and excitation, it appears a hotter, denser environment than that of our model at 3 days after merger may be required.

\begin{figure*}
    \centering
    \includegraphics[trim={0.0cm 0.1cm 0.4cm 0.3cm},clip,width = 0.49\textwidth]{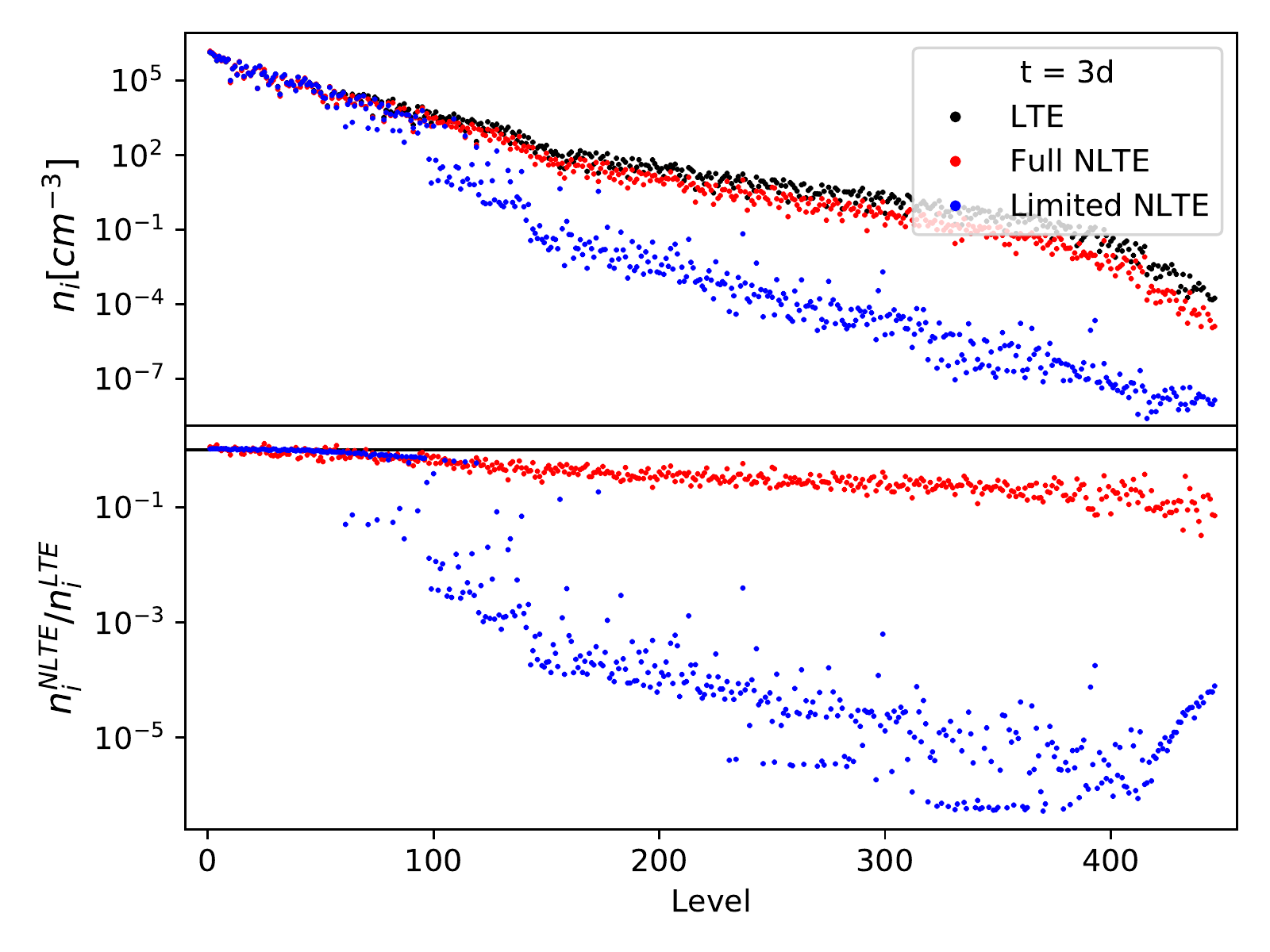} 
    \includegraphics[trim={0.0cm 0.1cm 0.4cm 0.3cm},clip,width = 0.49\textwidth]{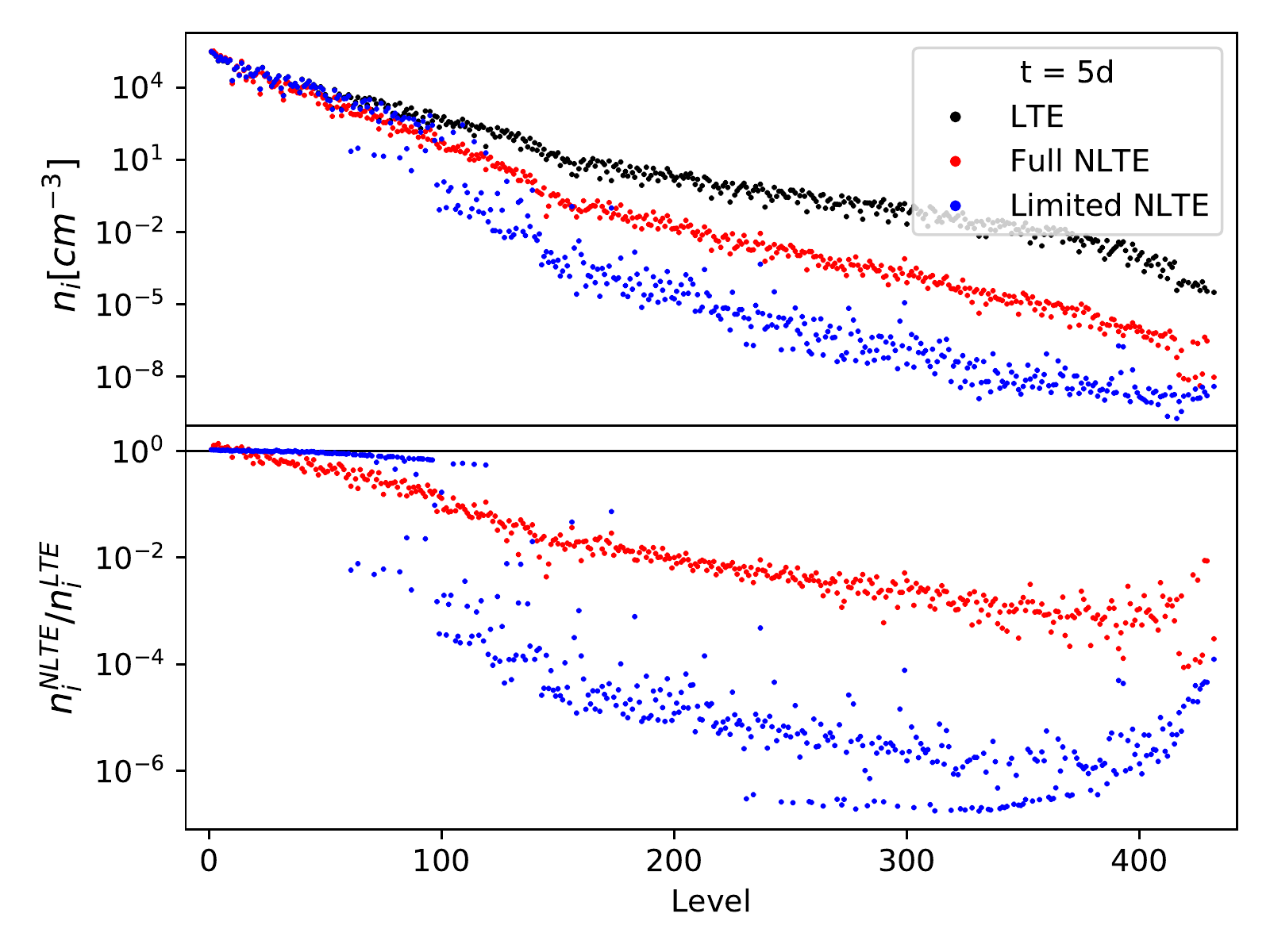}
    \includegraphics[trim={0.0cm 0.1cm 0.4cm 0.3cm},clip,width = 0.49\textwidth]{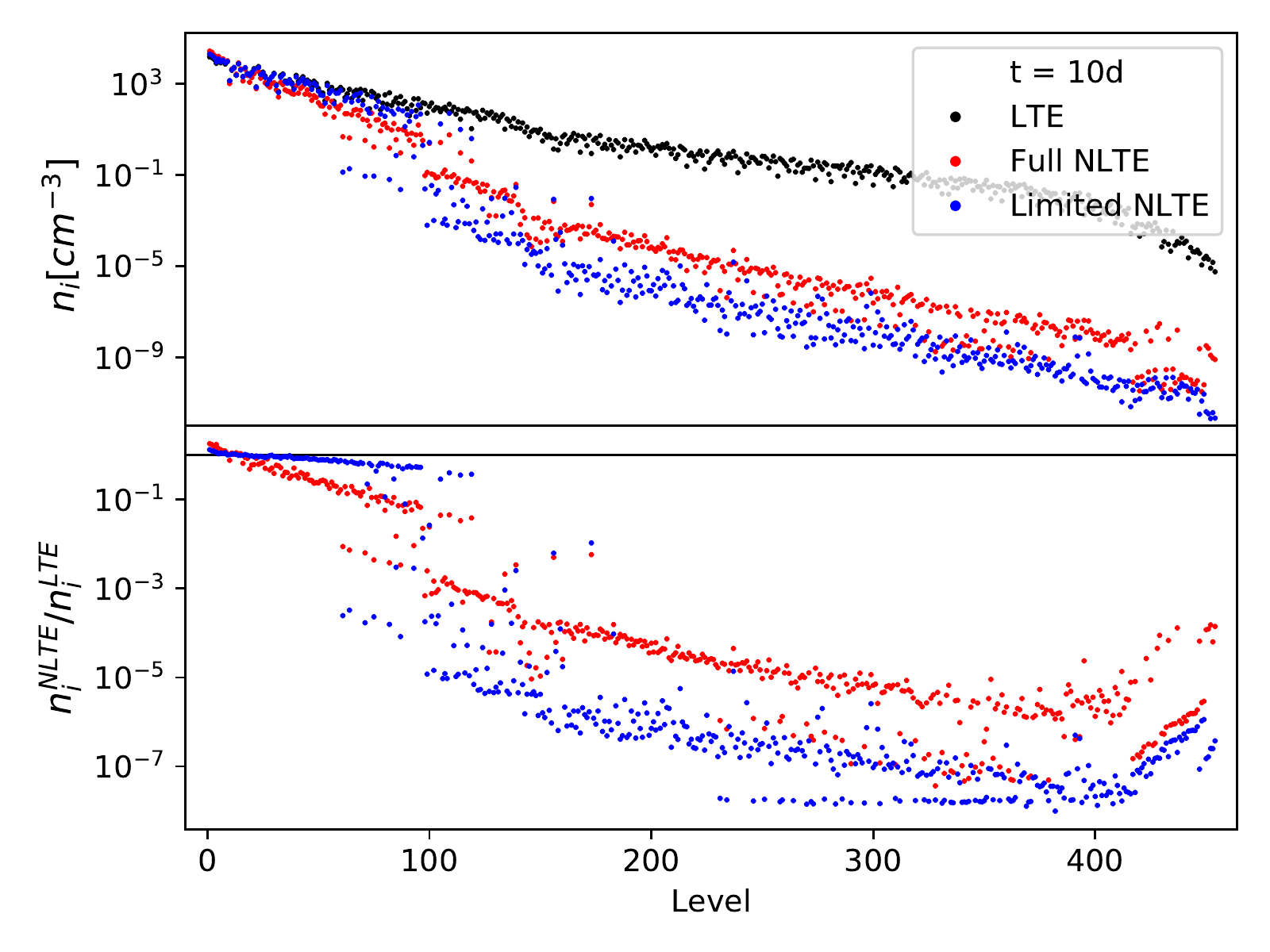}
    \includegraphics[trim={0.0cm 0.1cm 0.4cm 0.3cm},clip,width = 0.49\textwidth]{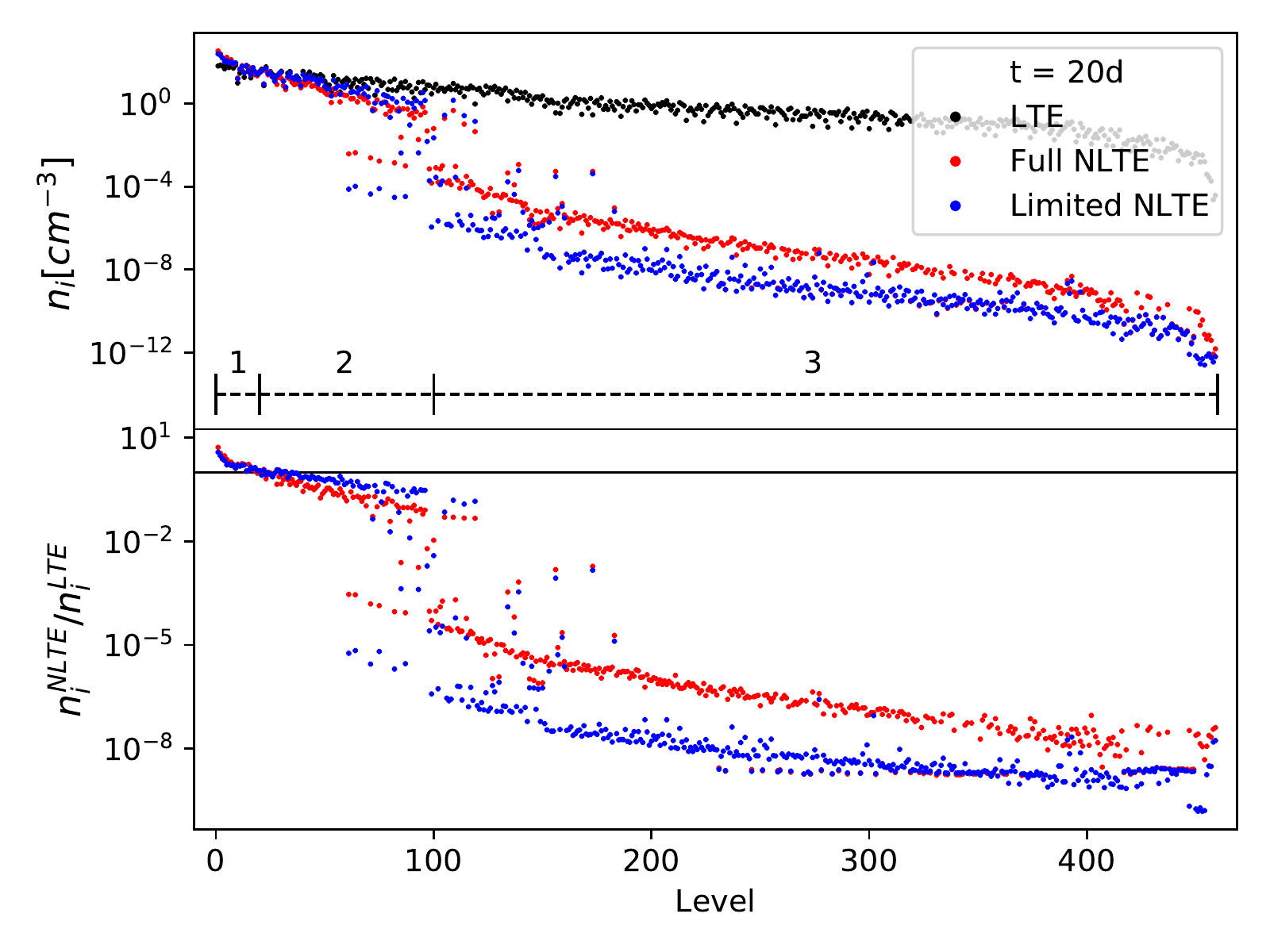}
    \caption{Level populations of Ce II, the most important contributor to opacity for wavelengths $\lambda \gtrsim 1500 \ang$, among the ions included in this study, at 3, 5, 10, and 20 days after merger (top left to bottom right). States with an LTE level population smaller than the most populated state by a factor of $10^{10}$ are cut from these plots. The NLTE level populations with (red) and without (blue) the radiation field are compared to the pure LTE populations (black). The sharp drop seen at later times around $n \sim 100$ is due to levels having allowed spontaneous decay transitions. The level groups are marked out in the 20 day panel.}
    \label{fig:CeII_levelpops}
\end{figure*}

\begin{figure*}
    \centering
    \includegraphics[trim={0.2cm 0.1cm 0.4cm 0.3cm},clip,width = 0.49\textwidth]{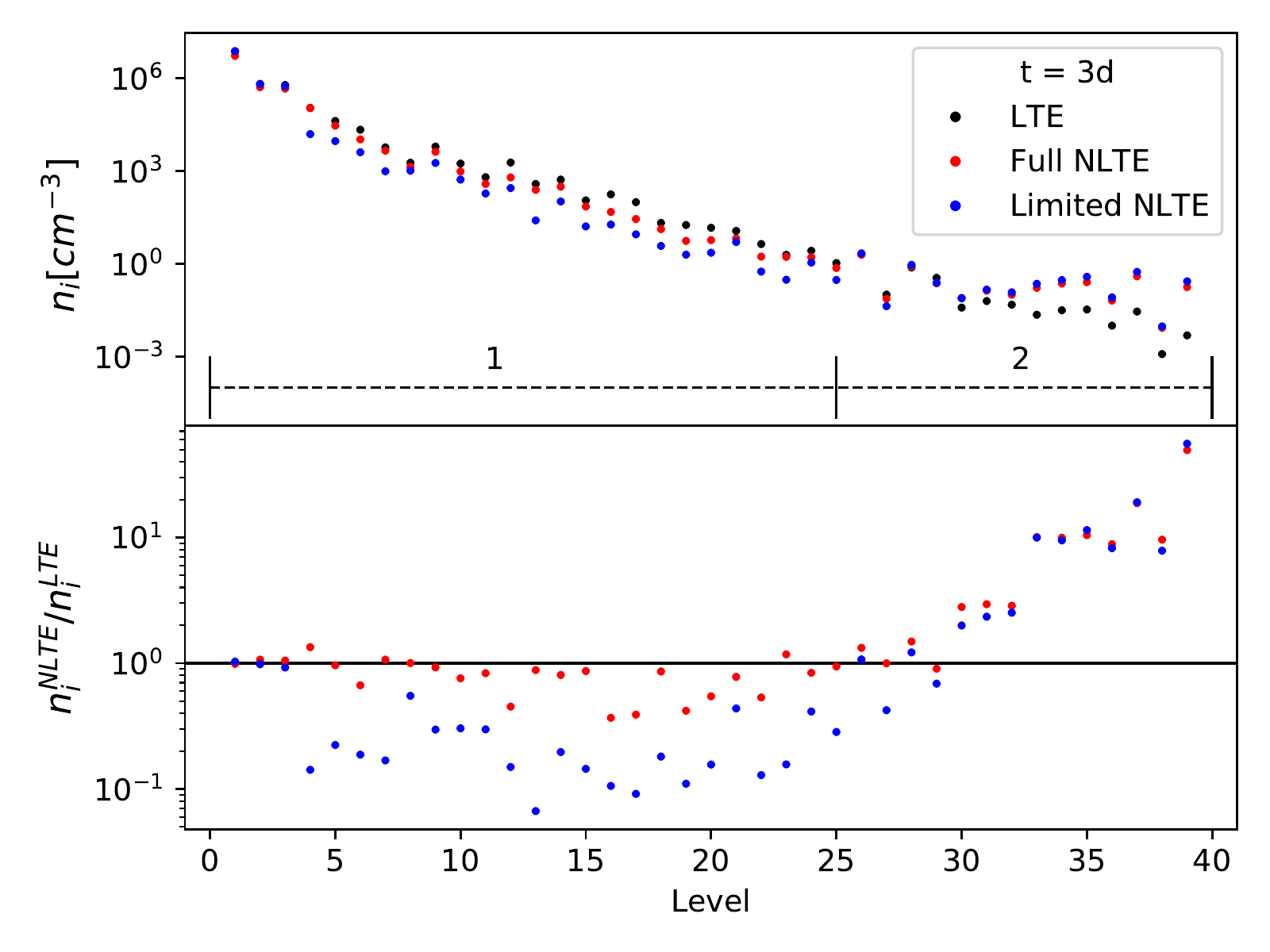}
    \includegraphics[trim={0.2cm 0.1cm 0.4cm 0.3cm},clip,width = 0.49\textwidth]{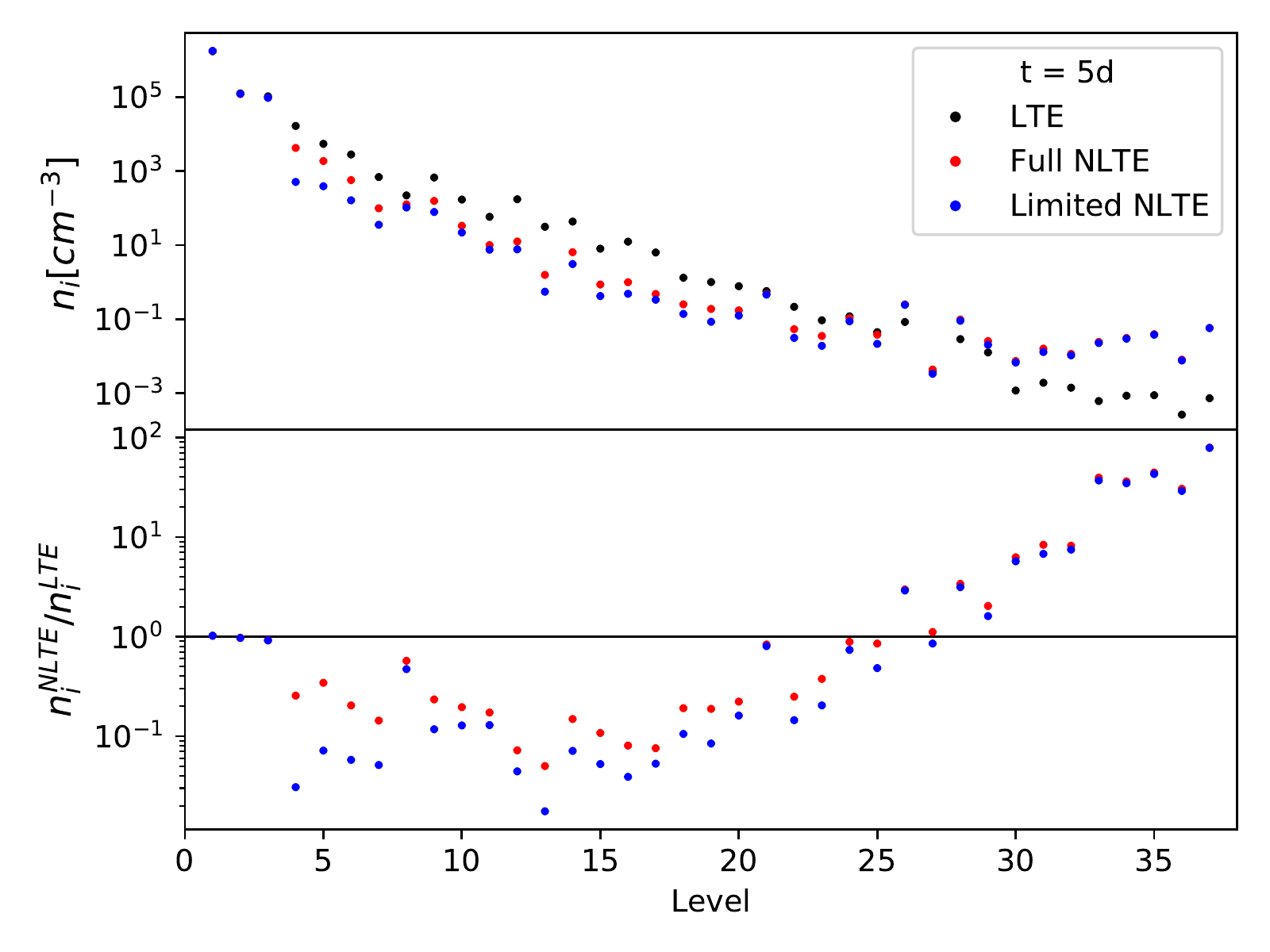} 
    \includegraphics[trim={0.2cm 0.1cm 0.4cm 0.3cm},clip,width = 0.49\textwidth]{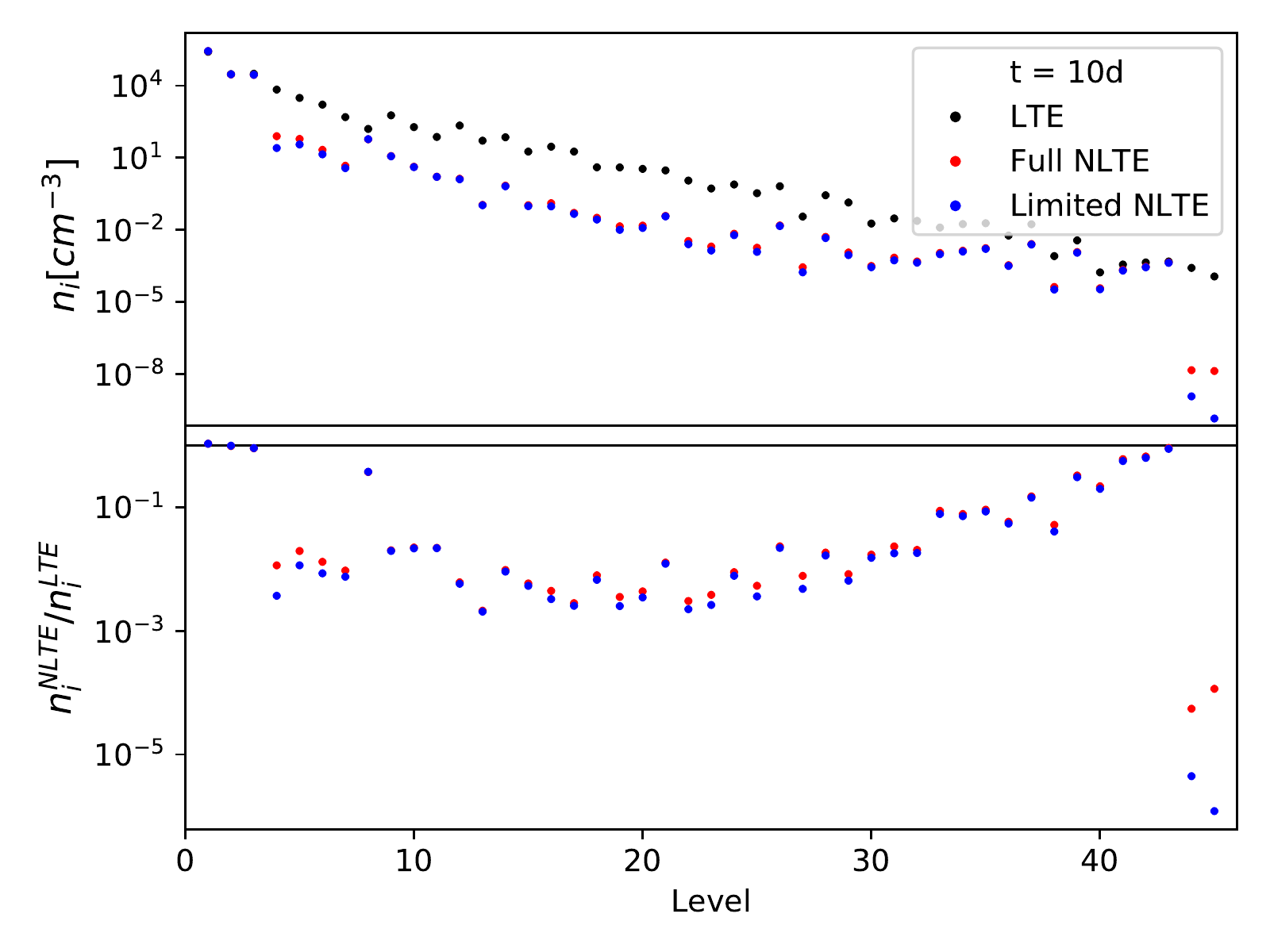} 
    \includegraphics[trim={0.2cm 0.1cm 0.4cm 0.3cm},clip,width = 0.49\textwidth]{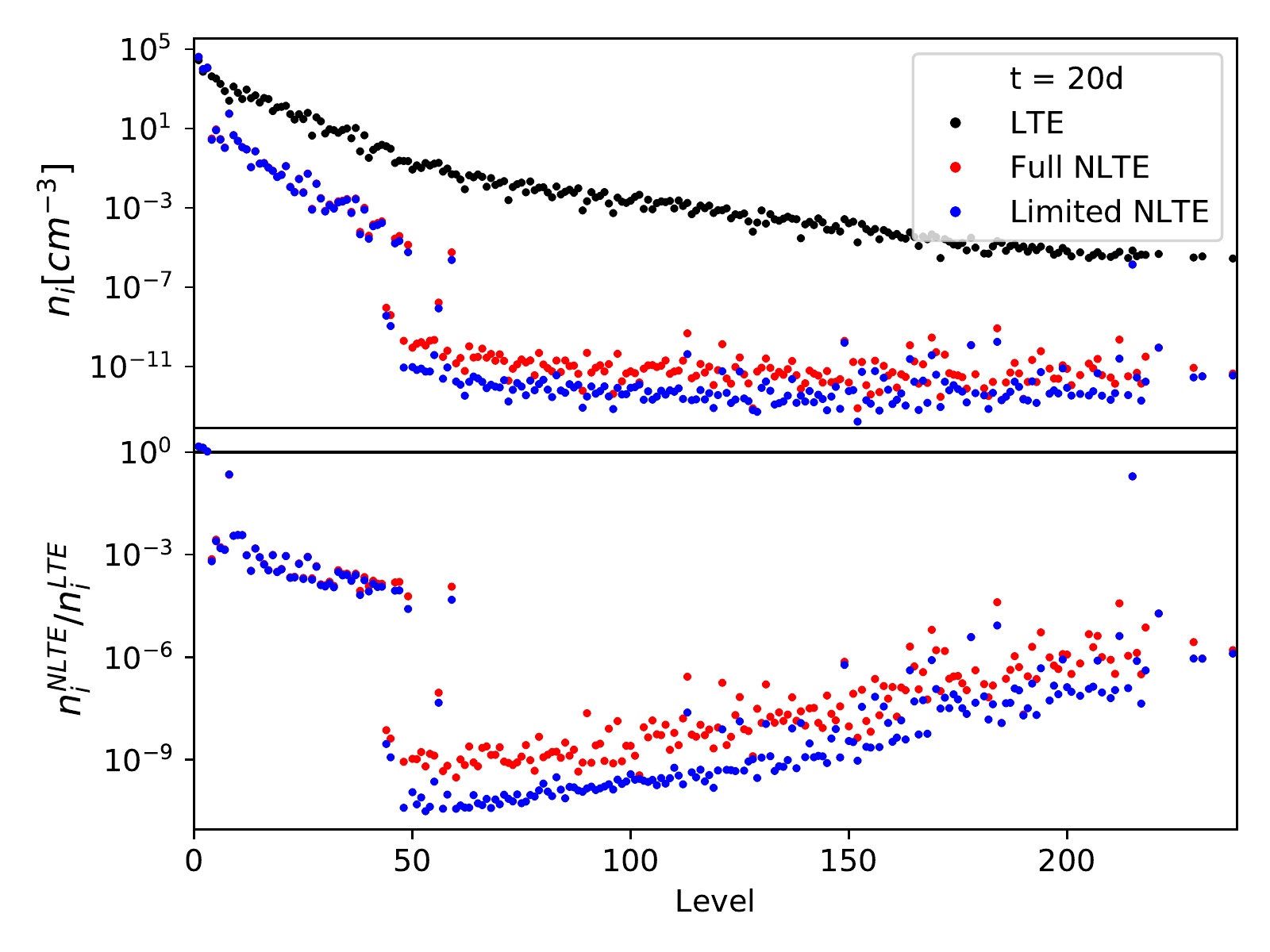} 
    \caption{Level populations of Pt III, the most important contributor to opacity for wavelengths $\lambda \lesssim 1500$ \AA, at 3, 5, 10, and 20 days after merger (top left to bottom right). States with an LTE level population smaller than the most populated state by a factor of $10^{10}$ are cut from these plots. The NLTE level populations with and without the radiation field are compared to the pure LTE populations. The sharp drop seen at later times around $n \sim 45$ is due to levels beginning to have allowed transitions along the forbidden ones. Level groups are marked out in the 3 day panel, though we note that group 3 ($n \gtrsim 40$) is only visible in the 20 day plot.}
    \label{fig:PtIII_levelpops}
\end{figure*}

\section{Excitation Structure}
\label{sec:levelpops}

We begin by comparing the level populations of the first three models in our study, that is the full NLTE, limited NLTE and limited LTE models. The limited NLTE excitation structure is calculated in a stand-alone program, using rate equations including thermal collisional rates, spontaneous radiative decay rates (using Sobolev escape probabilities), and recombination rates at the ion densities, electron density and temperature of the full NLTE field. This solver omits photoexcitation/deexcitation and photoionisations. 

Figures \ref{fig:CeII_levelpops} and \ref{fig:PtIII_levelpops} show the level population structures of Ce~II and Pt~III respectively, while the rest of the ions are shown in Appendix \ref{app:levelplots}. These particular two ions are chosen due to their strong contributions to the opacity, as described later in Section \ref{sec:opacity}. These two plots show the evolution of the various level populations from 3 to 20 days, with the bottom half of each panel showing the ratios of NLTE to LTE solutions, in which the horizontal line demarks identical solutions. Level populations smaller than 10 orders of magnitude from the maximum LTE value are cut, as these sparsely populated states are unlikely to contribute much to the opacity. 

In the following analysis, we define three groups of levels according to their behaviour described below. For Ce II, group 1 consists of levels $n \sim 1 - 20$, group 2 $n\sim 20 - 100$, and group 3 $n \gtrsim 100$. For Pt III, they are $n \sim 1 - 25$, $n\sim 25 - 40$ and $n \gtrsim 40$ respectively.

\subsection{Boltzmann LTE and Full NLTE Excitation Structures}
\label{subsec:boltz_pop}

We begin by comparing the limited LTE level populations to the full NLTE ones. For most ions, the first few excited states, e.g. group 1 and 2 populations, are quite well represented by the LTE approximation. At early times ($t \sim 3$d), LTE even approximates the less populated, group 3 states quite well, as can be seen in the top left panel of Figure \ref{fig:CeII_levelpops}. However, it is clear that the LTE solutions generally tend to overestimate the population of these high-lying states, with the approximation worsening with higher excitation energies. 

The difference increases with time, and the LTE level populations are seen to be relatively poor approximations of the full NLTE solutions for large ions from around 5 days after merger. A steep drop in population in the full NLTE values occurs for the group 3 states after several days, arising from allowed radiative transitions depopulating those levels. As densities decrease with time, collisional rates drop, though spontaneous emission from these high-lying states remains unaffected, thereby decreasing their population as time goes on. The Boltzmann equation has no knowledge of spontaneous emission, however, thus leading to an increase in the divide between LTE and NLTE solutions.

There are also some cases, however, where the LTE solutions appear to \textit{underestimate} the level population relative to the full NLTE solution. An example of this can be seen for the group 2 levels of Pt III at 3 and 5 days after merger, shown in the top panels of Figure \ref{fig:PtIII_levelpops}. The LTE solutions here underestimate the populations of these levels by up to two orders of magnitude. The explanation here comes from the inclusion of recombination inflows in the NLTE solutions. Recombination from higher ions, in this example Pt IV to Pt III, leads to significant inflows into these levels, which are otherwise difficult to populate by collisional excitation. The key difference between Ce II and Pt III is that the Pt III excitation energies are significantly higher. The $n = 30$ level of Pt III, where we begin to see this effect, has an excitation energy which corresponds approximately to $n = 408$ for Ce II in our atomic data set. As such, these high excitation energy states become difficult to populate from collisional excitation. Other processes such as recombination however, have no such problem reaching those levels, and thus push the NLTE populations over the LTE estimate in this case.

Group 1 populations for Ce II are also underestimated by LTE at late times, by up to an order of magnitude at 20 days after merger (bottom right panel of Figure \ref{fig:CeII_levelpops}). In this case however, recombination is not the dominating process. Rather, the depopulation of group 3 (and to some extent group 2) states in NLTE by radiative emission leads to significant inflows into the group 1 states, thereby leading LTE to underestimate the correct populations. Such an effect seems to require particularly large ions however, as we do not see an LTE under-population of group 1 states for Pt III at 20 days after merger. As such, it is not an effect that can be broadly ascribed to every ion's excitation structure. 

Generally speaking, it appears that the Boltzmann equation gives deviating level populations compared to NLTE as early as $t \gtrsim 5$ days for these ejecta conditions, in particular for species with complex structures and many energy levels. Notably, higher excitation states are harder to correctly approximate using an LTE approach, as thermal collisional excitation struggles to populate them. Other processes not included in LTE, such as recombination/ionisation in/out of states, as well as all radiative processes, can also play an important role, and contribute significantly to deviation from LTE solutions. However, it is possible that these high-lying states do not contribute significantly to emission or opacity. As such, the effect of this overestimation may be small when considering derived quantities such as opacity, an aspect which is investigated further in Section \ref{sec:opacity}.

\subsection{The Importance of the Radiation Field}
\label{subsec:radfield_pop}

We now turn our attention to the impact of radiation field terms on the excitation structure of the ejecta, using the limited NLTE model. As mentioned previously, in order to investigate this, we calculate level populations in NLTE (holding ionisation and temperature fixed), but omit radiative absorption and stimulated emission, i.e. equations (\ref{eq:R_abs}) and (\ref{eq:R_stim}), as well as photoionisations. This leads to level population solutions given by the blue points in Figures \ref{fig:CeII_levelpops} and \ref{fig:PtIII_levelpops}.

While these solutions for Pt III appear to generally follow the full NLTE solutions quite well for most epochs, the large level count of Ce II reveals a noticeably different story. Considering the solution for Ce II at 3 days after merger (top left panel of Figure \ref{fig:CeII_levelpops}), we see that the group 1 and 2 full NLTE level populations are relatively well matched by the limited NLTE solutions, implying that radiative terms play little roles here. However, a steep drop in level population is experienced in the limited NLTE solutions for the group 3 states. Since the full NLTE solution at this early time is almost identical to the LTE solution, this implies that the radiation field plays a key role in pushing the populations of high-lying states towards LTE. Considering the relatively low ionisation energies of Ce ions ($\chi \sim 11$~eV for Ce~II, $\sim 20$~eV for Ce~III), energetic photons from a radiation field at the right temperature ($T = 3817$~K here), are populating the higher lying states of Ce III in a way that collisional processes are incapable of. The steep drop at higher levels in the limited NLTE solution is due to the presence of allowed radiative transitions out of those levels. These continue to depopulate the high-lying states, which are then not repopulated by radiative excitation, omitted in this solution. 

Another aspect of the radiation field's influence can be seen in the group 2 level populations of Ce II from about 5 days onwards. While the group 3 populations are pushed towards LTE values by radiative excitation, group 2 populations are pushed \textit{away} from LTE values. This can be seen most clearly in the bottom left panel of Figure \ref{fig:CeII_levelpops}, on the bottom half where the ratios of NLTE to LTE level populations are calculated. Here, we see that the limited NLTE populations of group 2, in blue, are seen to be closer to the LTE values than the full NLTE values, up to an order of magnitude for the higher lying states of group 2 around $n \sim 90$. Although the limited NLTE populations are already lower than the LTE ones for these states, we see with the full NLTE populations that the radiation field also plays a role in depopulating these states.

As time progresses, we see that the effect of the radiation field becomes less significant. At 20 days after merger, the limited NLTE solutions are closer to the full NLTE solutions than the LTE solution. However, group 1 and 2 states remain slightly overestimated compared to the full NLTE values, while group 3 states remain underestimated. Though the radiation field is now dilute and its effect mitigated, it still plays a noticeable role even at this epoch. From the consideration of excitation structures, it appears that a limited NLTE approach omitting radiative transfer is actually a worse approach than using the LTE Boltzmann equation for very early times $\sim 3$ days after merger. At later times, $t \gtrsim 10$ days, however, the limited NLTE calculation better recovers the full NLTE solution than the LTE approach. The effect this has on expansion opacity is investigated in Section \ref{sec:opacity}.

\begin{figure*}
    \centering
    \includegraphics[trim={0.cm 0.cm 0.4cm 0.4cm},clip,width = 0.49\textwidth]{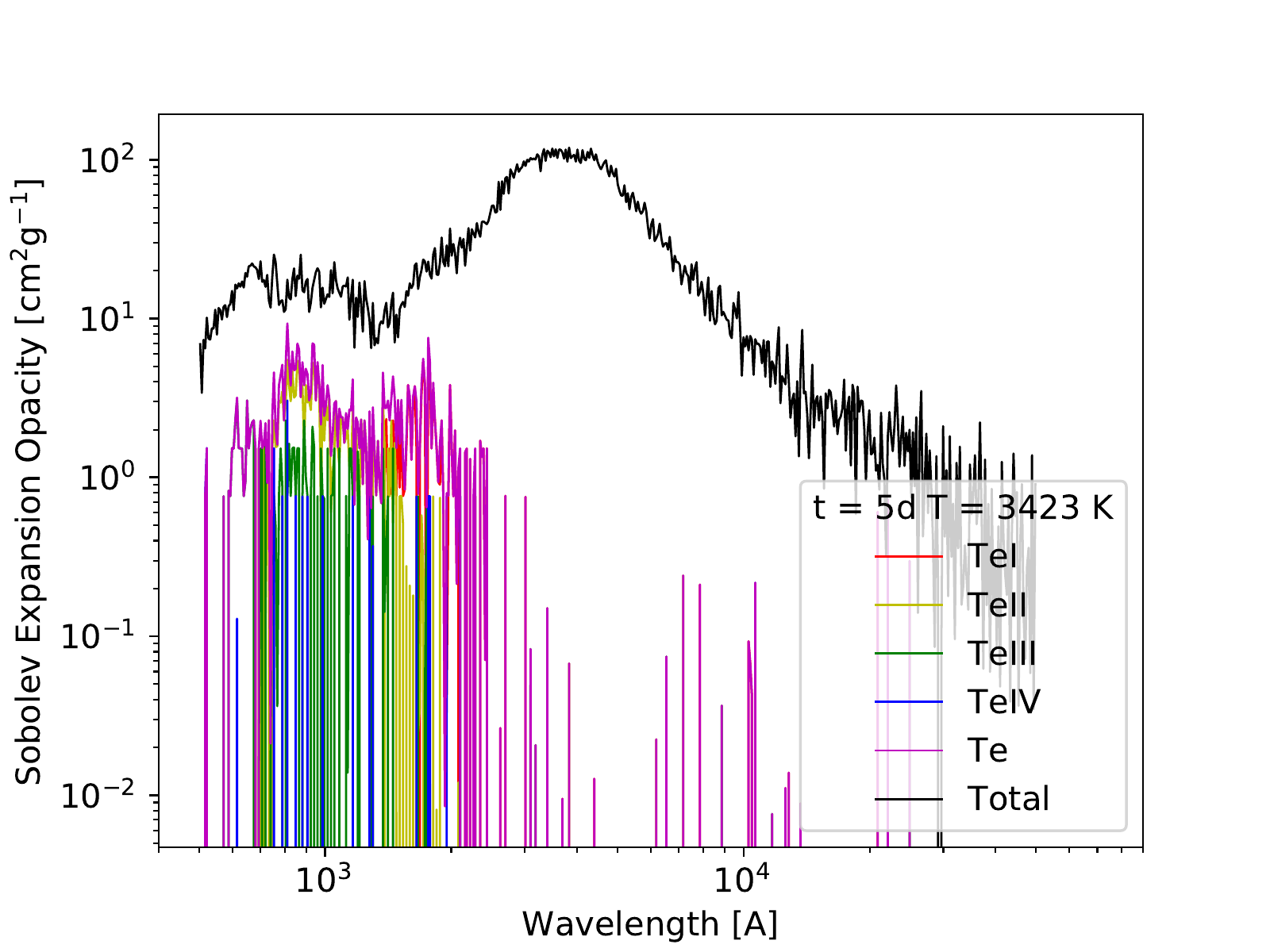}
    \includegraphics[trim={0.cm 0.cm 0.4cm 0.4cm},clip,width = 0.49\textwidth]{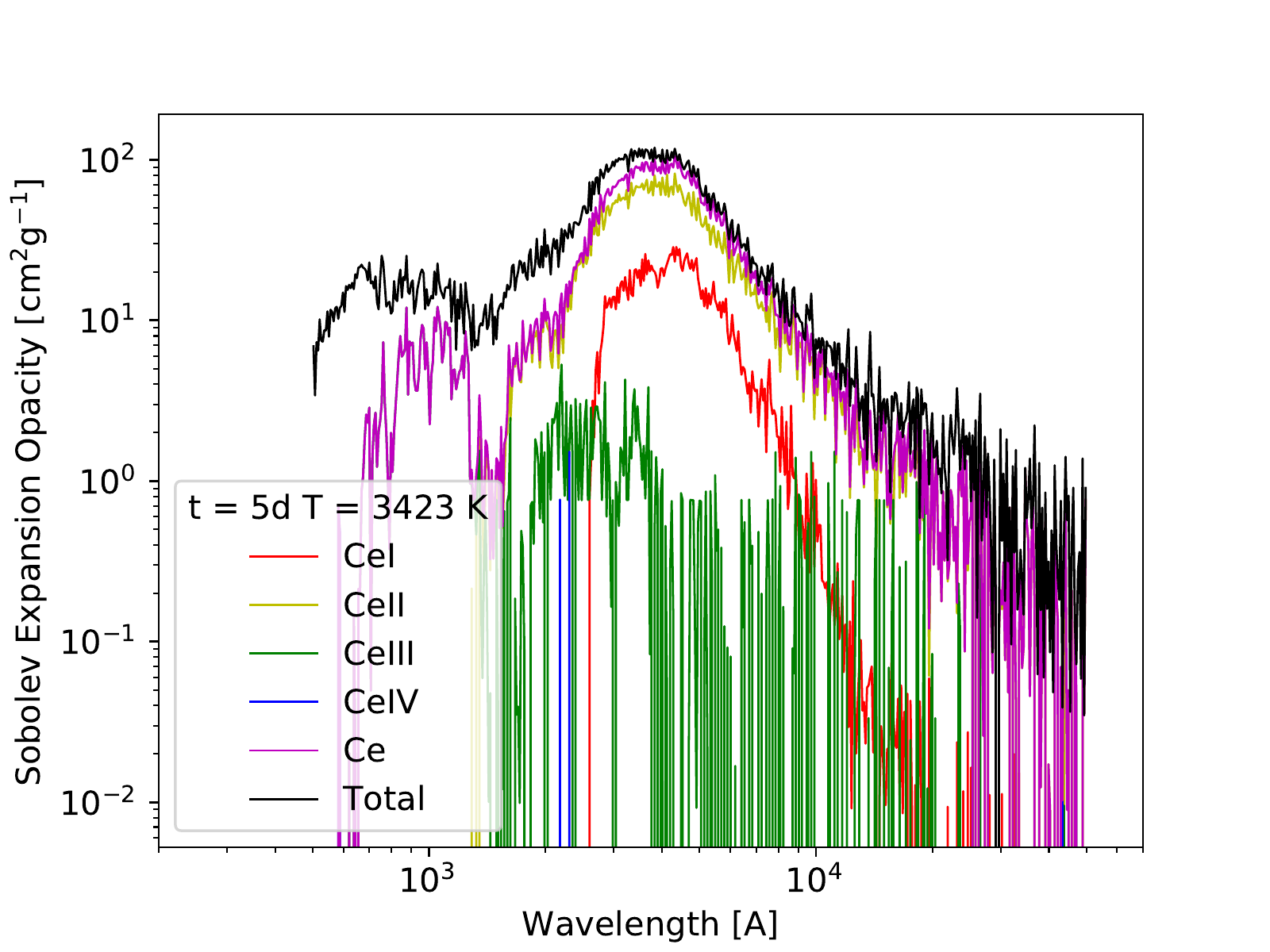}
    \includegraphics[trim={0.cm 0.cm 0.4cm 0.4cm},clip,width = 0.49\textwidth]{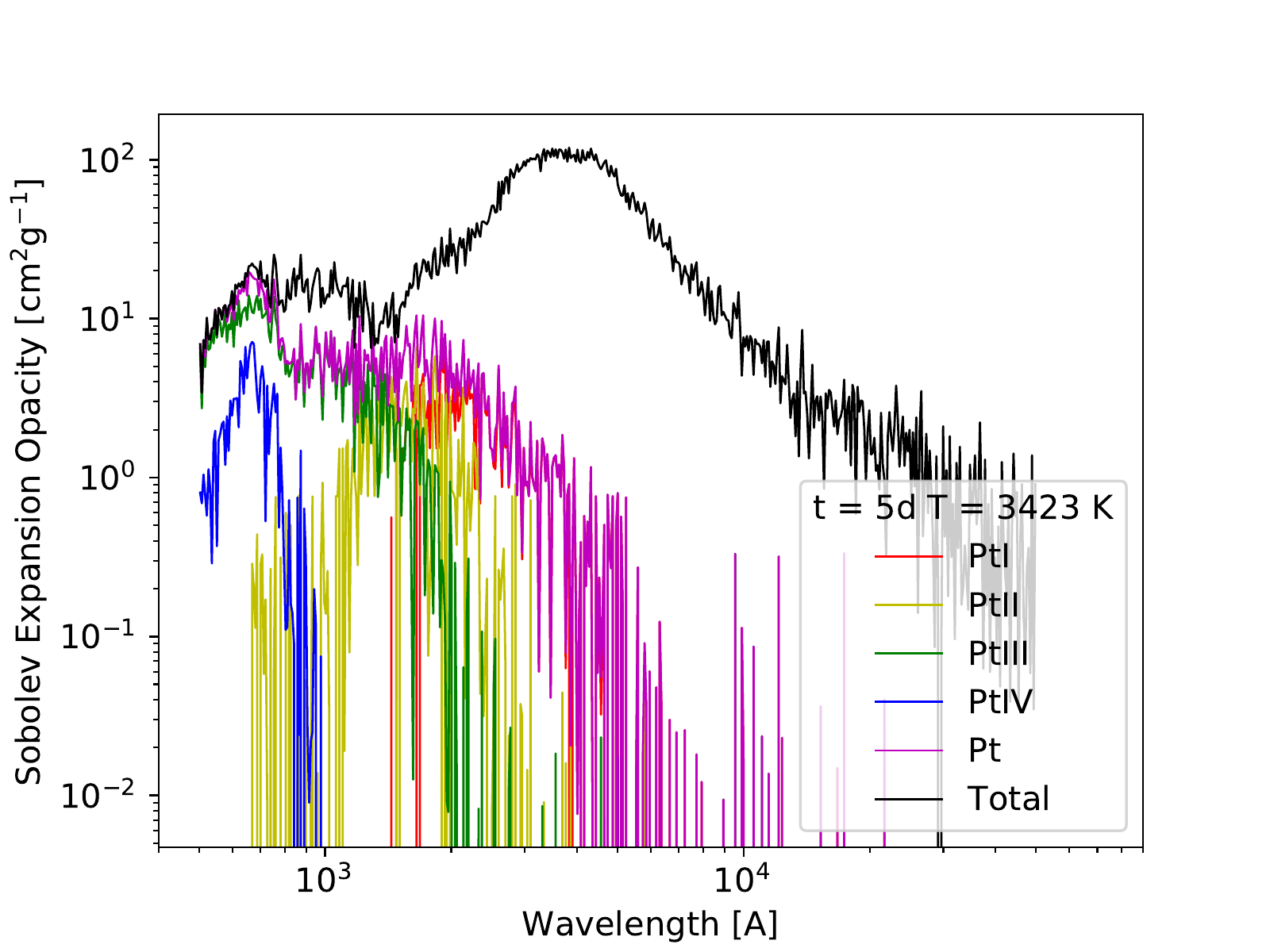}
    \includegraphics[trim={0.cm 0.cm 0.4cm 0.4cm},clip,width = 0.49\textwidth]{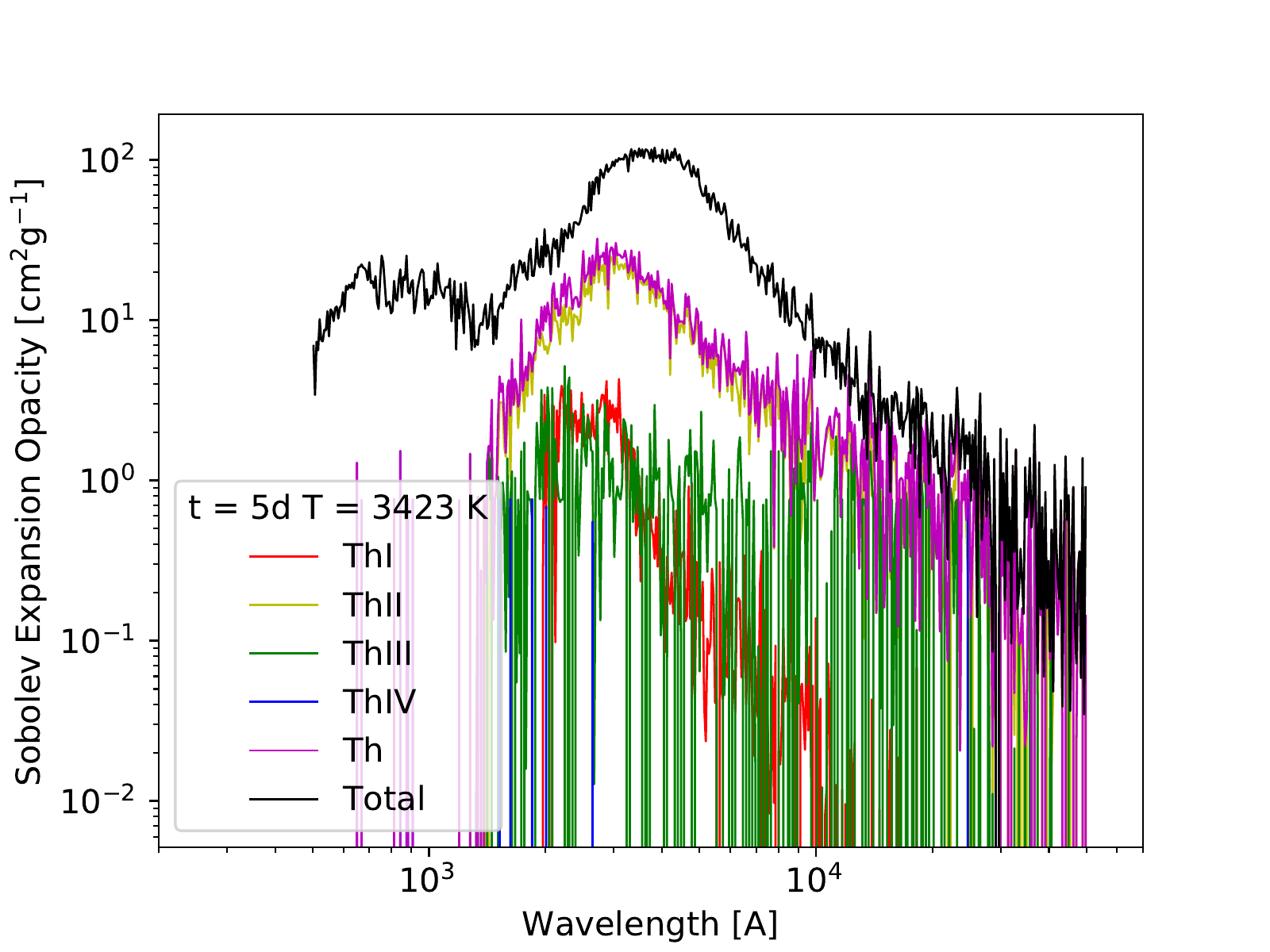}
    \caption{Expansion opacities calculated from the Boltzmann level populations shown in Section \ref{sec:levelpops} for each individual ion, and summed up for each element, at 5 days after merger. The total opacity of the entire limited LTE model is also shown for reference. In this particular model, the ion fractions for each ion, from neutral to triply ionised respectively were: Ce = [0.0026, 0.753, 0.212, 0.034], Te = [0.284, 0.485, 0.196, 0.035], Pt = [0.109, 0.617, 0.238, 0.035], Th = [0.0018, 0.620, 0.294, 0.084].}
    \label{fig:opacities_individual}
\end{figure*}

\section{Expansion Opacities}
\label{sec:opacity}

Expansion opacities are important as they are key to LC and SED modelling, and the inferral of ejecta properties from these. As LTE calculations of expansion opacities are thus far the only ones applied in the literature, it is of interest to investigate the accuracy of this approach, both in terms of ionisation and excitation structure. From the consideration of level populations, it is clear that the LTE approximation is generally only accurate at early times, though deviations rapidly appear for high-lying states in large ions. The full NLTE solution then appears to be better approximated by the limited NLTE structure without the radiative terms. It is interesting to see if this trend is matched by expansion opacity.

In this section, we first calculate the expansion opacities stemming from the excitation structures described in Section \ref{sec:levelpops}. That is, using the full NLTE solution from \textsc{sumo}, the limited NLTE calculation without radiative transfer, and the limited LTE solution using the Boltzmann equation. For each of these calculations, the temperature and ionisation structure are taken from the full NLTE solutions yielded by \textsc{sumo}. As such, this first comparison focuses solely on the effect of the method used to calculate excitation structure in the ejecta.

We then consider a full LTE approach, making use of the Saha and Boltzmann equations to calculate ionisation and excitation structures. The temperature and densities used to obtain the final number densities are from the \textsc{sumo} simulations, in order to allow a direct comparison between the full NLTE and full LTE solutions. The inclusion of a Saha ionisation structure calculation is intended to highlight that LTE conditions not only affect excitation structure, but also ionisation structure. 
Notably, different ions will contribute differently, both in magnitude and wavelength, to the expansion opacity. As such, significant differences in NLTE and LTE ionisation structure may have important consequences for opacity.

In order to understand the evolution of the models' total opacities presented below, and how these relate to the level populations presented in the previous section, it is beneficial to first consider the opacity contributions of individual elements and ions. To this end, the expansion opacities of the individual ions at 5 days after merger, calculated using the limited LTE solutions, are presented in Figure \ref{fig:opacities_individual}. We find that Ce II dominates at $\lambda \gtrsim 1500 \ang$, with some contributions from Ce I and Th II. At shorter wavelengths $\lambda \lesssim 1500 \ang$, we find that Pt III is the most important contributor, though Pt II, Pt IV and Ce III also provide some opacity.

\begin{figure*}
    \centering
    \includegraphics[trim={0.4cm 0.cm 0.4cm 0.4cm},clip,width = 0.49\textwidth]{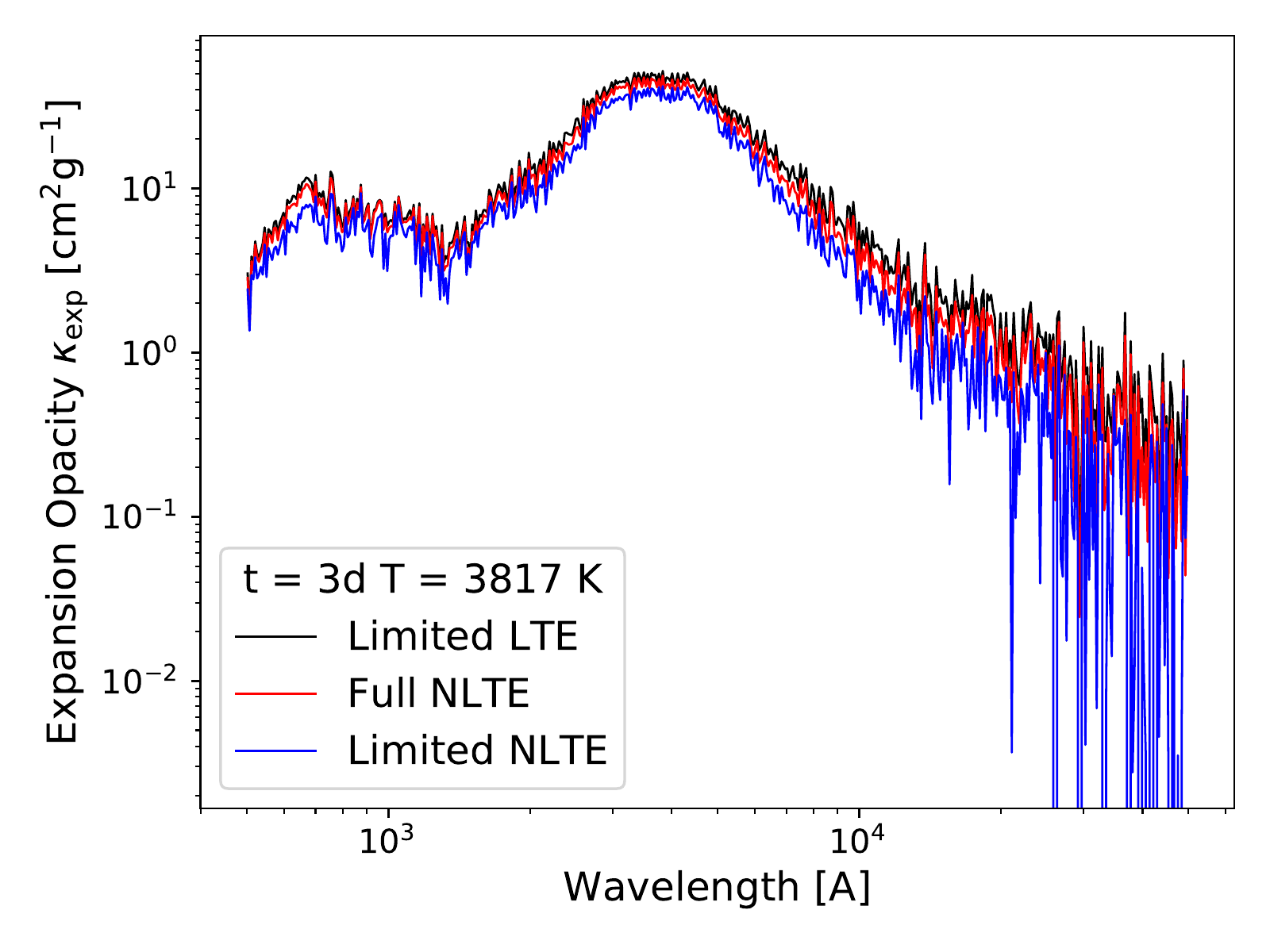} 
    \includegraphics[trim={0.4cm 0.cm 0.4cm 0.4cm},clip,width = 0.49\textwidth]{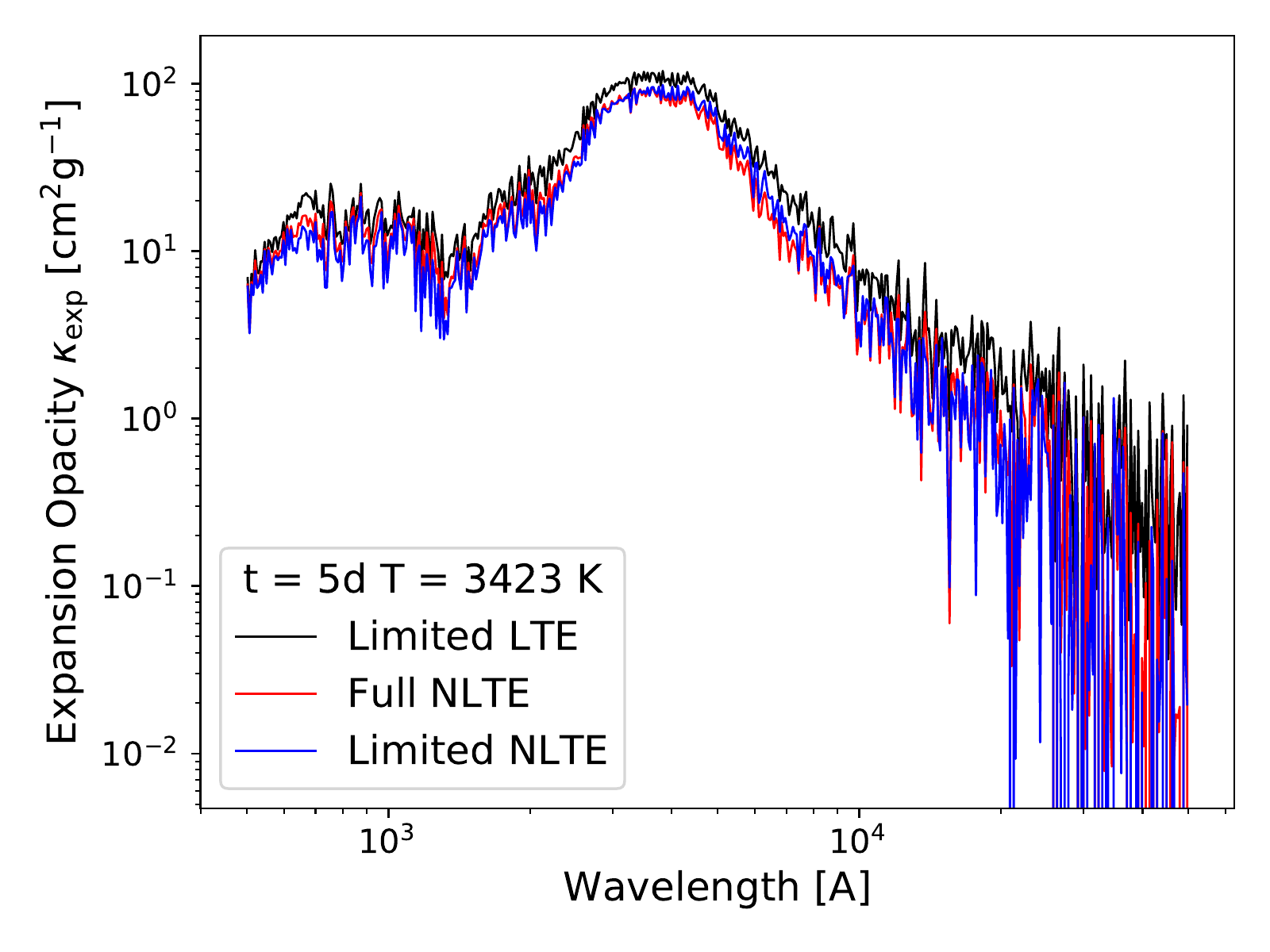}
    \includegraphics[trim={0.4cm 0.cm 0.4cm 0.4cm},clip,width = 0.49\textwidth]{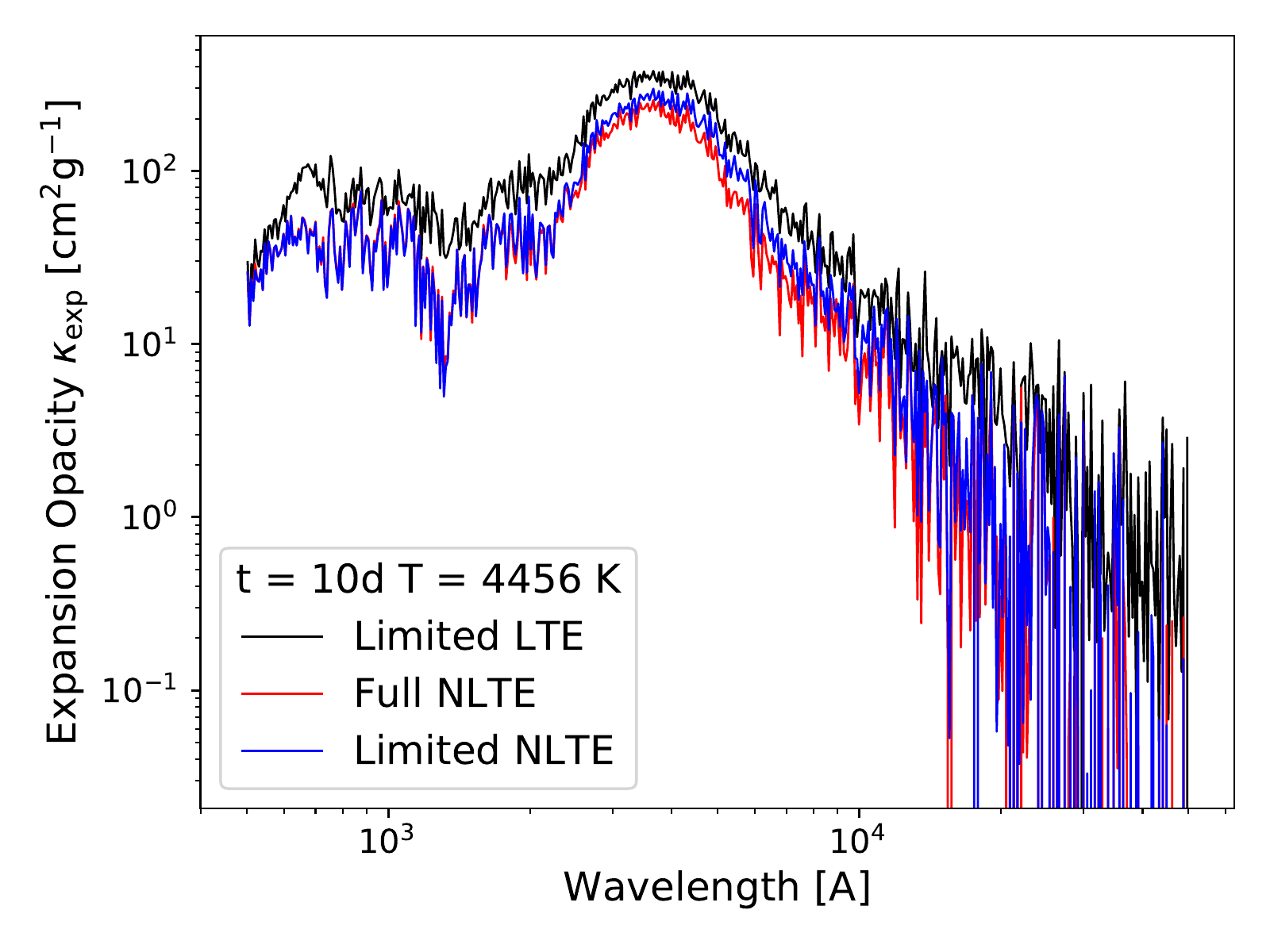} 
    \includegraphics[trim={0.4cm 0.cm 0.4cm 0.4cm},clip,width = 0.49\textwidth]{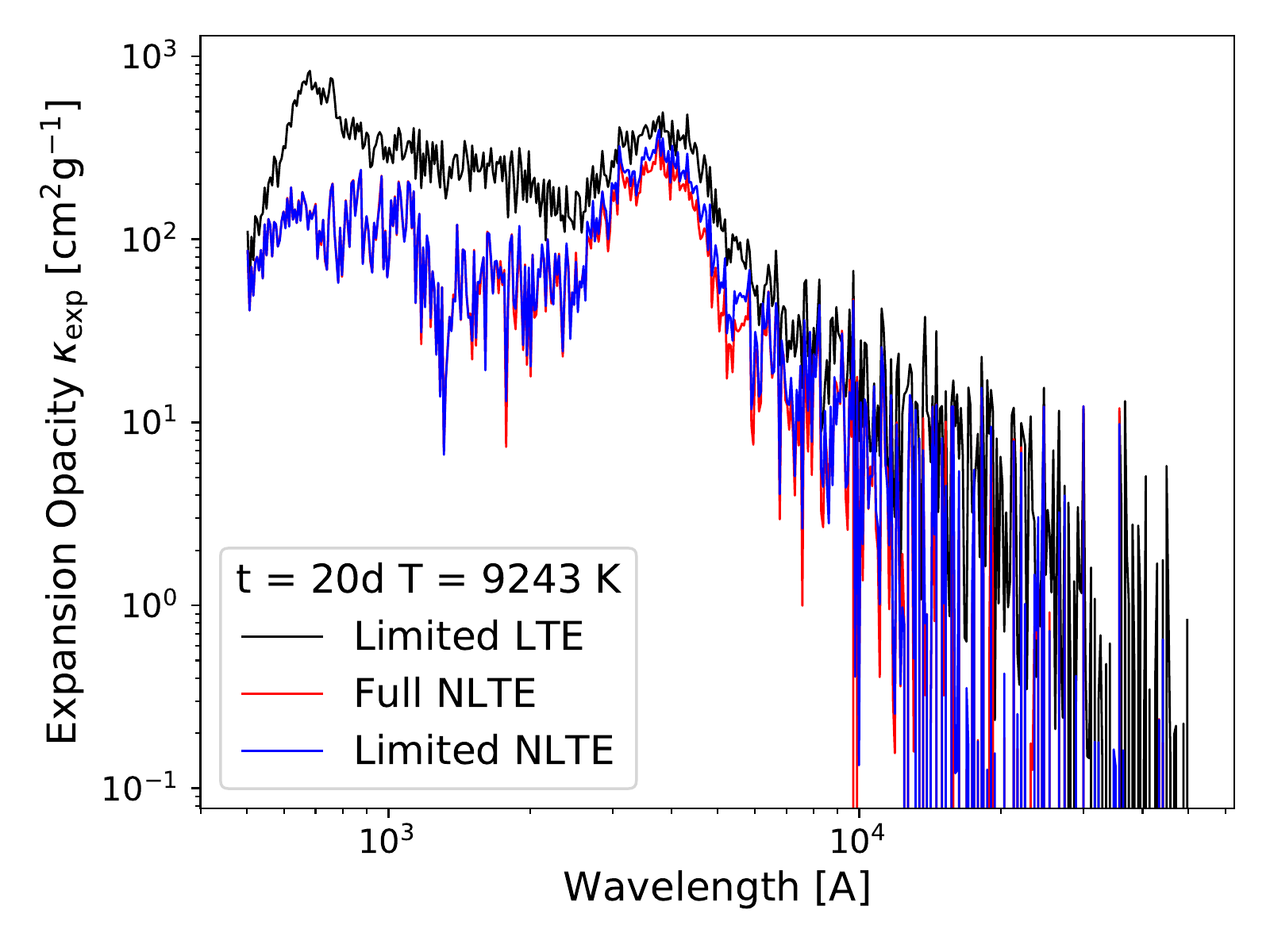}
    \caption{Expansion opacities for the standard model when excitation structure is calculated using the Boltzmann equation (black), the full NLTE calculations (red), and the limited NLTE calculations (blue). The panels show the increasing overestimate of the limited LTE opacity as time moves forwards. The limited NLTE approach better approximates the full NLTE opacities compared to when solely using the Boltzmann equations, especially at late times.}
    \label{fig:opacities_boltzmann}
\end{figure*}

Our elemental opacities appear consistent to those in the literature \citep[see e.g.][]{Kasen.etal:13,tanaka:opacities:2020}. In particular, we compare our exact values for Te, Ce and Pt in full LTE to those of \citet{tanaka:opacities:2020} using the parameters described therein ($t = 1$ day, $\rho = 10^{-13} \mathrm{g \; cm^{-3}}$, $T = 5000 $K), and find that our values to be consistent, within a factor of two at every wavelength considered, well within the intrinsic uncertainty stemming from atomic data.

\subsection{Effect of Excitation Structure on Expansion Opacity}
\label{subsec:boltz_opacity}

We consider first the effect the choice of excitation structure calculation has on expansion opacity, using the calculations presented in the previous section. Here, only excitation structure changes; temperature and ionisation structure are identical between all calculations at the same epoch. Figure \ref{fig:opacities_boltzmann} shows how the total expansion opacities of the model evolve with time, when calculated from the limited LTE, limited NLTE and full NLTE approaches respectively.

\subsubsection{Limited LTE and Full NLTE Opacities}

Considering first the general evolution of the limited LTE and full NLTE opacities, we find that they are quite similar, with both increasing with time, driven by the $1/\rho \propto t^3$ term in Equation \ref{eq:opacity}. As somewhat expected from the evolution of the excitation structures, the opacities are all essentially identical at 3 days after merger, but then slowly diverge as time goes on. The limited LTE opacities are higher than the full NLTE opacities over the whole wavelength range, increasingly so as time progresses. Notably, from 10 days onwards, the limited LTE values are greater than the full NLTE values by a factor of 2-3 over the range of wavelengths considered. By 20 days after the merger, the LTE values are almost an order of magnitude greater than the full NLTE opacities at certain wavelengths.

We now consider how the expansion opacities' evolution is linked to that of the level populations, taking Ce II as the sample ion due to its large size and important opacity contribution at $\lambda \gtrsim 1500 \ang$. Initially at 3 days after merger, the full NLTE excitation structure is essentially that of LTE, with high-lying states, i.e. group 3, being slightly less populated due to spontaneous radiative decay. Correspondingly, the LTE opacities are just slightly higher than the NLTE values (top left panel of Figure \ref{fig:opacities_boltzmann}). As time progresses, the NLTE excitation structure depopulates the group 3 states, in favour of the group 1 levels, which can be seen clearly in the bottom right panel of Figure \ref{fig:CeII_levelpops}. We then have group 1 states being overpopulated relative to LTE, and groups 2 and 3 being underpopulated. We thus have a competing effect on opacity from group 1 against groups 2 and 3. Since we see that the full NLTE opacity at $\lambda \gtrsim 1500 \ang$ decreases relative to the LTE values as time progresses, we find that the overestimation of group 2 and 3 populations by LTE is more important than the underestimation of group 1 populations.

The relative importance of group 2 vs 3 is difficult to ascertain clearly here, as although group 2 is more populated, group 3 has many more levels and thus one could theorise that its contribution to opacity may sum up to be equal to or greater than group 2's contribution. In reality, group 3 appears to have very little contribution to opacity due to the sparseness of the level populations, though this is better seen when considering the limited NLTE opacities, which we do below in the next subsection.

\begin{figure*}
    \centering
    \includegraphics[trim={0.2cm 0.cm 0.4cm 0.4cm},clip,width = 0.49\textwidth]{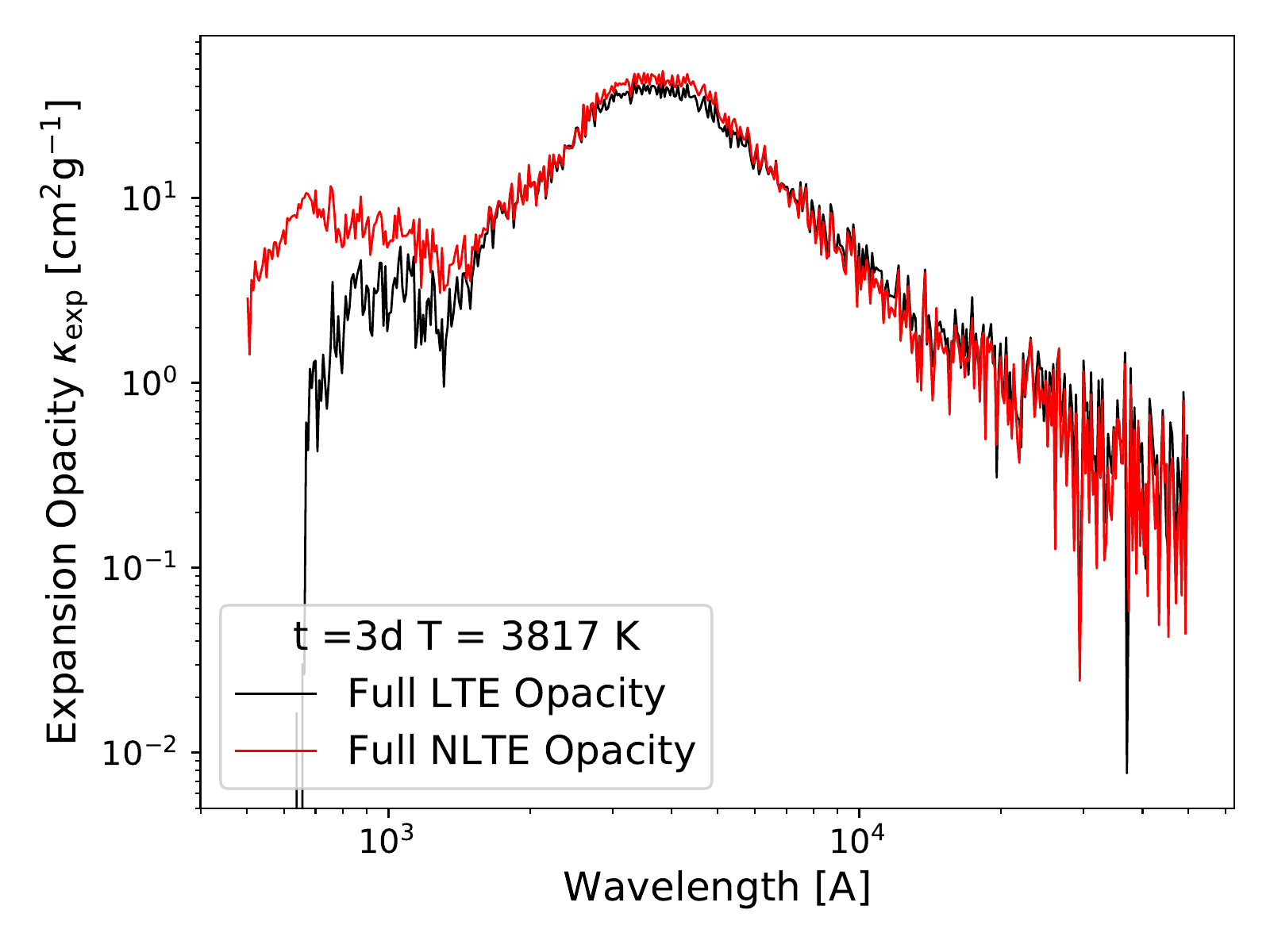} 
    \includegraphics[trim={0.2cm 0.cm 0.4cm 0.4cm},clip,width = 0.49\textwidth]{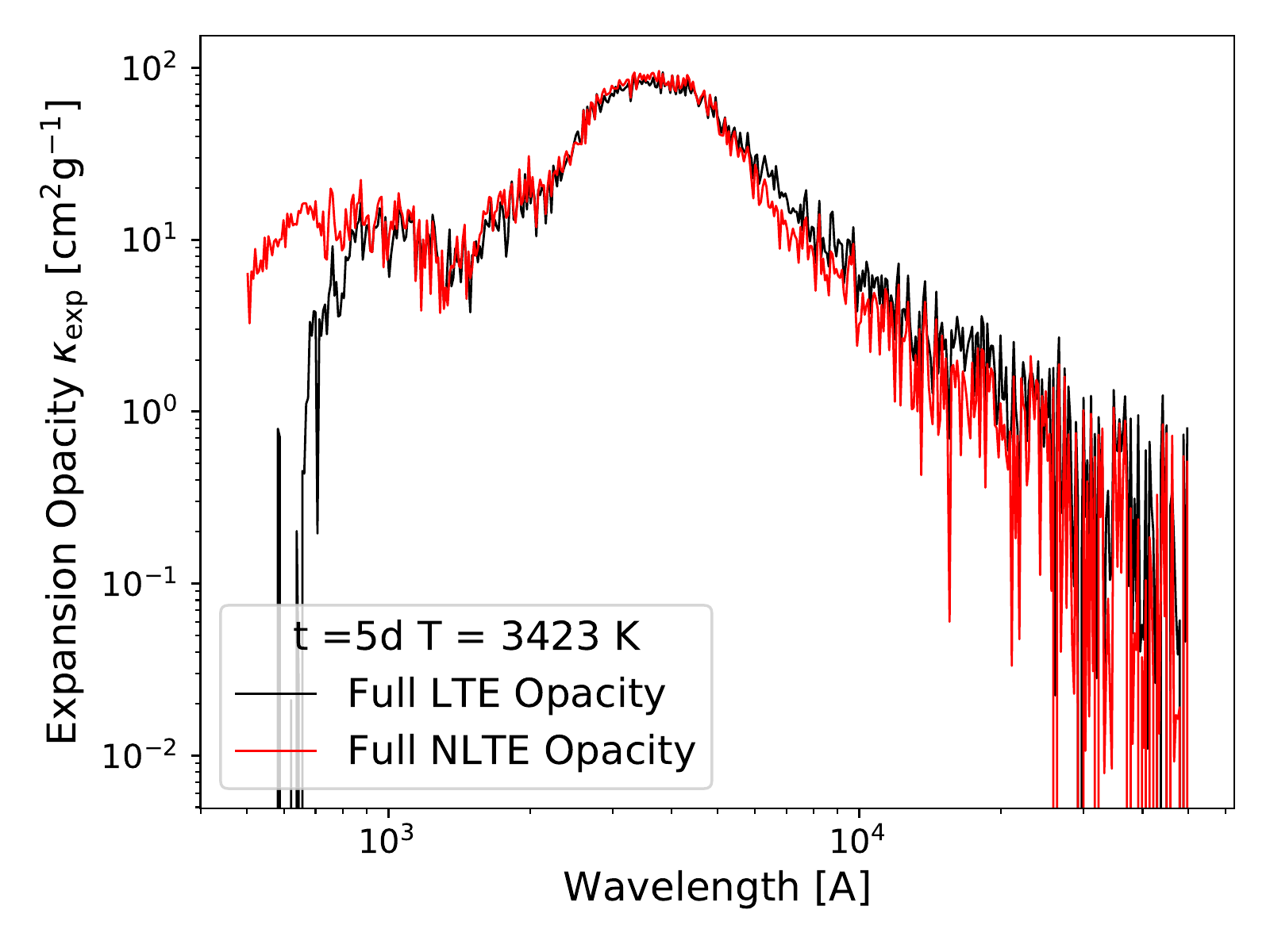}
    \includegraphics[trim={0.2cm 0.cm 0.4cm 0.4cm},clip,width = 0.49\textwidth]{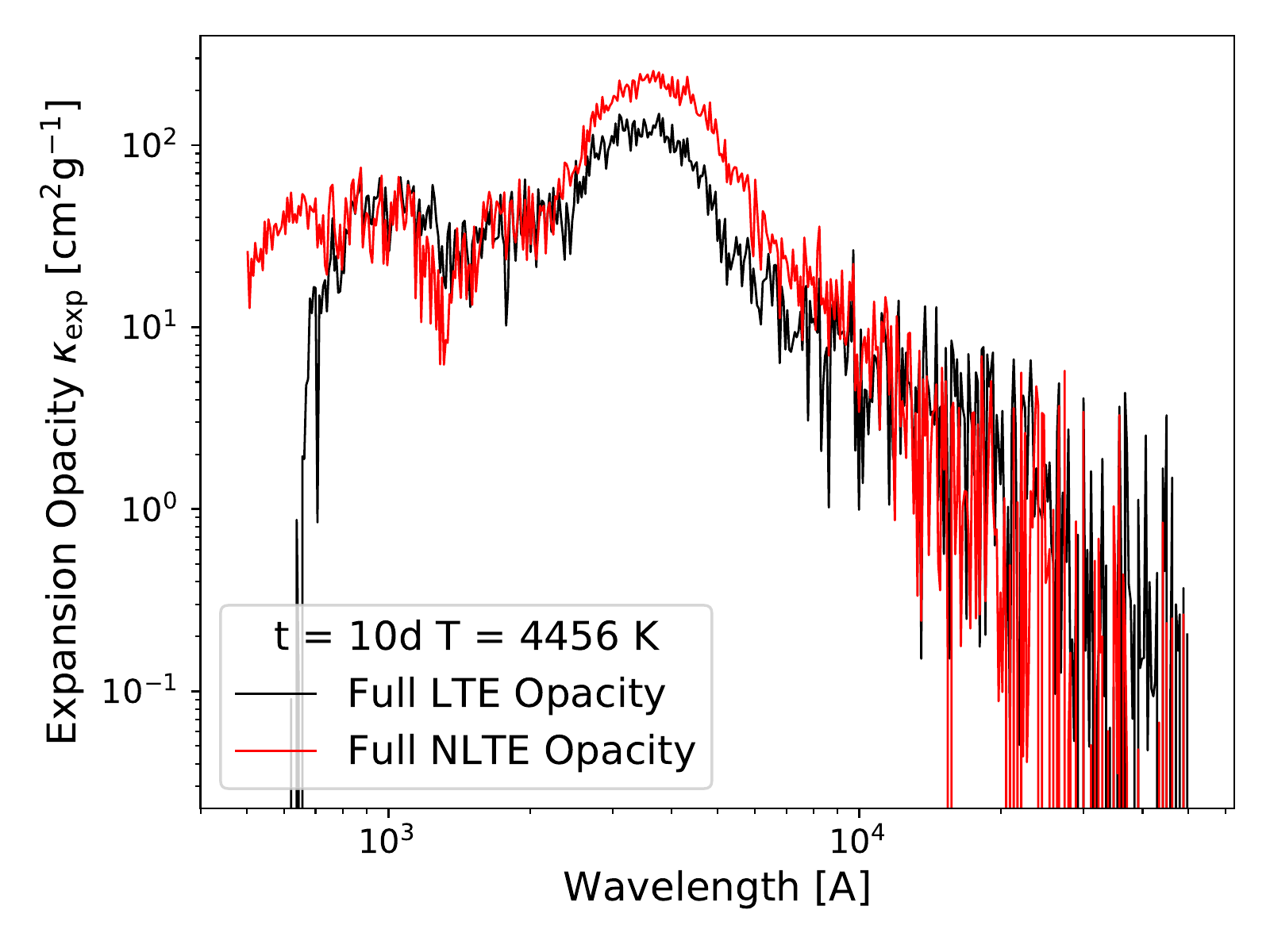}
    \includegraphics[trim={0.2cm 0.cm 0.4cm 0.4cm},clip,width = 0.49\textwidth]{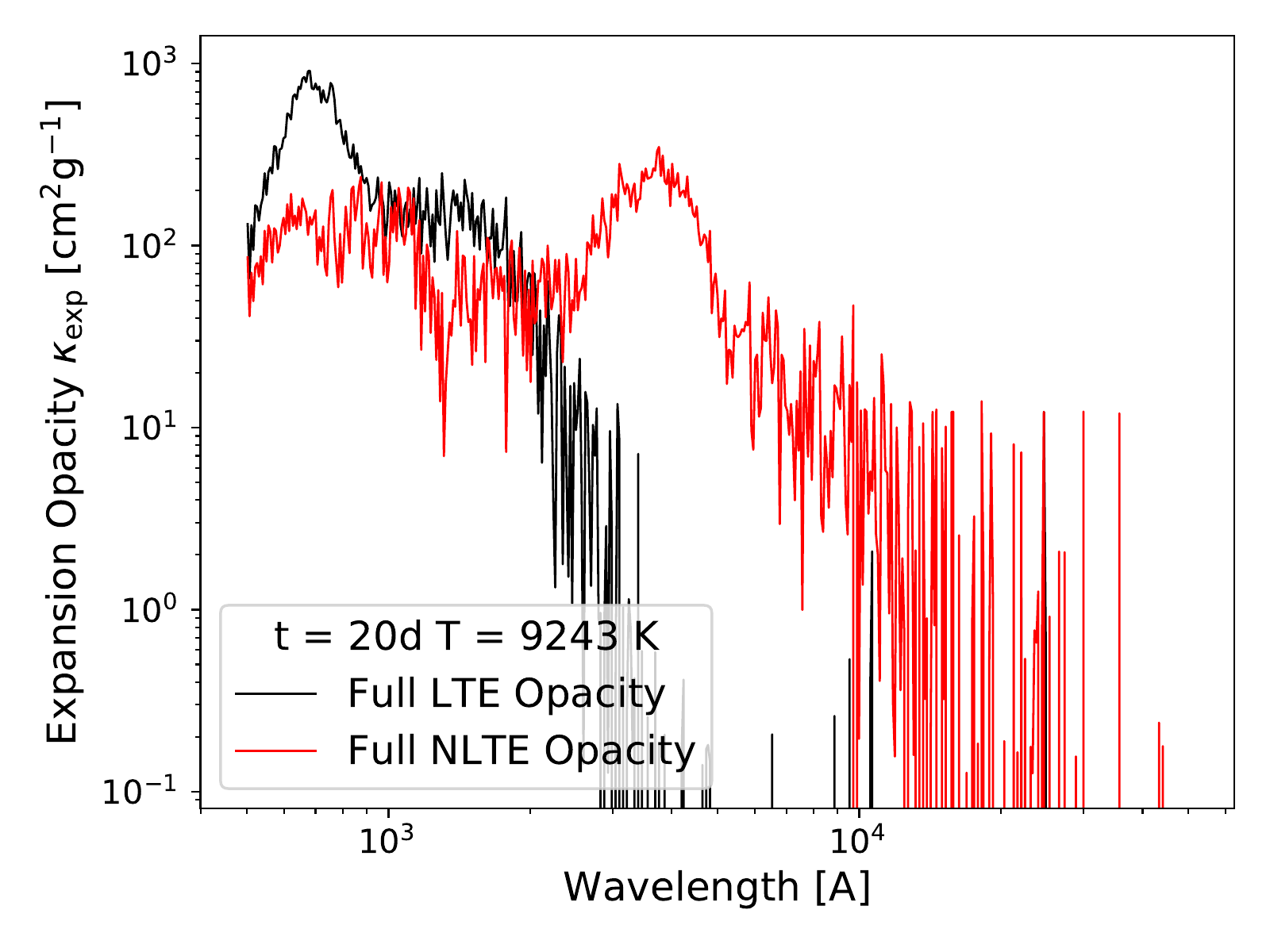}
    \caption{Expansion opacities for the models when employing a full NLTE approach compared to that of a full LTE approach using the Saha and Boltzmann equations. Note that the temperature solutions are from the full NLTE solutions.}
    \label{fig:opacities_saha}
\end{figure*}

A similar analysis can be conducted on Pt III for $\lambda \lesssim 1500 \ang$, though the group 3 states ($n \gtrsim 40$) are so sparsely populated that they only appear in figure (\ref{fig:PtIII_levelpops}) at 20 days after merger (bottom right panel). Thus, considering solely groups 1 ($n \sim 1 - 25)$ and 2 ($n \sim 25 - 40$) for Pt III, we see a somewhat different story than for Ce II. Initially the group 1 states are very close to the limited LTE values, if slightly lower at 5 days after merger. Conversely, the group 2 populations in full NLTE are greater than the LTE values up to 5 days after merger. We therefore again have a competing effect on opacity by these two level groups. Looking at the small wavelength opacity at 5 days in figure (\ref{fig:opacities_boltzmann}), we see that the limited LTE values are only slightly greater than the full NLTE values, implying that for Pt III, the group one levels are slightly more dominant than the group 2 (or sparsely populated group 3) states. At 10 and 20 days after merger, we now have the vast majority of group populations severely overestimated by the limited LTE solution, with the exception of the first $\sim 3$ levels. Correspondingly, we see a large jump in the limited LTE opacity at $\lambda \lesssim 1500 \ang$ at these epochs, which is not seen in the full NLTE opacity. 

From the above considerations, we find that the expansion opacity is highly dependent on the population of lower lying states (i.e. groups 1 and 2) with large populations, even should they represent only a small fraction of the ions' total levels. While group 3 has the most levels, these are typically sparsely populated compared to groups 1 and 2, and thus contributes the least to opacity. As such, the effect of the limited LTE approach to expansion opacities will depend on the balance of over/underestimation between these two groups. 

\subsubsection{NLTE Opacities without Radiation Field}

We now turn our attention to the expansion opacities calculated from the limited NLTE level populations, given by the blue curves in Figure \ref{fig:opacities_boltzmann}. The exact values of the limited NLTE opacities compared to the other two methods varies a bit depending on epoch, initially being slightly lower at 3 days, and moving in between full NLTE and LTE values by 10 days. However, the main result from this calculation is how similar the limited NLTE opacities are to the full NLTE opacities at every epoch tested here. For all wavelengths considered here, the limited NLTE values lie within less than a factor of two to the full NLTE values. Notably, the deviations seen in the LTE opacities for small wavelengths, stemming from Pt III, are not seen in the limited NLTE case. Therefore, we find that using limited NLTE opacities instead of LTE opacities at late times ($t \gtrsim 10$ days), for this ejecta model, yields a highly significant improvement to accuracy at all wavelengths.

\begin{figure*}
    \centering
    \includegraphics[trim={0.cm 0.cm 0.4cm 0.4cm},clip,width = 0.49\textwidth]{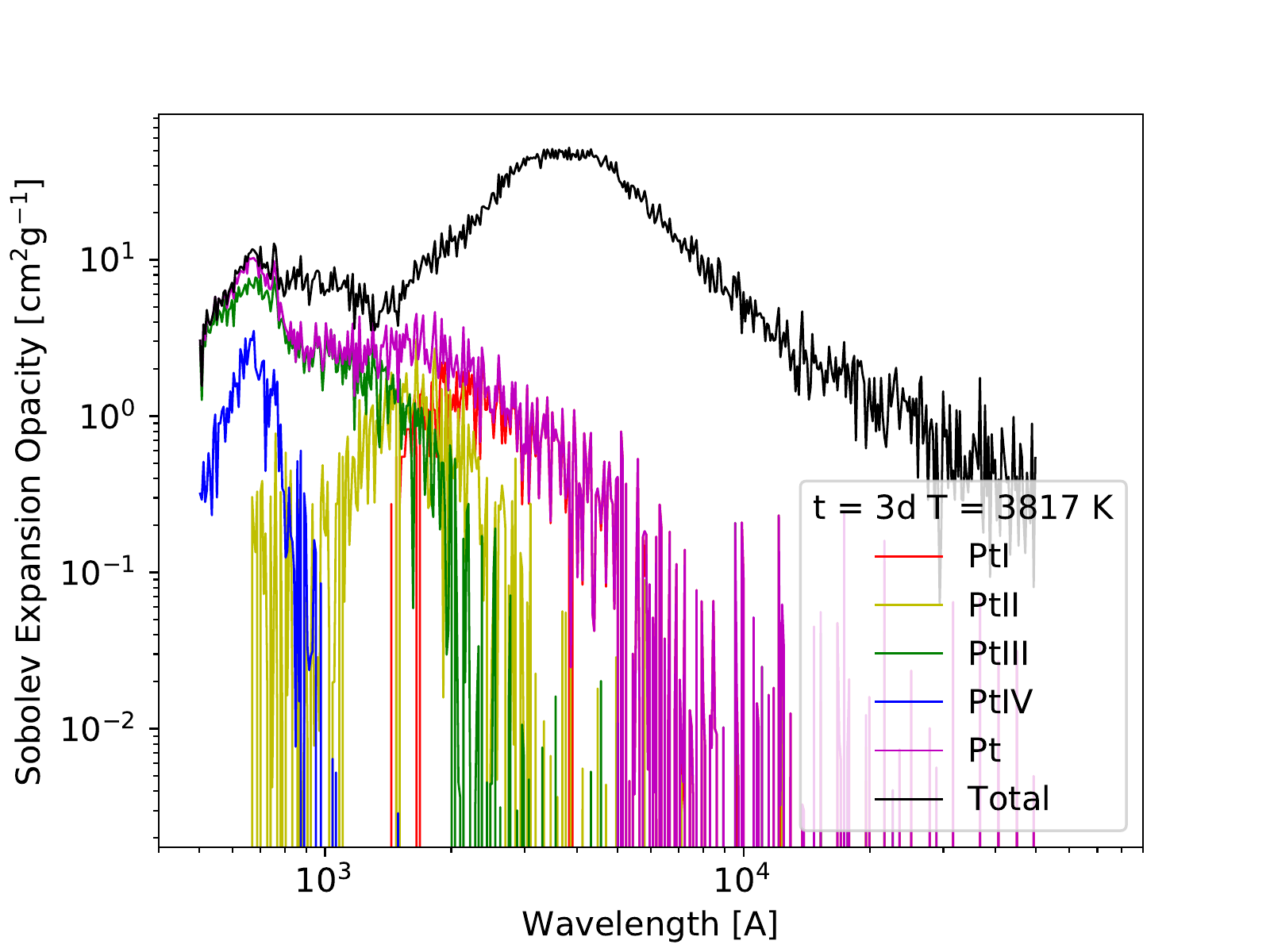}
    \includegraphics[trim={0.cm 0.cm 0.4cm 0.4cm},clip,width = 0.49\textwidth]{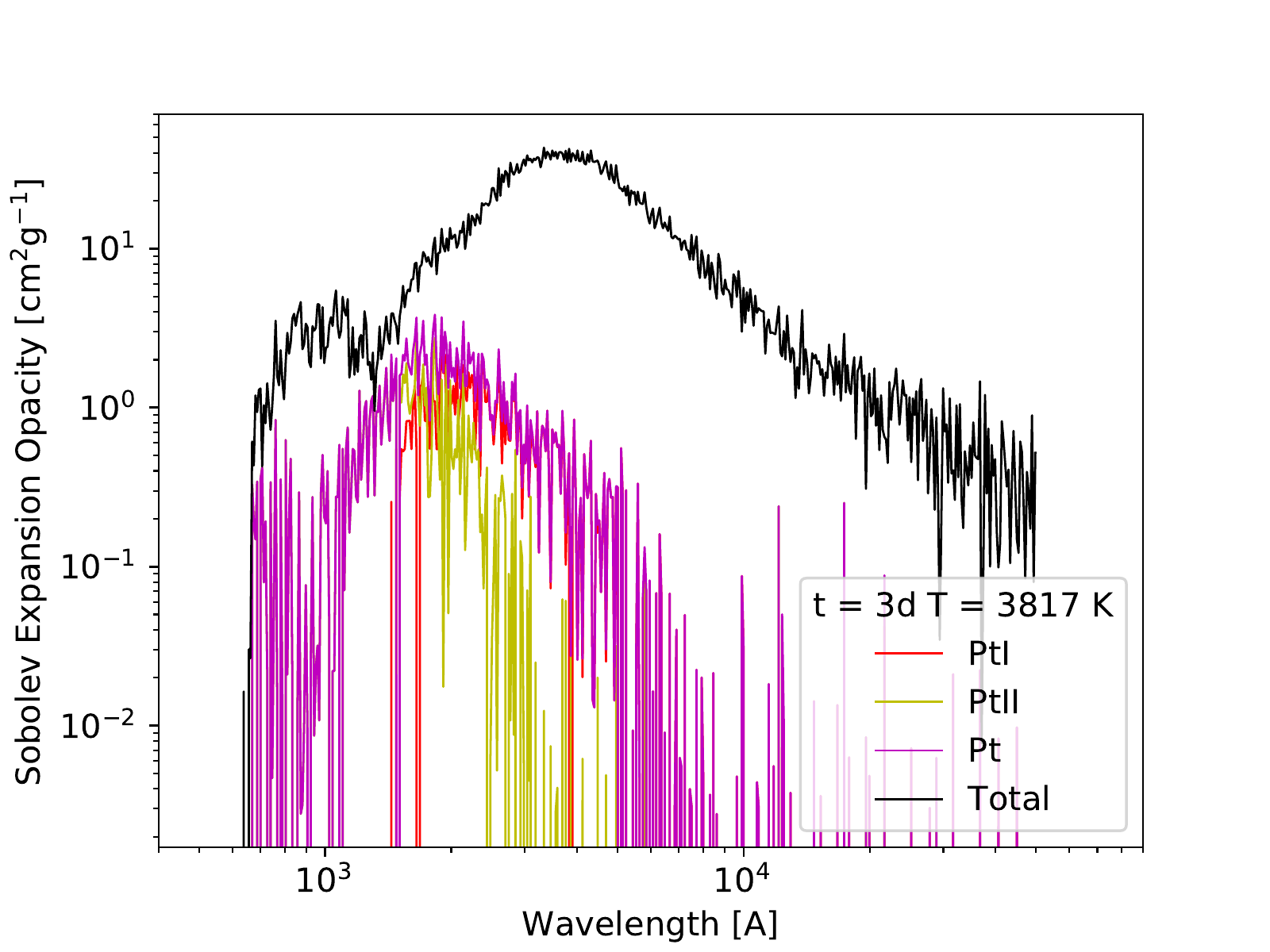}
    \includegraphics[trim={0.cm 0.cm 0.4cm 0.4cm},clip,width = 0.49\textwidth]{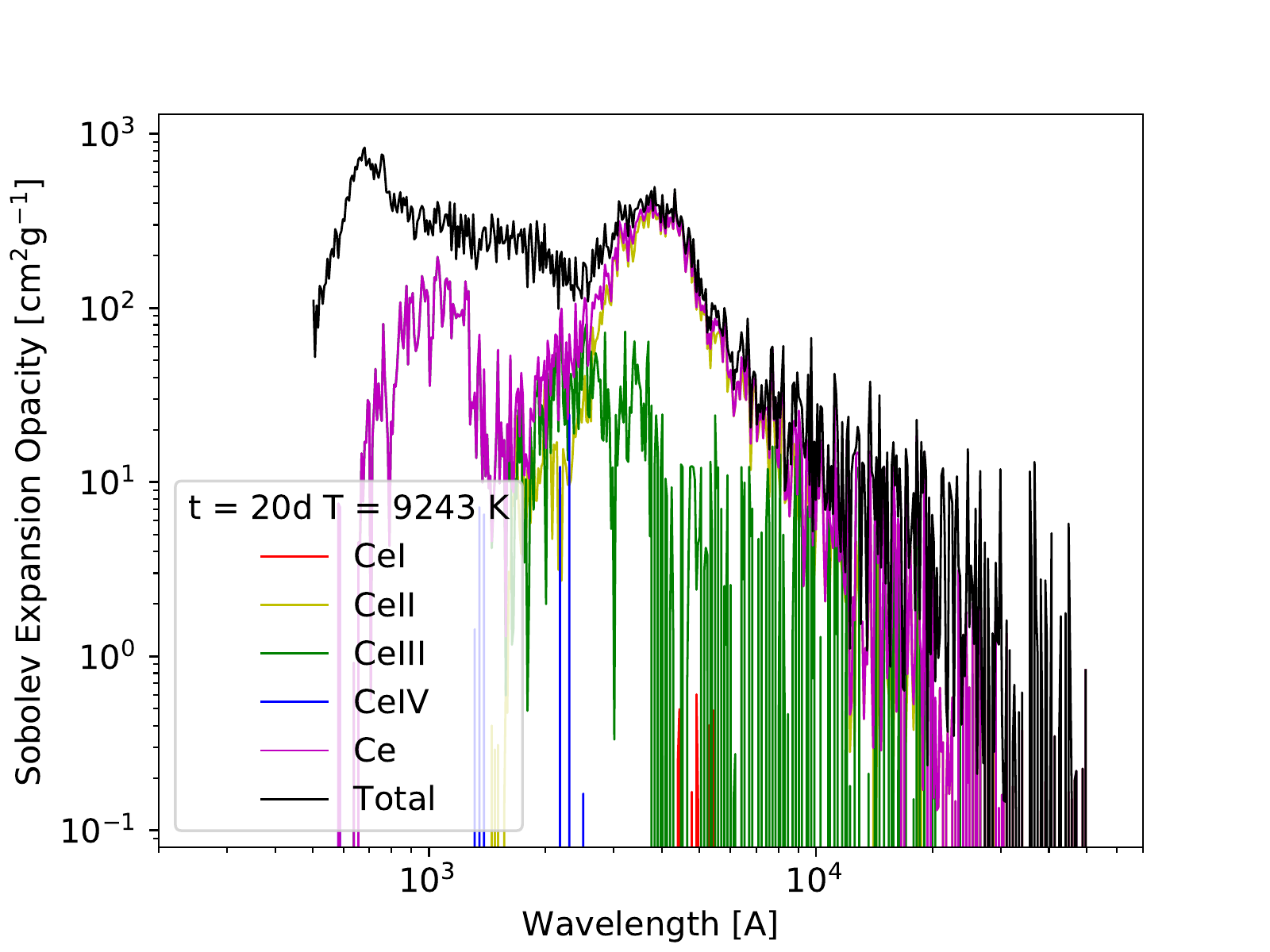} 
    \includegraphics[trim={0.cm 0.cm 0.4cm 0.4cm},clip,width = 0.49\textwidth]{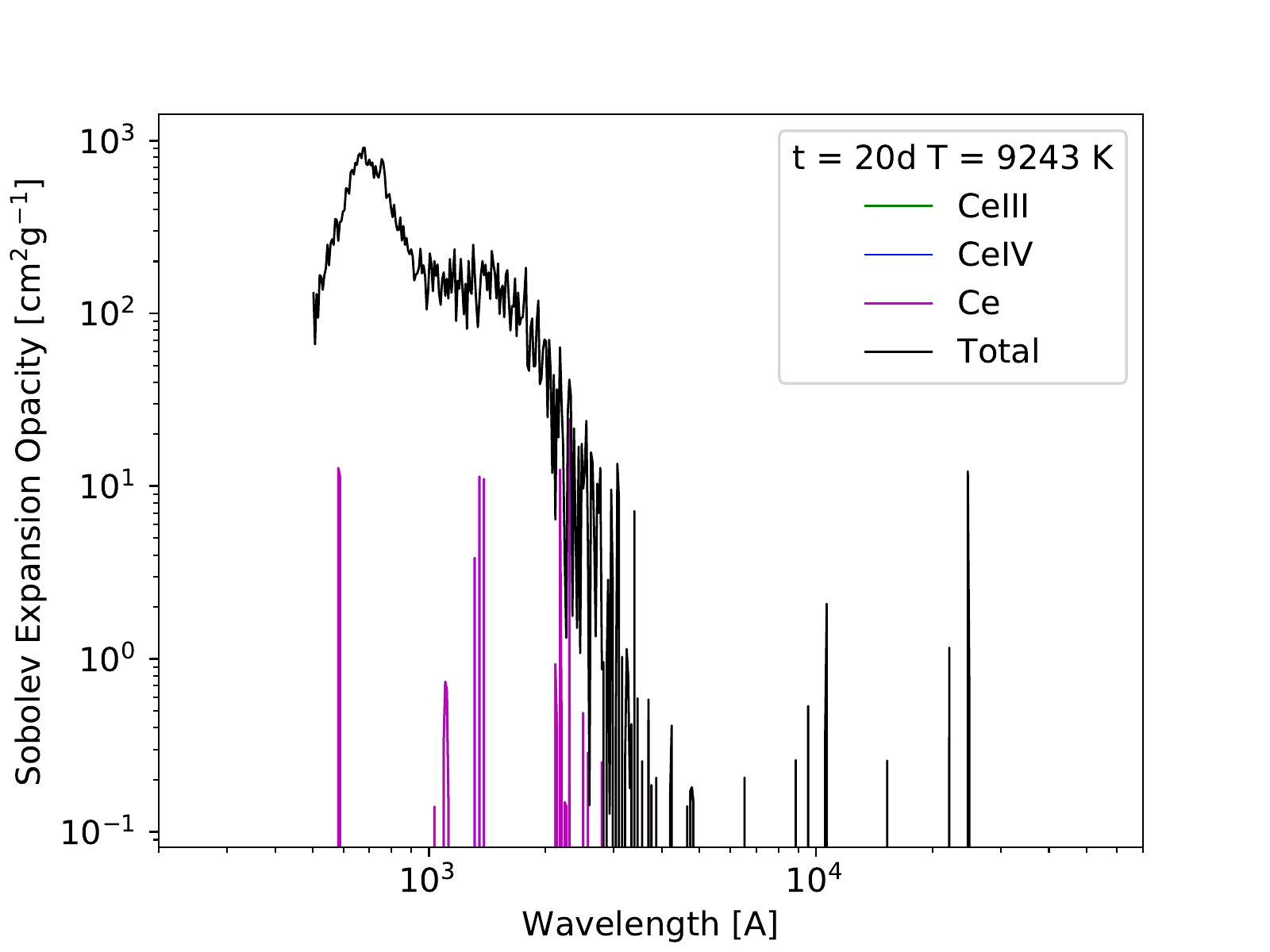}
    \caption{Expansion opacities of Pt (top) and Ce (bottom) at 3 days and 20 days after merger respectively. The left column shows the values from the calculations using the limited LTE model, and the right column shows the values when using the full LTE model. Ions with no abundance are not plotted.The total curves in the left panels are the limited LTE opacities, and the total curves in the right panels are the full LTE opacities.}
    \label{fig:opacities_select}
\end{figure*}

The strong similarity between full and limited NLTE opacities can be explained by following the same analysis as before with level population groups. We consider first the level populations at 3 days, when the limited NLTE opacity slightly underestimates the true values. Looking Ce II in Figure \ref{fig:CeII_levelpops}, top left panel, we see that the group 1 and 2 populations of the limited and full NLTE solutions are practically identical. However, the limited NLTE group 3 populations are drastically smaller; collisional excitation struggles to compete with allowed radiative transitions downwards. Group 3 contributes relatively little to opacity however, and thus we only see a slight underestimation for $\lambda \gtrsim 1500 \ang$ dominated by Ce II. The Pt III populations at 3 days after merger are shown in the top left panel of Figure \ref{fig:PtIII_levelpops}, and we see that the limited NLTE populations for group 1 are slightly lower than in the full NLTE solution. Thus, the limited NLTE opacity at early times and small wavelengths is also slightly underestimated. 

As time progresses however, we see from Figures \ref{fig:CeII_levelpops} and \ref{fig:PtIII_levelpops} that the limited NLTE approach becomes a better approximation of full NLTE than the LTE approach, especially from 10 days onwards. At 20 days after merger, the limited NLTE opacity in the Ce II dominated regime is very slightly larger than the full NLTE values, corresponding to the higher group 2 population seen in the bottom right panel of Figure \ref{fig:CeII_levelpops}. The small wavelengths dominated by Pt III however are almost identical, showing that group 3 populations are too small to significantly influence opacity, since group 1 and 2 populations are essentially the same for the full and limited NLTE solutions, as seen in the bottom right panel of Figure \ref{fig:PtIII_levelpops}. 

\subsection{Ionisation Structure Effect on Opacity}
\label{subsec:saha_opacity}

In the previous section, we examined the effect of excitation structure solutions on the expansion opacity, finding that the limited LTE solution yielded opacities similar to the full NLTE values up to about 5 days after merger. We now consider a full LTE approach, where the ionisation structure is calculated from the Saha equation, and the excitation structure from the Boltzmann equation. As shown in Figure \ref{fig:opacities_individual}, different ions of an element contribute very differently to the opacity. As such, should the Saha equation yield an ionisation structure notably different to that of the full NLTE solution, the expansion opacity may be significantly impacted.

Figure \ref{fig:opacities_saha} shows the evolution of the full LTE opacities compared to the full NLTE opacities from 3 to 20 days after merger. In broad terms, the full LTE opacity is more similar to the NLTE values at the early epochs when compared to the later epochs, though large differences at select wavelengths are noticeable at every epoch. Starting at 3 days after merger, we see that the LTE opacities closely track the NLTE values for wavelengths $\lambda \gtrsim 1000 \ang$. This agreement is quite remarkable, given that thermal collisional ionisation and three-body recombination play no role in the \textsc{sumo} simulations. We will return to this point below.

\begin{figure*}
    \centering
    \includegraphics[trim={0.2cm 0.1cm 0.2cm 0.2cm},clip,width = 0.49\textwidth]{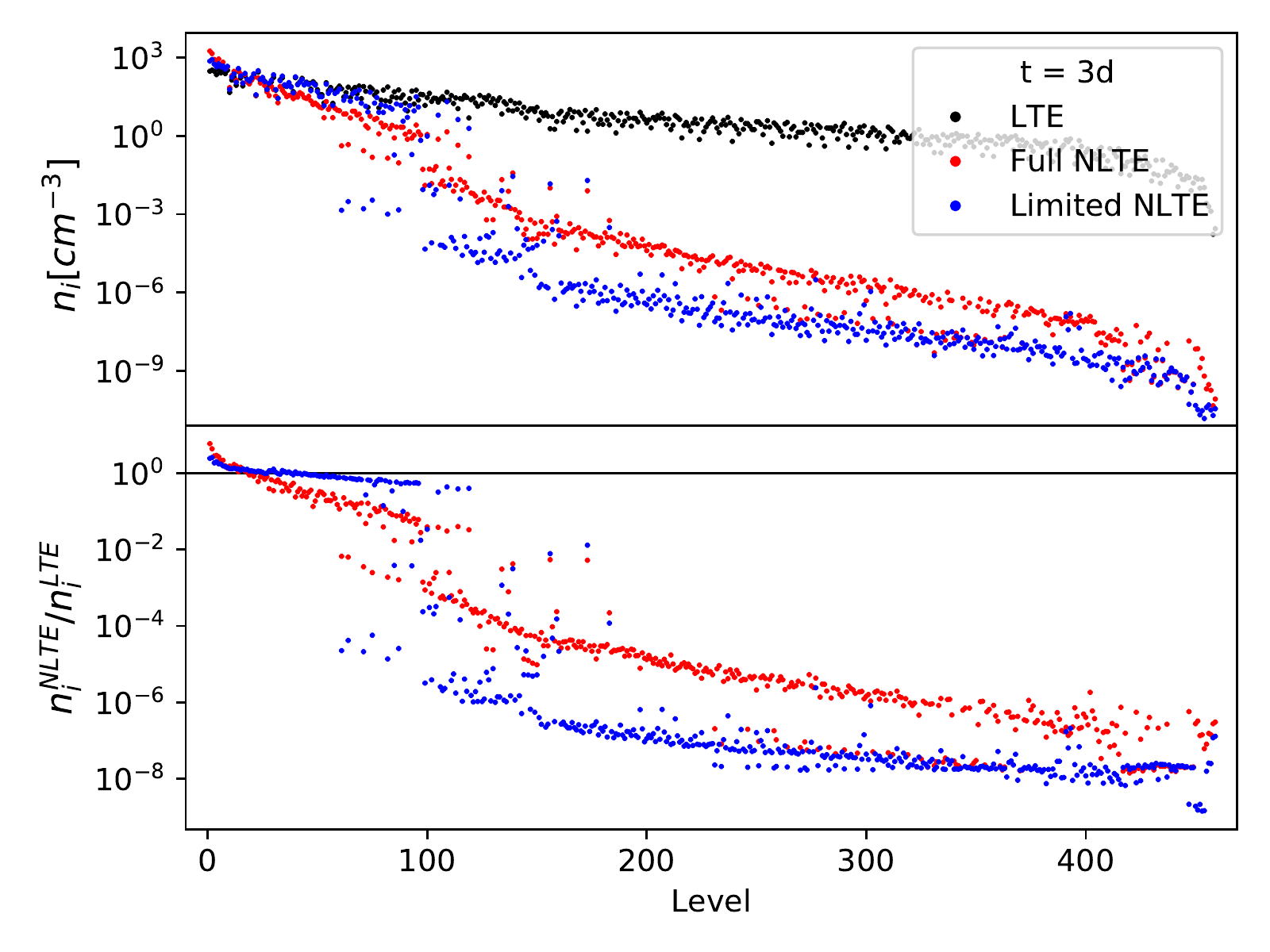} 
    \includegraphics[trim={0.2cm 0.cm 0.4cm 0.4cm},clip,width = 0.49\textwidth]{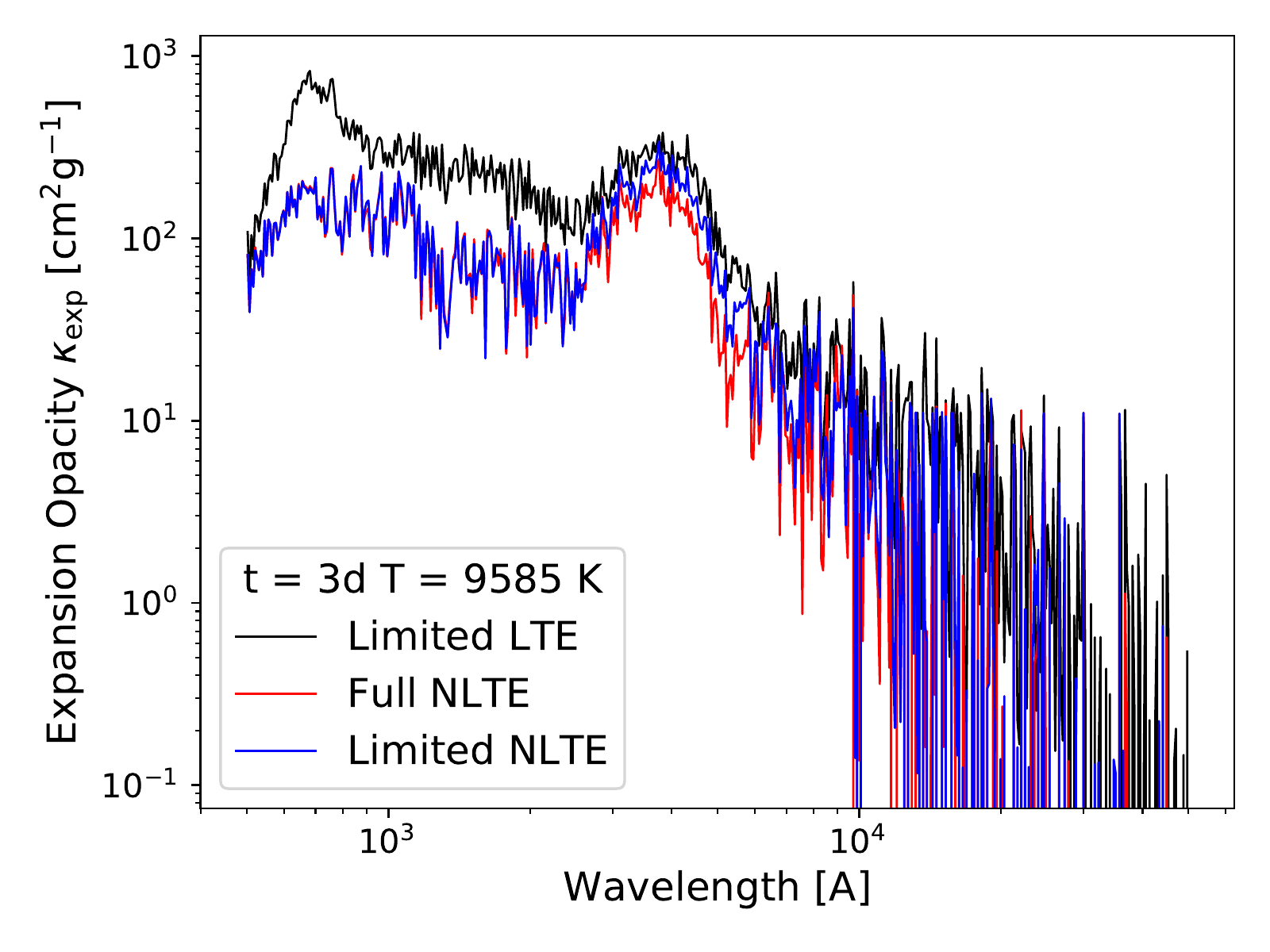}
    \caption{Left panel: Level populations of Ce II for the low density model at 3 days after merger. Right panel: expansion opacities for the low density model at 3 days after merger.}
    \label{fig:lowdens}
\end{figure*}

Under $1000 \ang$ the opacities are first underestimated, and then at later epochs highly overestimated by the full LTE solution. At 10 days, we see the LTE opacity underestimates the full NLTE opacity between $2000 \ang \lesssim \lambda \lesssim 7000 \ang$ by a factor of $\sim 2-3$. Finally at 20 days after merger, we see a complete lack of LTE opacity for wavelengths $\lambda \gtrsim 4000 \ang$. Since the level populations are still calculated from the Boltzmann equation as presented in Section \ref{sec:levelpops}, these radical differences in opacity must come from the difference in ionisation structure.

Indeed, the Saha equation does not truly reproduce the full NLTE ionisation structure at any epoch, nor is it expected to (see Section \ref{subsec:analytic}). Notably, certain key ions are entirely absent over ranges of temperature solutions, thereby yielding a large effect on opacity. At 10 days and prior, Pt III is missing from the Saha ionisation structure, and so the large amount of opacity this ion provides at short wavelengths is missing. At 20 days, however, Pt III is highly abundant according to the Saha equation, and so we recover the overestimation of LTE opacity at $\lambda \lesssim 1500 \ang$ as seen in the previous section. The underestimation of opacity in the central wavelength range seen at 10 days is due to the absence of Ce I, and lower fraction of Ce II in LTE compared to the NLTE solution. At 20 days after merger, Ce II is completely absent from the Saha ionisation solution, and Ce III is only present in traces, thus leading to a lack of opacity at longer wavelengths. The opacity of Pt at 3 days and Ce at 20 days is shown in Figure \ref{fig:opacities_select} to illustrate this explanation. 

The effect of an ionisation structure deviating from that of LTE is difficult to predict, since different ions contribute differently to opacity. As an example, in our models we find that Ce II provides a large amount of opacity at longer wavelengths, so a Saha solution over-predicting the abundance of Ce II may overestimate opacity. On the other hand, under-predicting the abundance, like at 20 days after merger, will lead to an underestimation of opacity. Naturally, the exact effect depends heavily on the elemental composition of the ejecta, as well as the atomic data on which all the calculations are based. A caveat that must be mentioned here is that the NLTE treatment of ionisation in \textsc{sumo} is for the moment quite coarse, with recombination rates given by a fixed value, $\alpha = 10^{-11} \; \mathrm{cm^{-3}}$, non-thermal collisional ionisation is estimated using the Lotz approximation for collisional cross sections \citep[][]{Lotz:67}, and PI cross sections are hydrogenic (see \citet{Pognan.etal:22} for details on ionisation and recombination treatment.) As such, the comparison here between Saha and NLTE ionisation should be taken as indicative, rather than as an absolute measure of deviation. More sophisticated NLTE ionisation calculations are needed before a full picture can be established.

\subsection{Importance of Ejecta Density}
\label{subsec:lowdens}

As shown in the previous sections, the conditions within the ejecta are expected to deviate further from LTE as time goes on, though the excitation structure appears to remain close to that of LTE for epochs $t \leq 5$ days. However, LTE conditions require thermal collisional processes, and/or radiative processes from a thermal background, both of which are sensitive to ejecta density. As ejecta velocities, masses and compositions can vary significantly depending on their origin in the merger, it is interesting to gauge whether LTE can be broadly applied to all ejecta types at early times, and how the length of validity depends on ejecta parameters. 

In a first order attempt to address this, we consider a lower density model with $M_{\rm{ej}} = 0.01 \Msol$ and $v_{\rm{ej}} = 0.2$c at 3 days after the merger. This corresponds to an ejecta density lower by a factor of 40 compared to the standard model with $M_{\rm{ej}} = 0.05 \Msol$ and $v_{\rm{ej}} = 0.1$c at the same epoch. We do not test higher densities, as these are expected to be pushed further towards LTE. The level populations of Ce II, and the corresponding expansion opacities are shown in Figure \ref{fig:lowdens}. We see there that the level populations and opacities resemble those of our standard model at 20 days after merger, which has a factor of 2 lower density. As expected, lower density ejecta will deviate away from LTE conditions at earlier times, since both collisional processes and the radiation field density are significantly reduced.

Low density ejecta such as that considered here are typically found in dynamical, tidal ejecta with low electron fractions. As such, opacity rich elements such as lanthanides, i.e. Ce in this study, can be synthesised in these ejecta. Since lanthanide elements typically contribute immensely to opacity, correctly modelling their excitation and ionisation structures, and resulting opacities, is key to correctly interpreting KN observations. From Figure \ref{fig:lowdens}, it is clear that such low density ejecta are already far away from LTE conditions even 3 days after merger, and opacities may be severely overestimated in LTE.

\section{Implications for KN Modelling}
\label{sec:discussion}

In the previous sections, we have shown that excitation structure is approximately in LTE at 3 days after merger, with an important contribution from the radiation field in pushing high-lying states towards LTE. The corresponding expansion opacity appears to be very close to that of the full NLTE solution up to 5 days after merger, though a full LTE approach with Saha ionisation can increase this difference. The key difference appearing is that limited LTE opacities tend to overestimate both the full and limited NLTE opacities across all wavelengths. The full LTE opacity on the other hand, underestimates opacity over certain wavelength ranges due to an effect stemming from a change in ionisation structure. We now discuss the implications of these over and underestimations. 

At post-diffusion times, the opacity in LTE codes determines the SED, and is thus widely used to describe different KN ejecta that give rise to models such as the 'two component' model \citep[see e.g.][]{Perego.etal:14,Cowperthwaite.etal:17,Tanvir.etal:17,Tanaka.etal:17}. Considering the models in \citet{Tanaka.etal:17} (their Figure 3), we see that higher opacity at fixed mass, velocity and epoch yields a typically redder SED. The enhanced opacity there is linked to the presence of heavy r-process elements, notably lanthanides (such as Ce in our study). The presence of such heavy elements itself implies that the ejecta has a relatively low electron fraction $Y_e$, which can then be used to infer qualities about the equation of state of the neutron star material, and potentially information about the progenitor system itself, or the nature of the remnant.

As such, an overestimation of opacity, such as that seen in our limited LTE model after 5 days, implies that inferences of ejecta composition from LTE model-data comparisons may give too low proportions of lanthanides and potentially 3rd r-process peak elements, which are expected to be produced alongside lanthanides at low enough $Y_e$. The underestimation of lanthanide fraction in the ejecta corresponds to fitting a too high value of $Y_e$. An underestimation of opacity, such as that seen in the full LTE model at 10 days onwards, would naturally have the opposite effect. As mentioned previously however, the large drop of full LTE opacity at 20 days is also dependent on our model composition, and so cannot be readily generalised to all ejecta models.

Considering the observed KN AT2017gfo, the presence of lanthanides in the ejecta was robustly inferred from the SED \citep[][]{Abbott.etal:17,Chornock.etal:17,Drout.etal:17,Kasen.etal:17,Perego.etal:17}, as well as lightcurves in the optical red and infra-red peaking at later times than the bluer wavelengths, e.g. up to $\sim 2-3$ days for i-band to K-band \citep[see e.g][figure 1]{Villar.etal:17}. The mass of the various KN components were inferred from combined lightcurves up to $t \sim 30$ days after merger. From our results presented here, it is possible that opacity from 10 days onwards is overestimated in LTE for our standard model of $M_{\rm{ej}} = 0.05 \Msol$, $v_{\rm{ej}} = 0.1c$. However, this model is over-dense compared to the estimates of individual ejecta components, e.g. a lanthanide rich component with $M_{\rm{ej}} \sim 0.01 \Msol$, $v_{\rm{ej}} = 0.15c$ \citep[][]{Villar.etal:17} (though see also \citet{Kasen.etal:17} with $M_{\rm{ej}} = 0.04 \Msol$, $v_{\rm{ej}} = 0.1c$). This lanthanide rich component may thus have ejecta parameters closer to that of our low density model with $M_{\rm{ej}} = 0.01 \Msol$, $v_{\rm{ej}} = 0.2c$, which is already well out of LTE conditions even at 3 days after merger, right after the IR LCs peak. This implies that the mass estimates of lanthanide rich components from AT2017gfo may in fact be slightly underestimated, especially if later time ($t \gtrsim 5$ days) observations are used under the assumption of LTE opacity.

In general, key quantities such as ejecta mass and velocity, as well as broad composition are inferred from observables such as LC peak time and luminosity, which are directly dependent on opacity \citep[e.g. equations 2 and 3 in][]{Kasen.etal:17}. Furthermore, the emergence of various KN components at different times, such as the 'red' and 'blue', or alternatively lanthanide 'rich/free', components is also directly dependent on the assumed opacity of the KN ejecta. The wavelength dependence and evolution of ejecta opacity therefore plays a central role in inferring the physical characteristics of the ejecta, and thus requires precise modelling in order to accurately derive these quantities. The results presented here highlight that the assumption of LTE conditions must be carefully considered, taking into account not only epoch after merger, but also ejecta parameters, notably with respect to ejecta density and composition.

\section{Conclusions}
\label{sec:conclusions}

We present the first study of the NLTE level populations in kilonova ejecta. We calculate these at 3-20 days post merger for an ejecta with parameters $M_{\rm{ej}} = 0.05 \Msol$, $v_{\rm{ej}}= 0.1c$, and composition Te, Ce, Pt and Th, using the \textsc{sumo} radiative transfer code. We compare the NLTE solutions to the LTE Boltzmann approximation for the same temperature, to delineate when significant LTE deviations occur, and for which levels.

The atomic levels can roughly be divided into three groups based on their behaviour in NLTE vs LTE. The first group consists of the lowest lying states, and are typically well represented by LTE populations at early to intermediate times, as most of the ions are in these states. A first order effect for higher lying states is that, already from a few days, thermal collisions are unable to maintain equilibrium, and the levels become underpopulated in NLTE due to rapid spontaneous decays. We define as group 3 those levels with allowed transitions to excited states (giving low optical depths) - the level populations of these are orders of magnitudes smaller than LTE. Group 2 are intermediate levels with only forbidden or optically thick deexcitation channels - the LTE deviations for these are smaller with typical populations one or two orders of magnitude away from LTE. 

While it has long been recognized that collisions cannot maintain LTE for more than a few days in KN ejecta, it has remained unknown whether a strong radiation field, experiencing a complex transfer through millions of bound-bound transitions, could maintain LTE for longer. With our simulations, we are able to provide a first answer to this question. By comparing NLTE solutions with and without diffuse radiation field terms, we demonstrate that the radiation field has a strong influence on the level populations. For high-lying levels (group 3) it pushes the populations closer to LTE, especially at early times. This effect is necessarily accompanied by some lower states (group 2) being pushed \textit{away} from LTE. This effect is often quite mild, though can become significant for large ions such as Ce II. Generally, the radiation-free NLTE excitation structure more closely resembles that of full NLTE than the LTE values, especially at late times.

From the excitation structures yielded by the Boltzmann equation (limited LTE), full NLTE (SUMO solutions), and limited NLTE (no diffuse radiation field), we calculate the expansion opacity, a key quantity used in KN lightcurve and SED modelling. For our 4-element composition, (Te, Ce, Pt, Th,) the lanthanide Ce plays a particularly important role, notably Ce II which dominates the expansion opacity at optical wavelengths and longer. 

We find that the limited LTE opacities for our standard model are similar (factor $\lesssim$ 2 error) to full NLTE ones to approximate 5 days after merger. This is the first verification for the applicability of LTE expansion opacities for the early KN phase. Up to about 3 days, the radiation field helps to maintain LTE opacity values by its influence on group 3 populations. By 5 days, however, the lowering of group 2 populations for Ce II offsets the gains for group 3 and the opacity is instead pushed away from its LTE value compared to collisional NLTE.

At later times, the NLTE opacities are systematically lower than the LTE ones by factor 2-10. For our particular model, the opacity at the most important wavelengths ($\lambda \gtrsim 3000 \ang$), is still accurate to factor 2-3 in LTE.
As we see under-population of the vast majority of states in NLTE compared to LTE, and Sobolev expansion opacities have the property of depending on the number of optically thick lines rather than on the values, we assess that an overestimate of opacity in LTE is a generic property of a Boltzmann approach.

The opacity yielded by the limited NLTE excitation structure is remarkably similar to that of the full NLTE solution at all epochs, notably being almost identical even at 20 days after merger. For our particular model, we find that they are better approximations of the full NLTE opacities than the limited LTE opacities from 5 days onwards. Since the limited NLTE opacities are calculated using an excitation structure that is defined without radiation terms, they may be calculated without any application of radiative transfer. This provides an interesting future direction for calculating NLTE expansion opacities for KN ejecta that would be applicable even to late times after the merger.

When it comes to ionisation, NLTE calculations are still at a relatively crude stage where \textsc{sumo} uses generic formulae for ionisation cross sections (Lotz), fixed and temperature-independent recombination rates, and hydrogenic photoionisation cross sections. Thus, NLTE ionisation solutions are far more uncertain than the excitation solutions for a given ion abundance, and the NLTE/LTE comparisons are therefore more tentative. Discrepancies between LTE and NLTE opacities will, naturally, be larger when NLTE ionisation effects are also considered. A generic effect is that the non-thermal ionisations occurring in NLTE give a broader range of ion stages compared to LTE, which tends to favor certain ions for a given temperature and density.

Somewhat by coincidence, our particular model gives almost identical optical and IR opacities up to 10 days compared to the full LTE values. This arises because a single ion, Ce II, dominates at these wavelengths, and the abundance of this ionisation stage is similar in LTE and NLTE. However, a complete breakdown of LTE ionisation is illustrated by the results at 20 days, where the full LTE opacities are off by two orders of magnitudes. Similarly large errors can probably occur earlier for other compositions than the one tested here (e.g. the UV opacity if off by factor 10 at 3 days where Pt III dominates). A complete understanding of full NLTE vs full LTE opacities requires both improved NLTE physics and a larger model grid exploring different densities and compositions.

Finally, we also study a lower-density model ($M_{\rm{ej}} = 0.01 \Msol$ and $v_{\rm{ej}} = 0.2$c). As these parameters yield a density lower by a factor of 40 compared to the original model at the same epoch, we see a significant deviation of the NLTE level populations from the LTE values even at 3 days, subsequently leading to an overestimation of opacities. Since certain key KN ejecta components, such as lanthanide rich, dynamical ejecta, may have masses and velocities similar to the low density model, this may have implications for inferred lanthanide masses in AT2017gfo.

\section*{Acknowledgements}

We acknowledge funding from the European Research Council (ERC) under the European Union's Horizon 2020 Research and Innovation Program (ERC Starting Grant 803189  -- SUPERSPEC, PI Jerkstrand).

Furthermore, JG thanks the Swedish Research Council for the individual starting grant with contract No. 2020-05467.

The computations were enabled by resources provided by the Swedish National Infrastructure for Computing (SNIC), partially funded by the Swedish Research Council through grant agreement no. 2018-05973. 

\section*{Data Availability}

The data underlying this article will be shared on reasonable request to the corresponding author.



\bibliographystyle{mnras}
\bibliography{biblio} 



\appendix

\section{Level Population Plots}
\label{app:levelplots}

\begin{figure*}
    \centering
    \includegraphics[trim={0.2cm 0.1cm 0.4cm 0.3cm},clip,width = 0.49\textwidth]{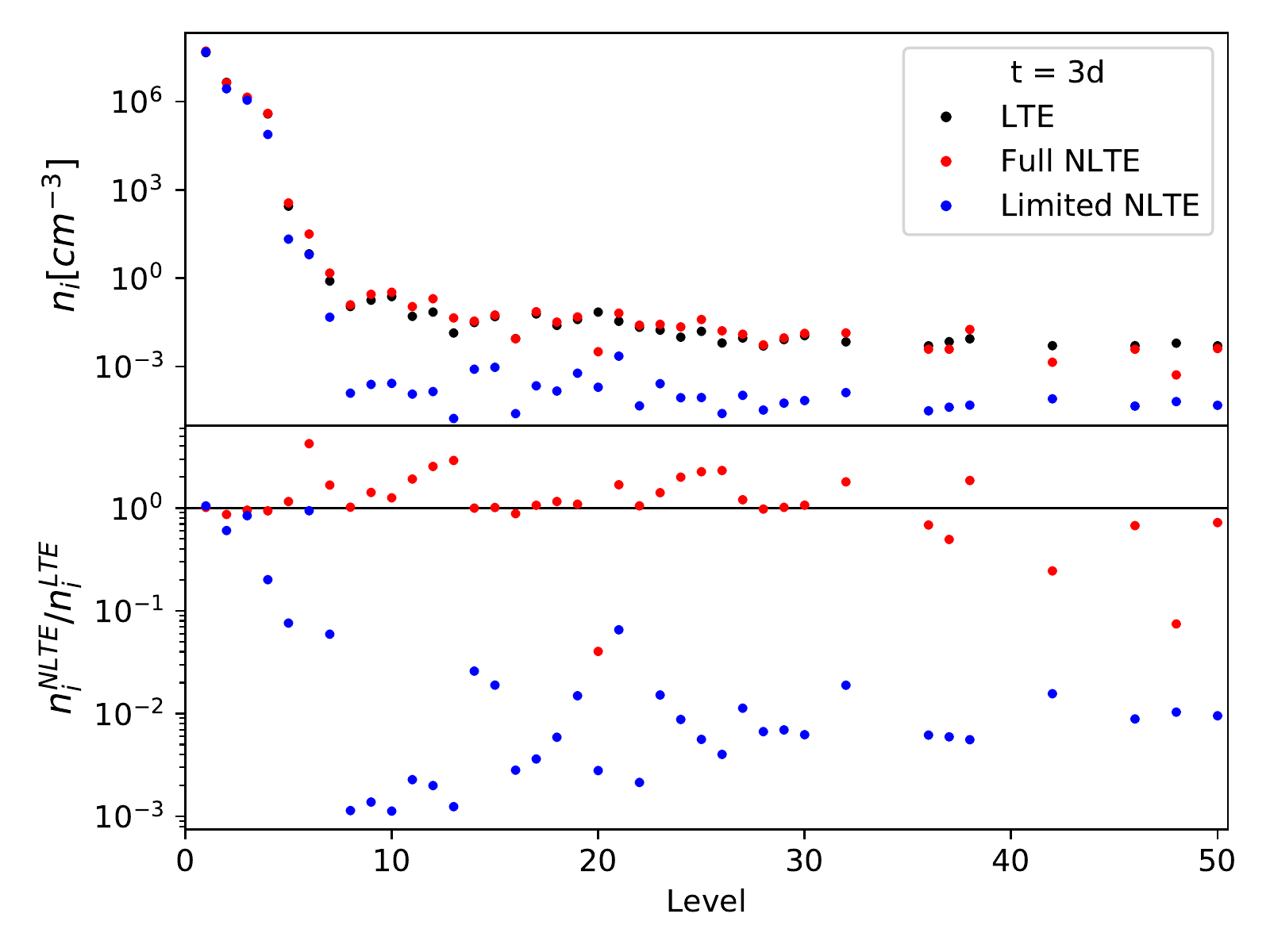}
    \includegraphics[trim={0.2cm 0.1cm 0.4cm 0.3cm},clip,width = 0.49\textwidth]{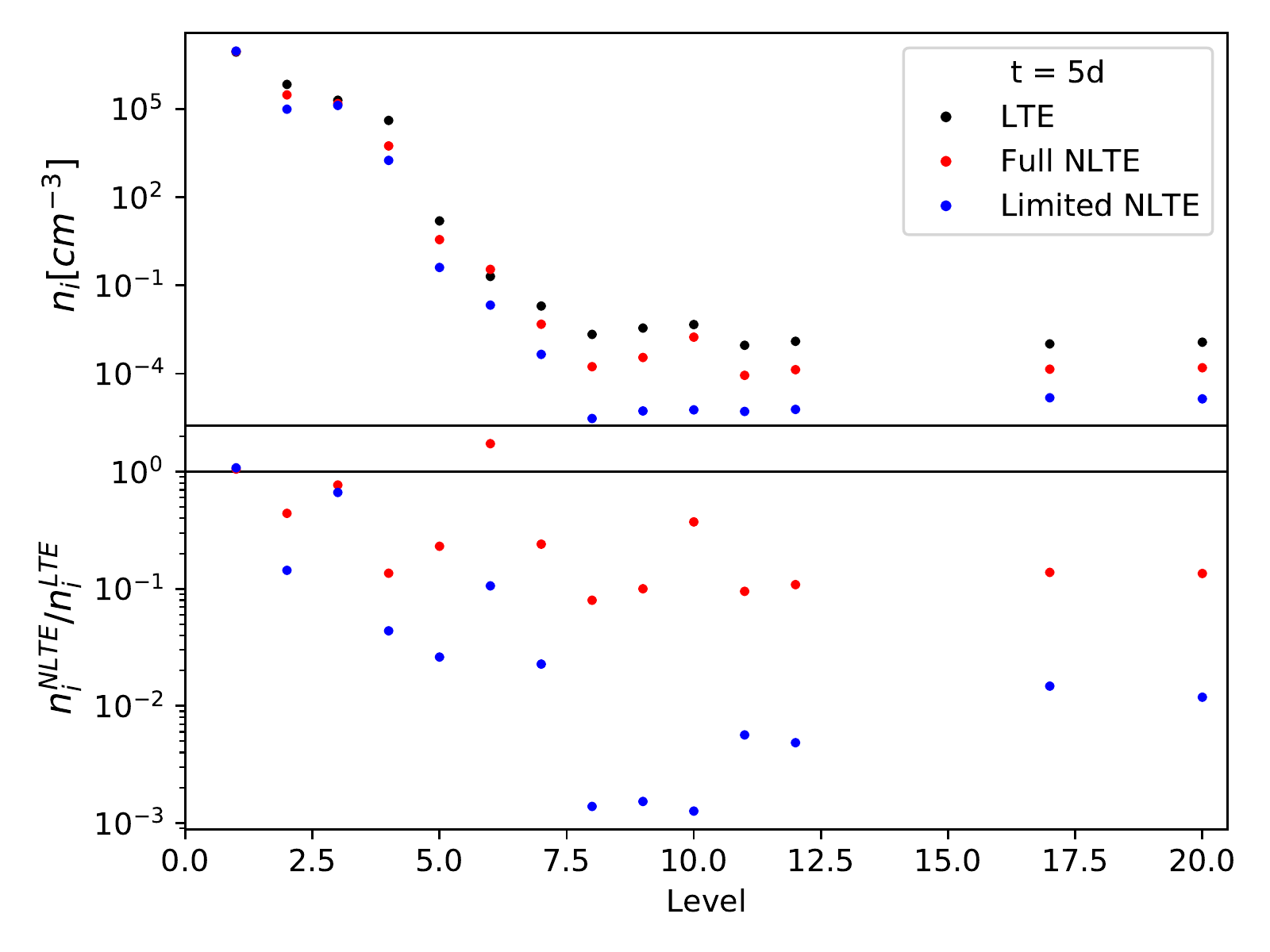} 
    \includegraphics[trim={0.2cm 0.1cm 0.4cm 0.3cm},clip,width = 0.49\textwidth]{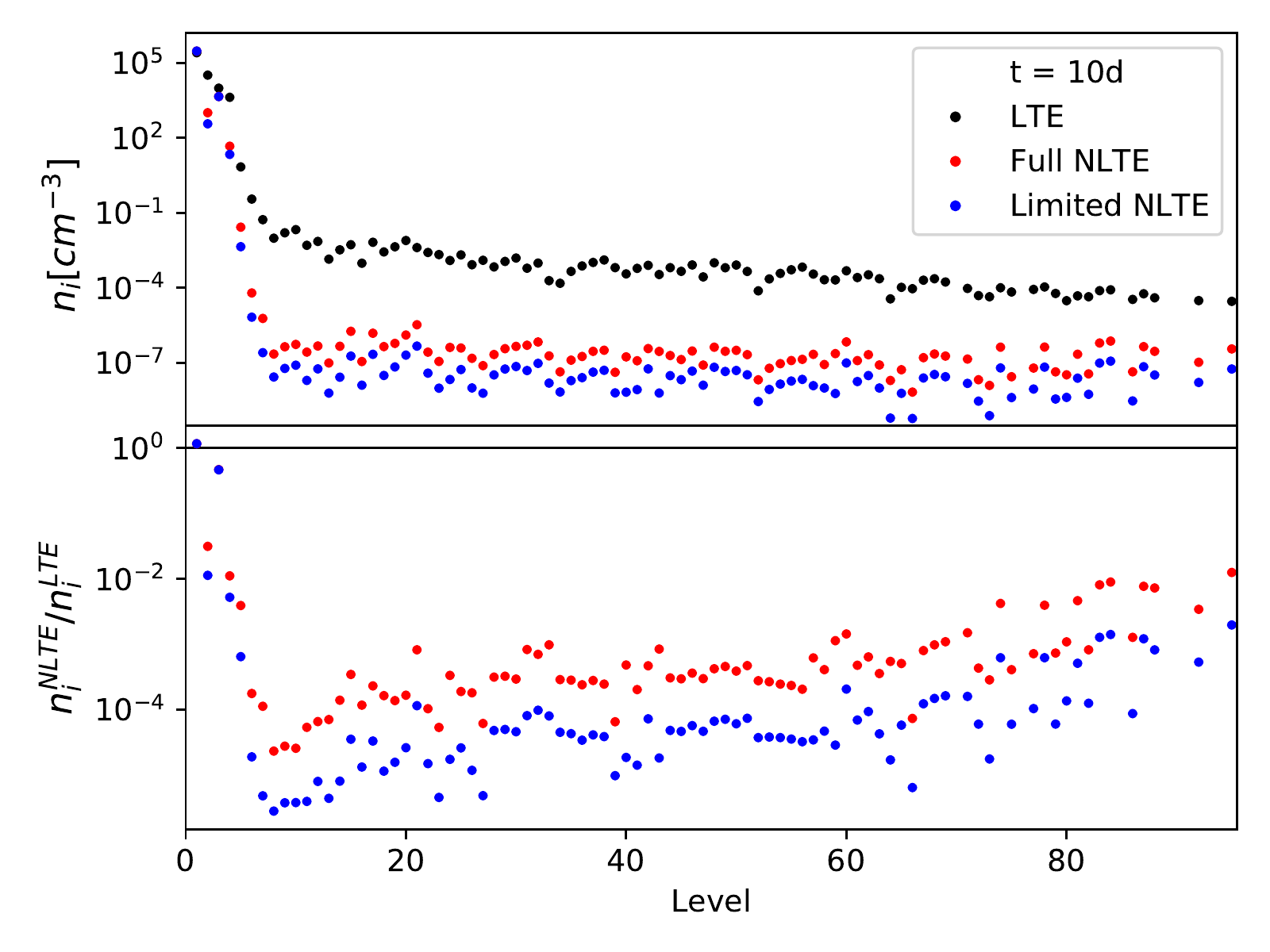} 
    \includegraphics[trim={0.2cm 0.1cm 0.4cm 0.3cm},clip,width = 0.49\textwidth]{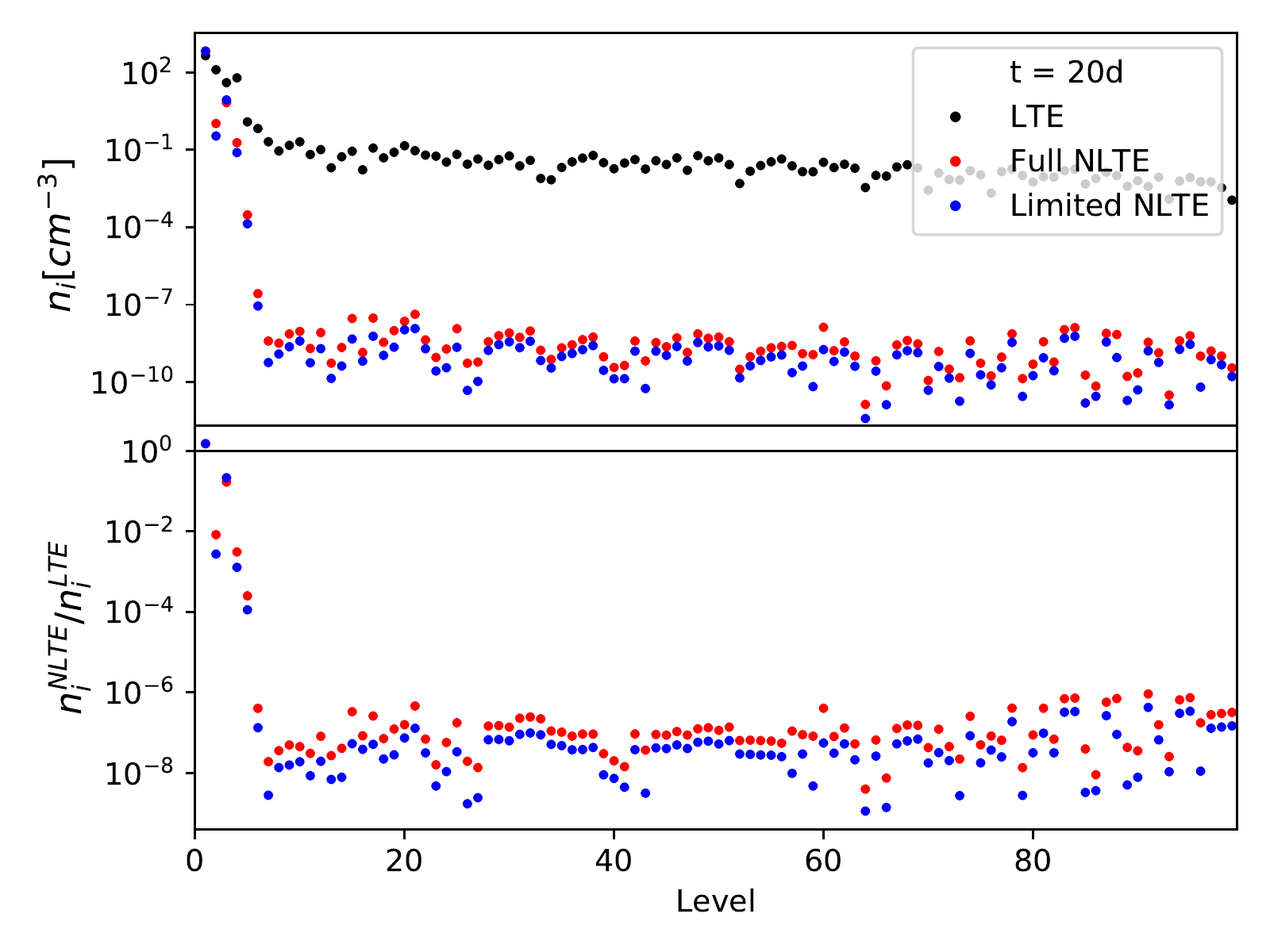} 
    \caption{Level populations of Te I at 3, 5, 10, and 20 days after merger (top left to bottom right). States with an LTE level population smaller than the most populated state by a factor of $10^{10}$ are cut from these plots.}
    \label{fig:TeI_levelpops}
\end{figure*}

\begin{figure*}
    \centering
    \includegraphics[trim={0.2cm 0.1cm 0.4cm 0.3cm},clip,width = 0.49\textwidth]{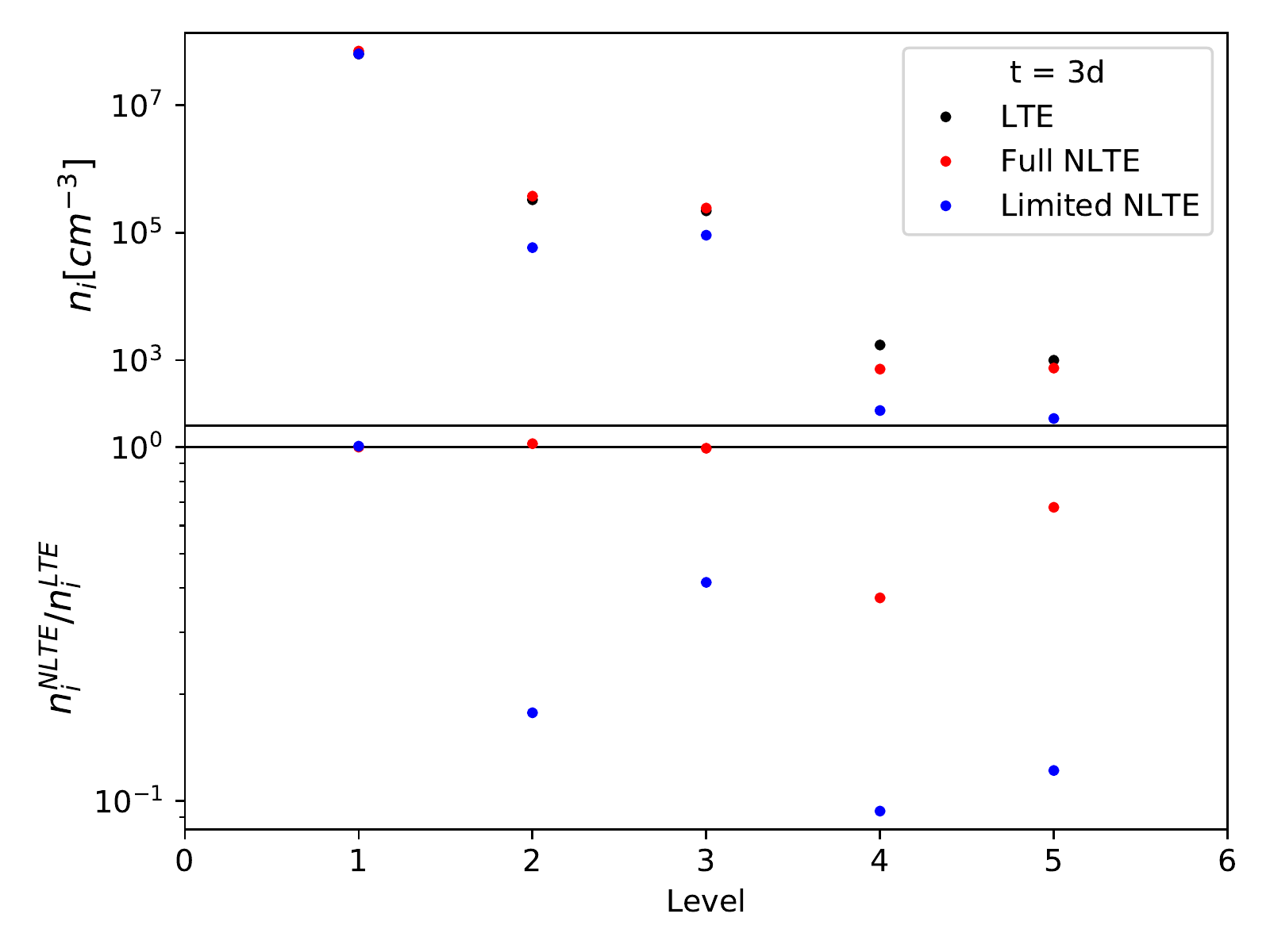}
    \includegraphics[trim={0.2cm 0.1cm 0.4cm 0.3cm},clip,width = 0.49\textwidth]{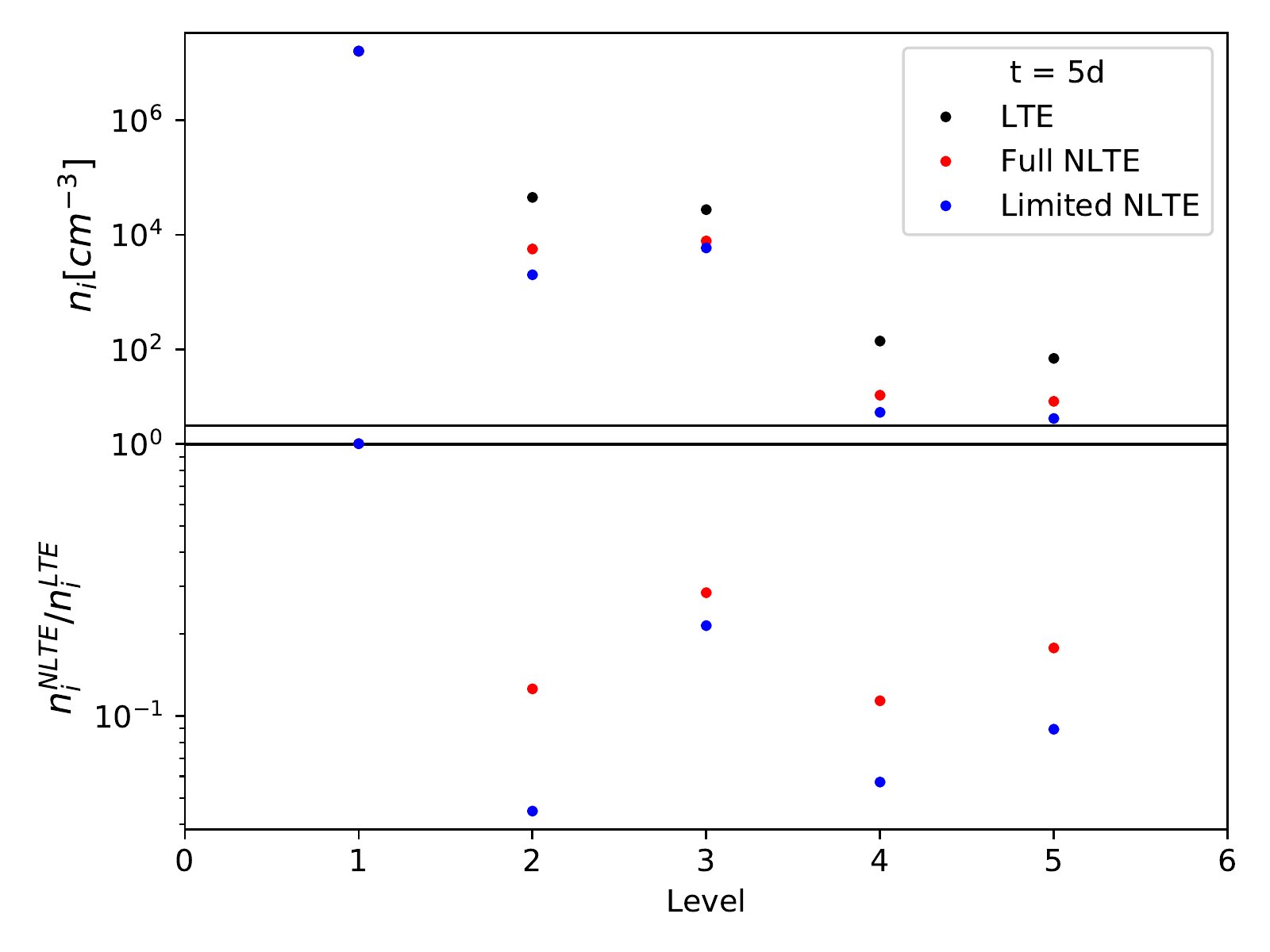} 
    \includegraphics[trim={0.2cm 0.1cm 0.4cm 0.3cm},clip,width = 0.49\textwidth]{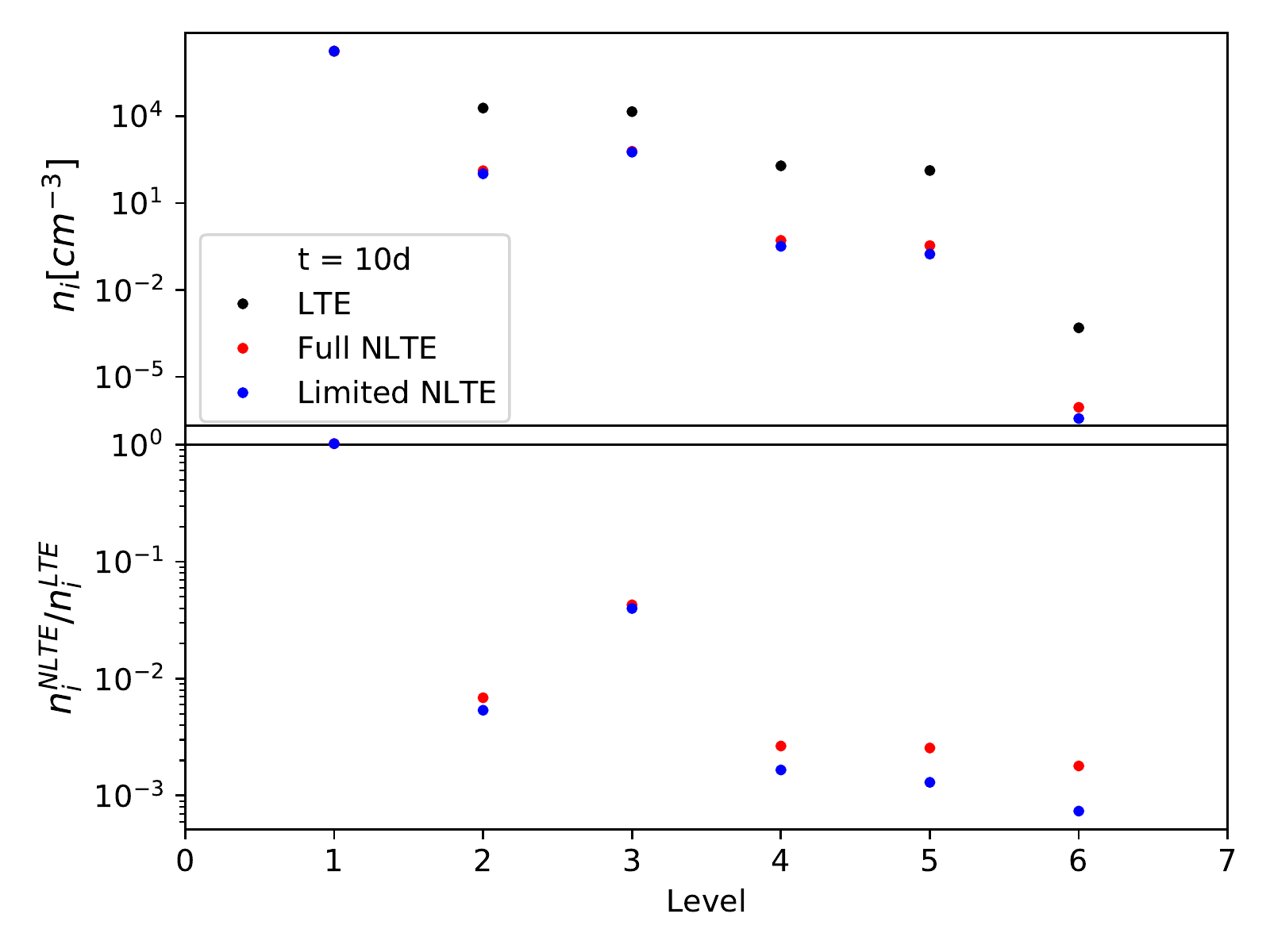} 
    \includegraphics[trim={0.2cm 0.1cm 0.4cm 0.3cm},clip,width = 0.49\textwidth]{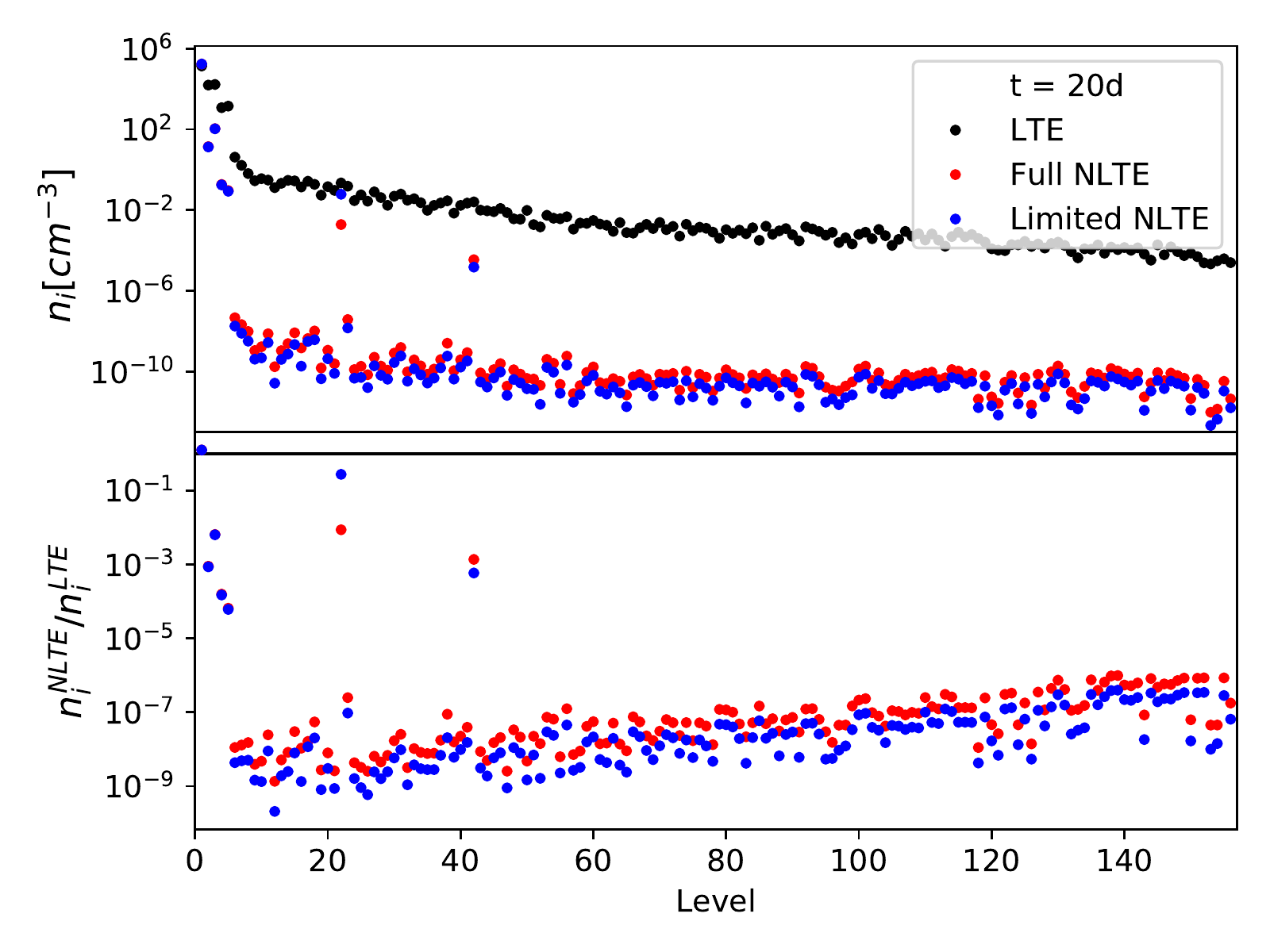} 
    \caption{Level populations of Te II at 3, 5, 10, and 20 days after merger (top left to bottom right). States with an LTE level population smaller than the most populated state by a factor of $10^{10}$ are cut from these plots.}
    \label{fig:TeII_levelpops}
\end{figure*}

\begin{figure*}
    \centering
    \includegraphics[trim={0.2cm 0.1cm 0.4cm 0.3cm},clip,width = 0.49\textwidth]{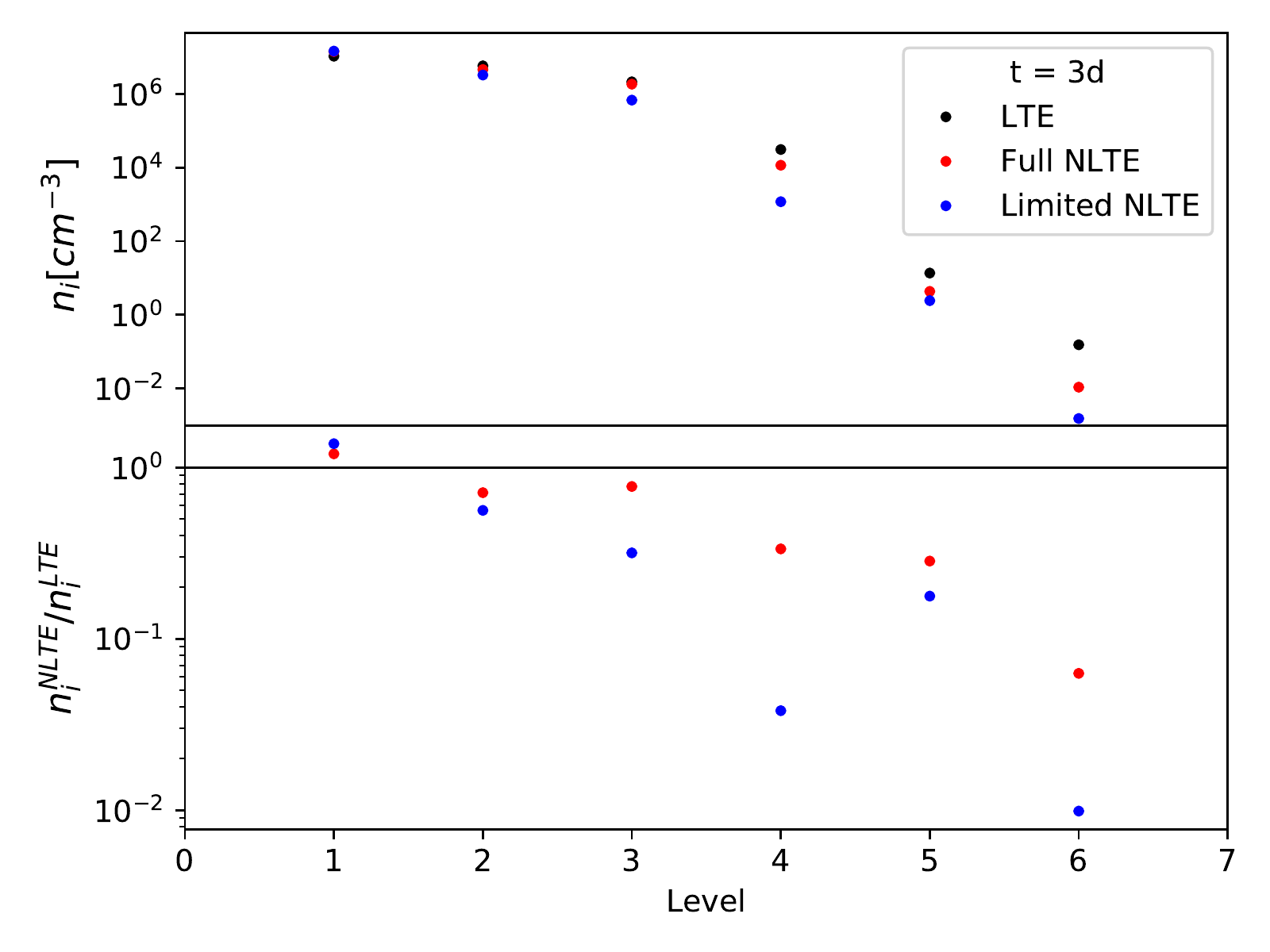}
    \includegraphics[trim={0.2cm 0.1cm 0.4cm 0.3cm},clip,width = 0.49\textwidth]{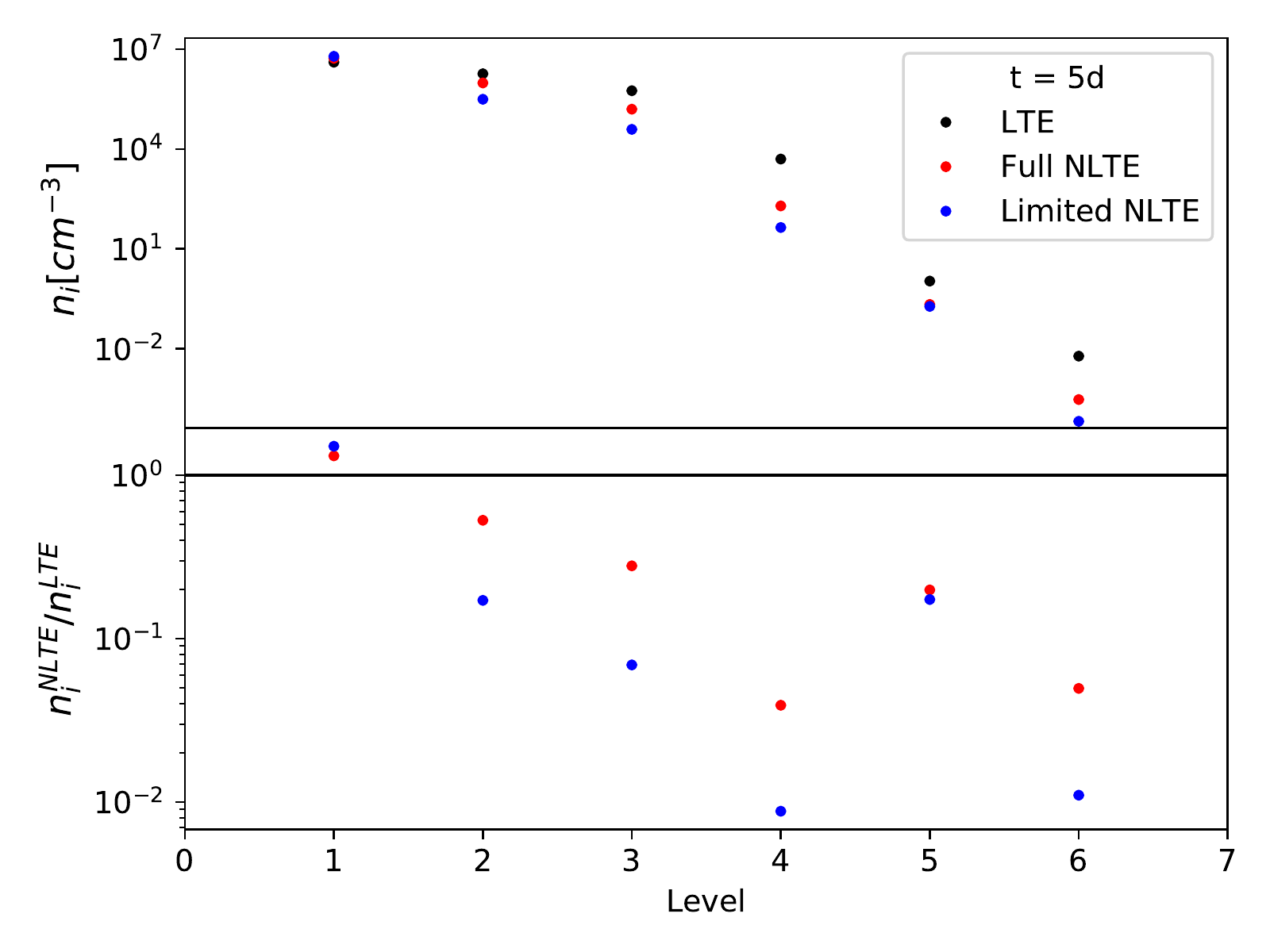} 
    \includegraphics[trim={0.2cm 0.1cm 0.4cm 0.3cm},clip,width = 0.49\textwidth]{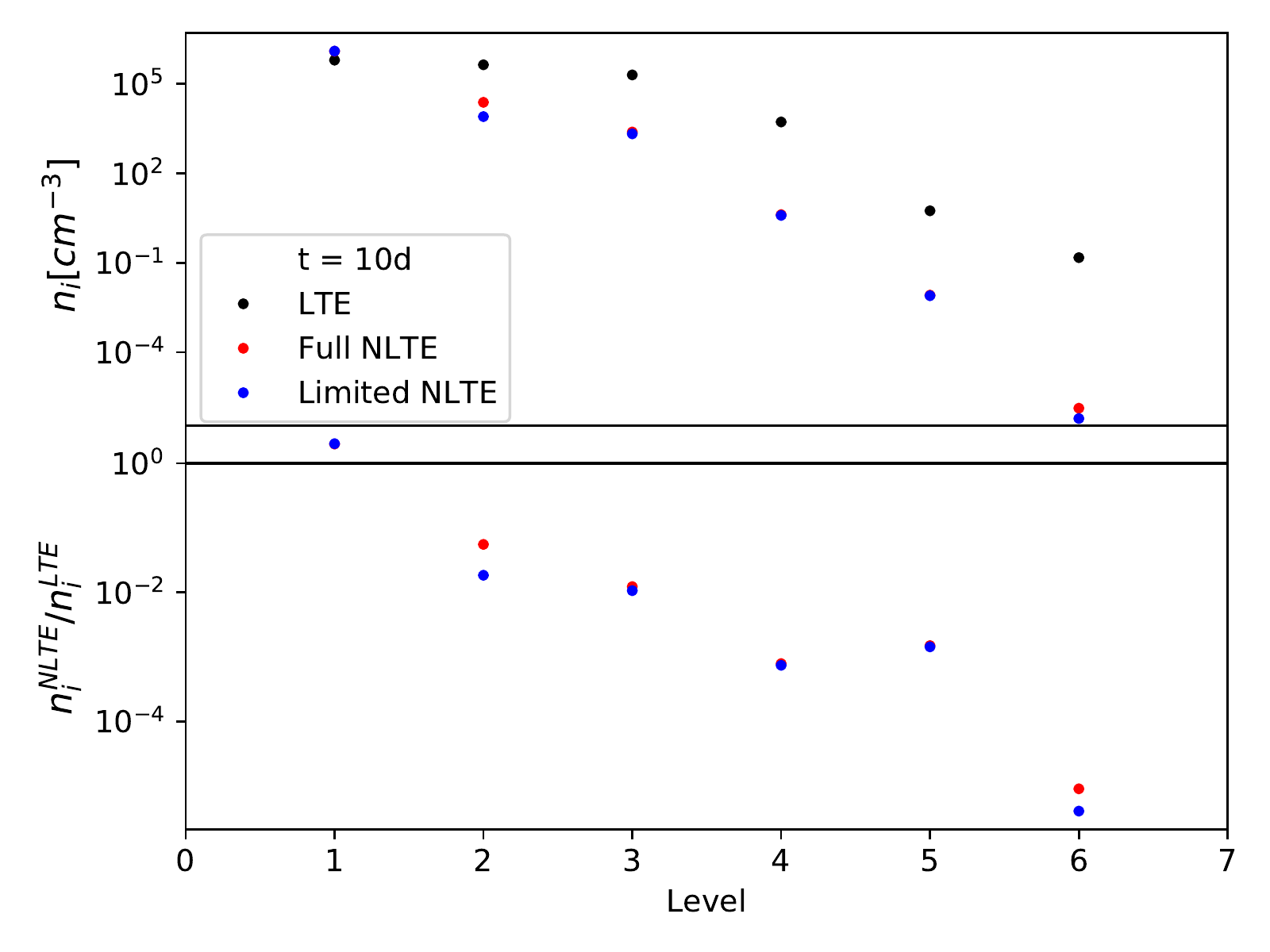} 
    \includegraphics[trim={0.2cm 0.1cm 0.4cm 0.3cm},clip,width = 0.49\textwidth]{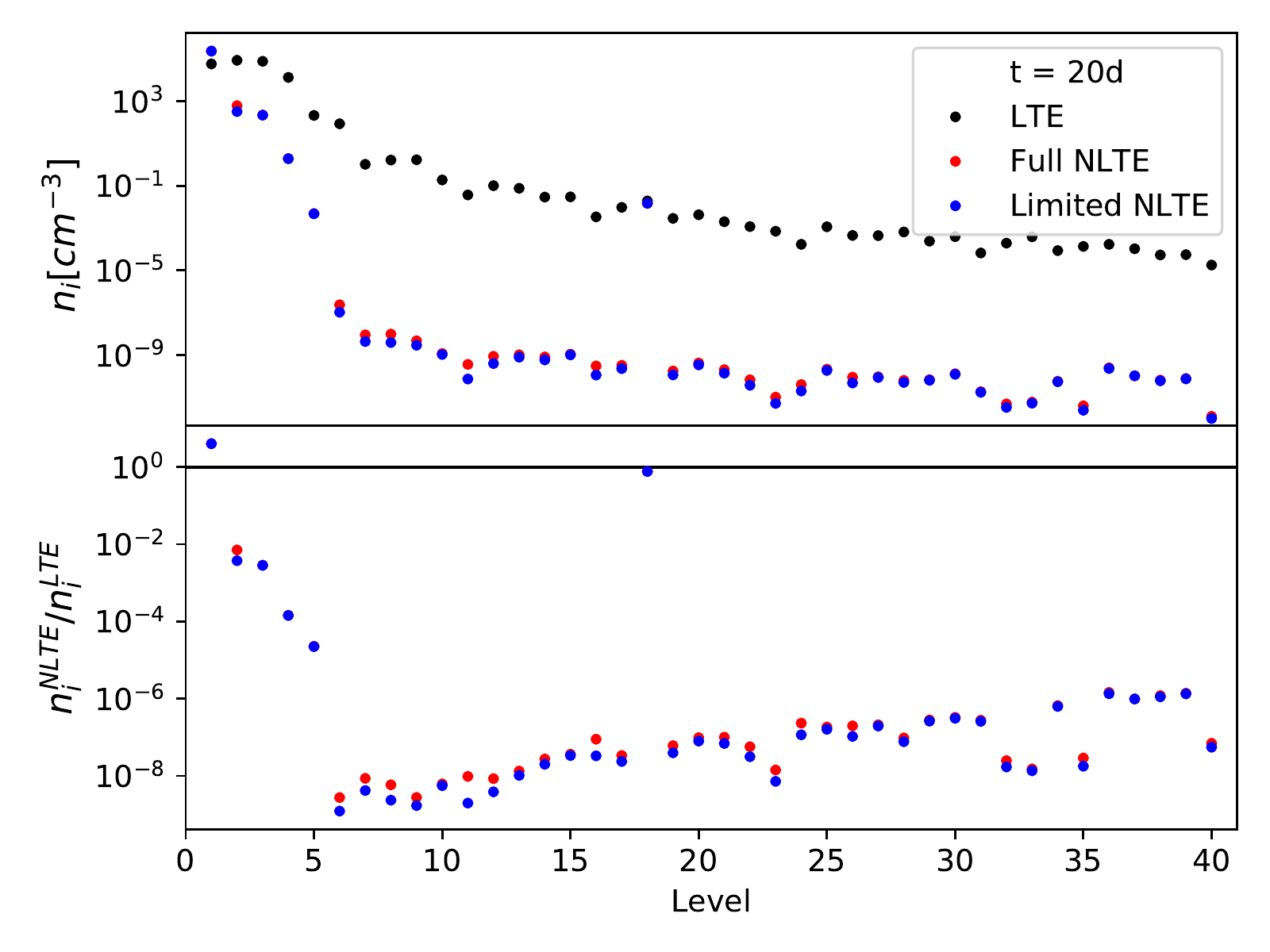} 
    \caption{Level populations of Te III at 3, 5, 10, and 20 days after merger (top left to bottom right). States with an LTE level population smaller than the most populated state by a factor of $10^{10}$ are cut from these plots.}
    \label{fig:TeIII_levelpops}
\end{figure*}

\begin{figure*}
    \centering
    \includegraphics[trim={0.2cm 0.1cm 0.4cm 0.3cm},clip,width = 0.49\textwidth]{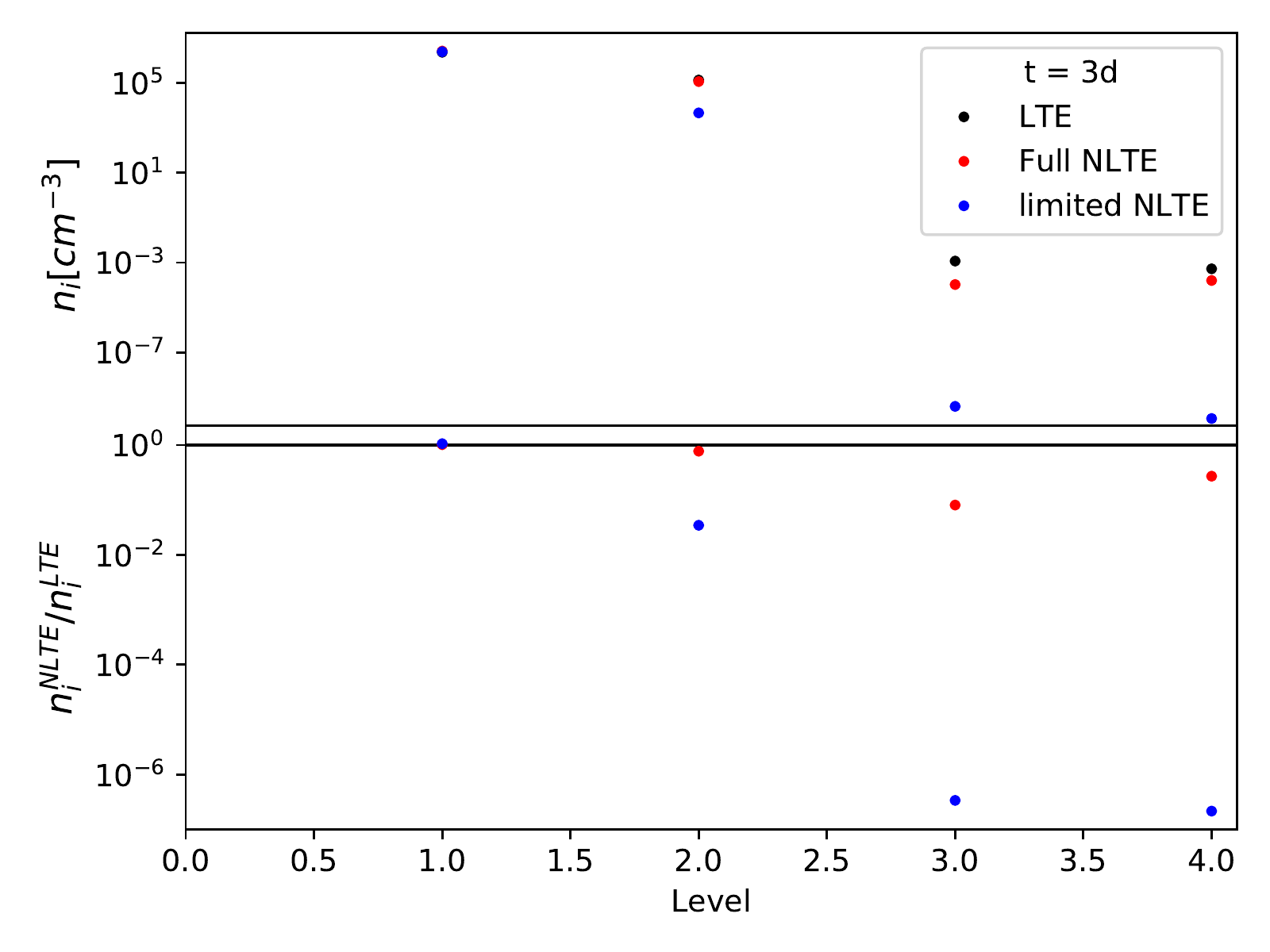}
    \includegraphics[trim={0.2cm 0.1cm 0.4cm 0.3cm},clip,width = 0.49\textwidth]{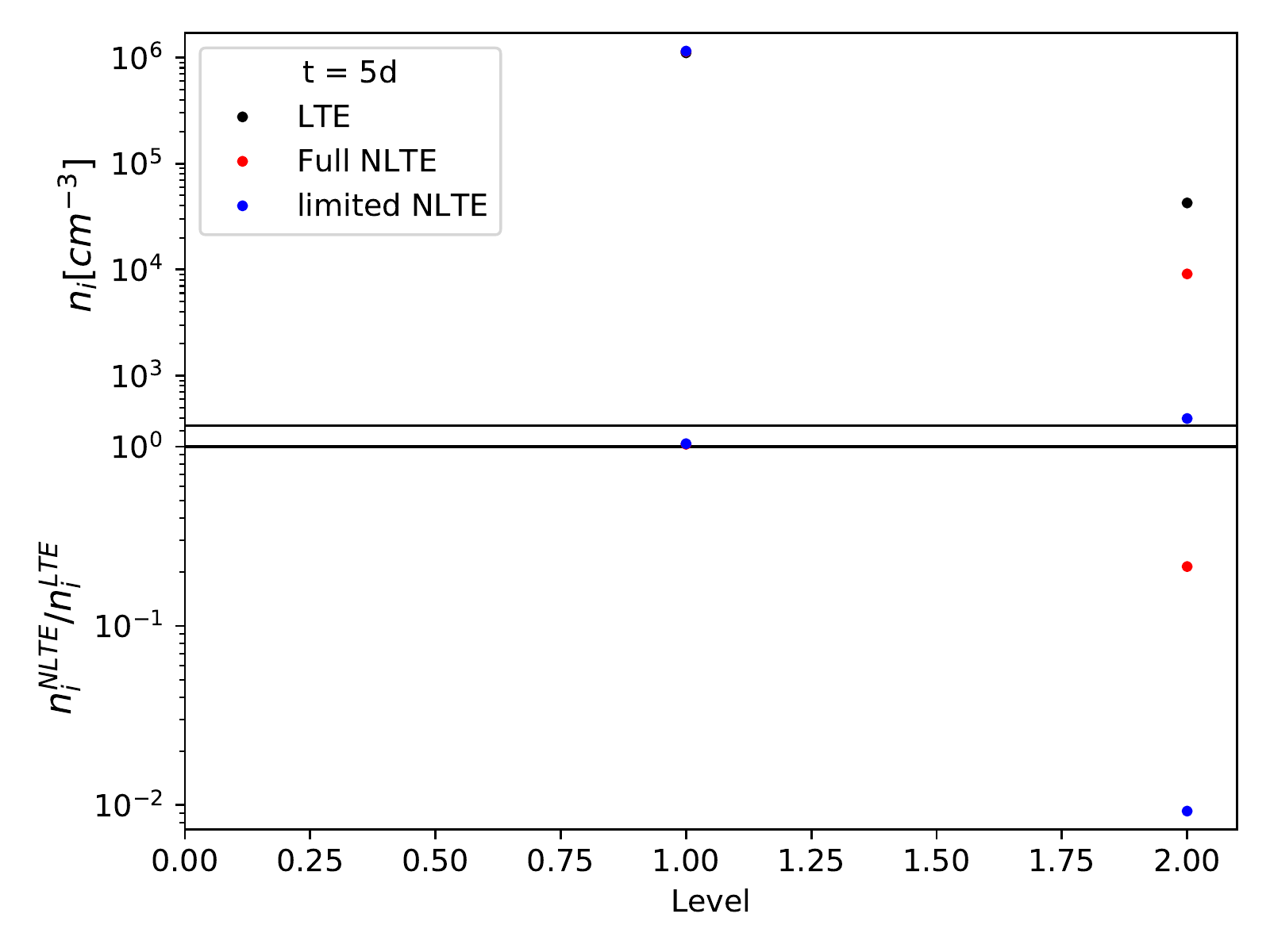} 
    \includegraphics[trim={0.2cm 0.1cm 0.4cm 0.3cm},clip,width = 0.49\textwidth]{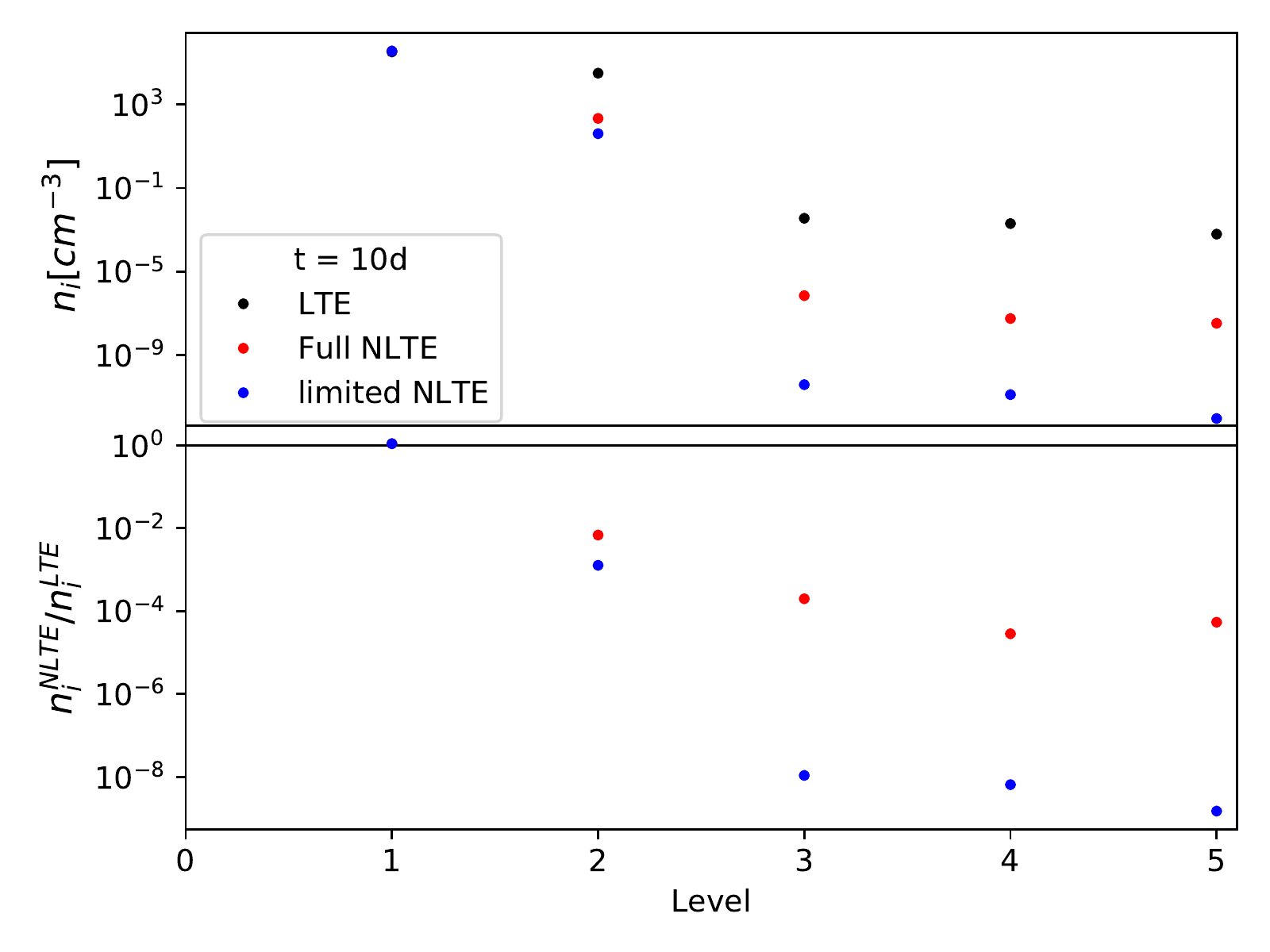} 
    \includegraphics[trim={0.2cm 0.1cm 0.4cm 0.3cm},clip,width = 0.49\textwidth]{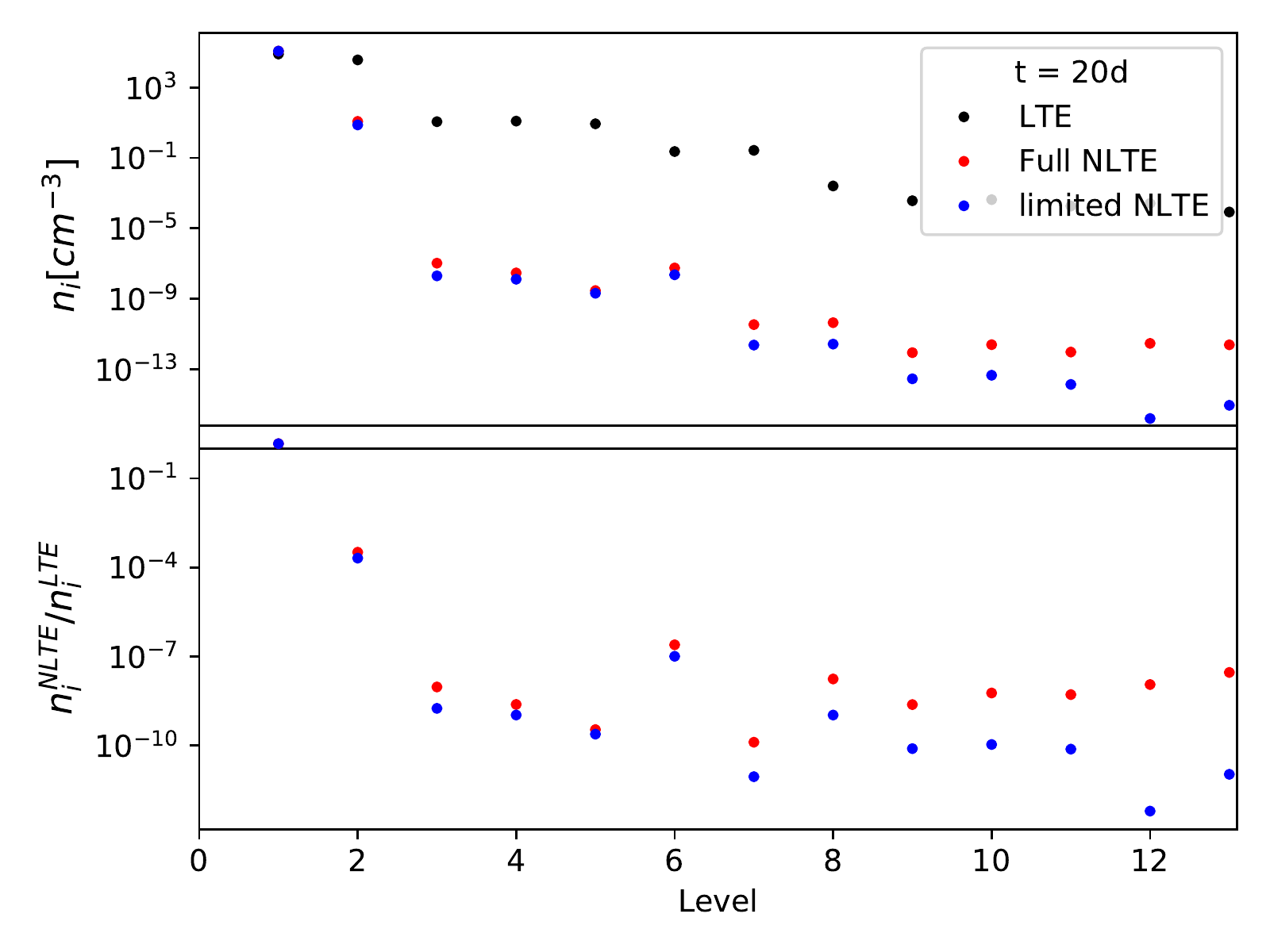} 
    \caption{Level populations of Te IV at 3, 5, 10, and 20 days after merger (top left to bottom right). States with an LTE level population smaller than the most populated state by a factor of $10^{10}$ are cut from these plots.}
    \label{fig:TeIV_levelpops}
\end{figure*}

\begin{figure*}
    \centering
    \includegraphics[trim={0.2cm 0.1cm 0.4cm 0.3cm},clip,width = 0.49\textwidth]{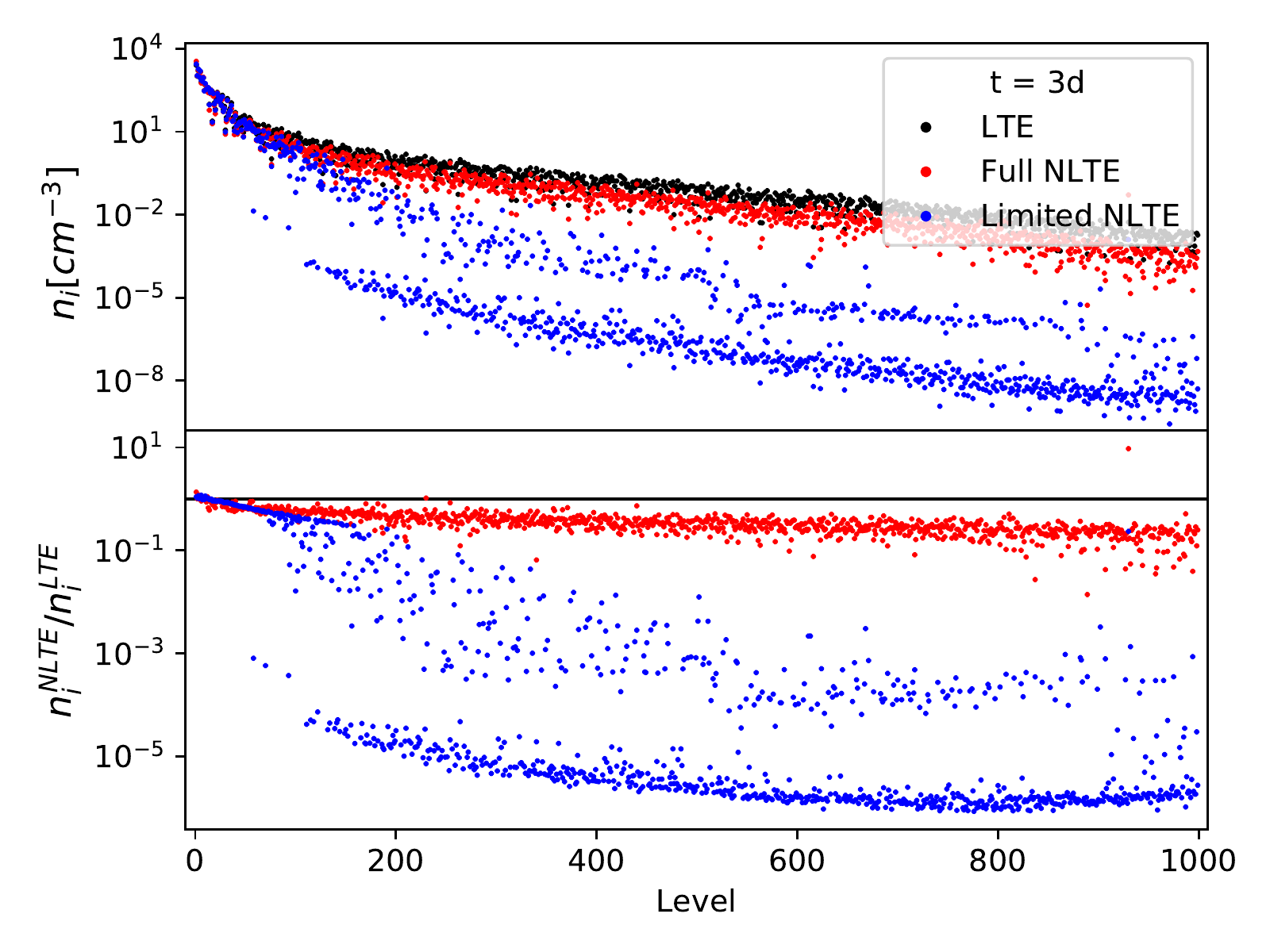}
    \includegraphics[trim={0.2cm 0.1cm 0.4cm 0.3cm},clip,width = 0.49\textwidth]{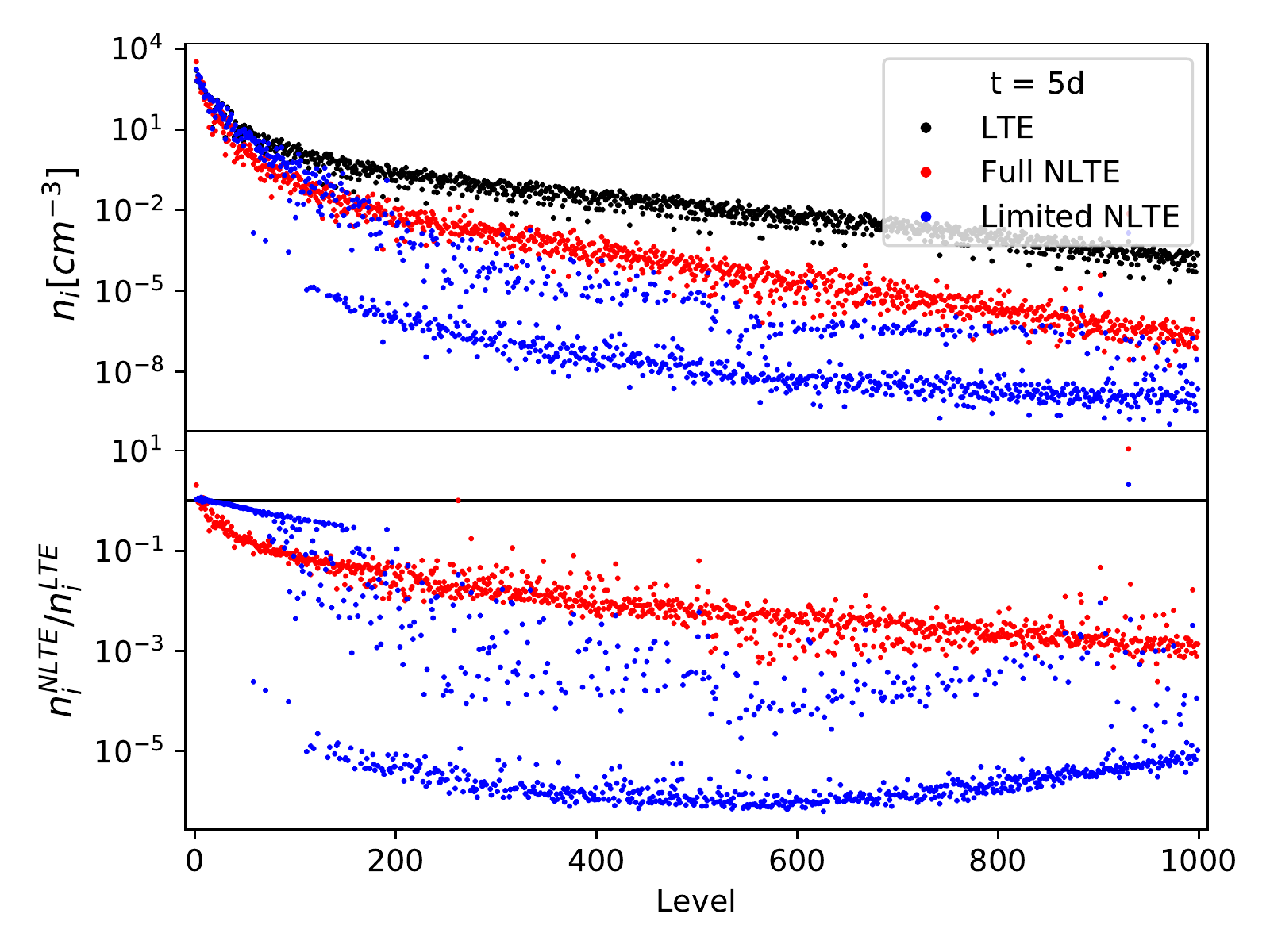} 
    \includegraphics[trim={0.2cm 0.1cm 0.4cm 0.3cm},clip,width = 0.49\textwidth]{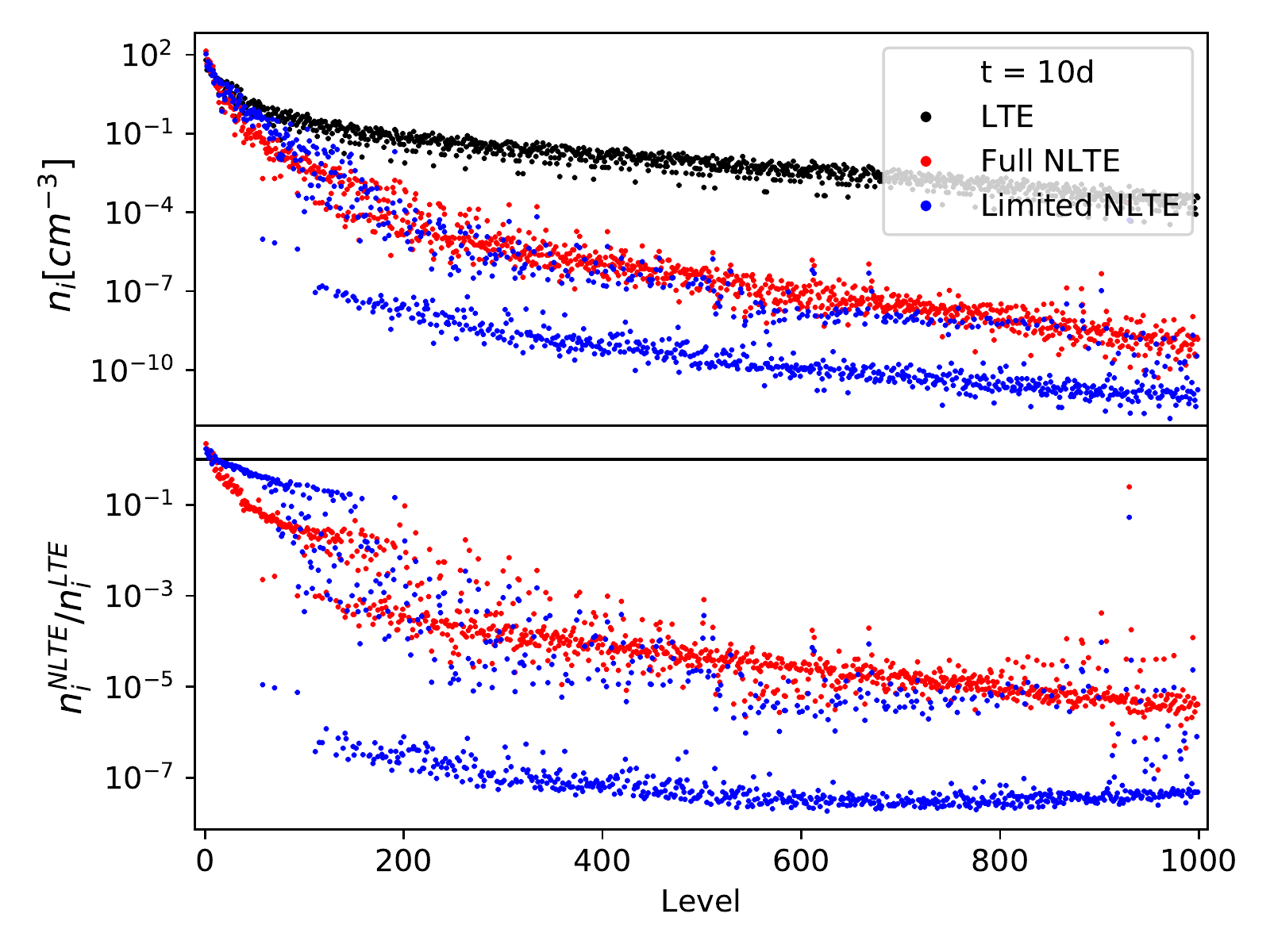} 
    \includegraphics[trim={0.2cm 0.1cm 0.4cm 0.3cm},clip,width = 0.49\textwidth]{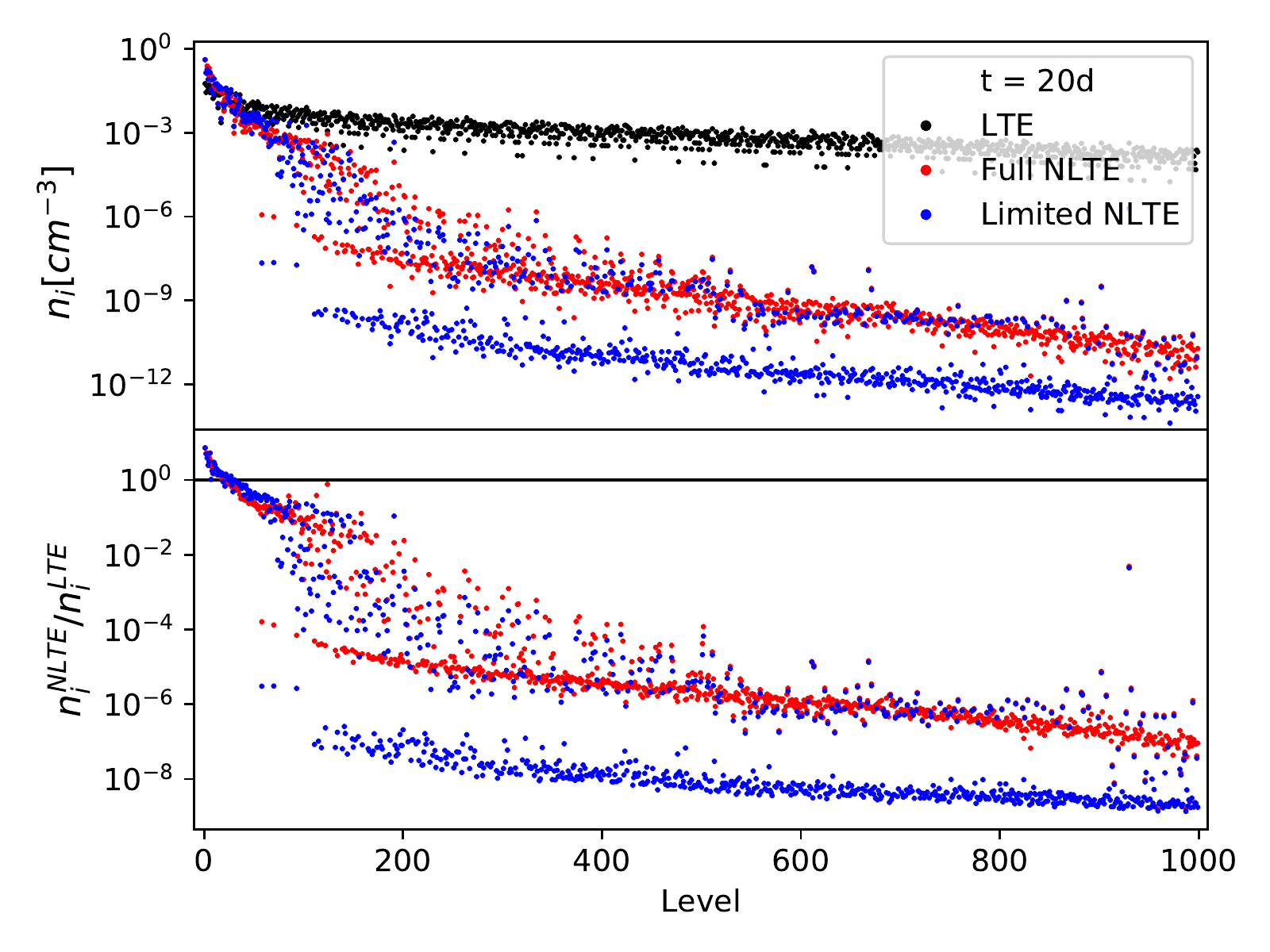} 
    \caption{Level populations of Ce I at 3, 5, 10, and 20 days after merger (top left to bottom right). States with an LTE level population smaller than the most populated state by a factor of $10^{10}$ are cut from these plots.}
    \label{fig:CeI_levelpops}
\end{figure*}

\begin{figure*}
    \centering
    \includegraphics[trim={0.2cm 0.1cm 0.4cm 0.3cm},clip,width = 0.49\textwidth]{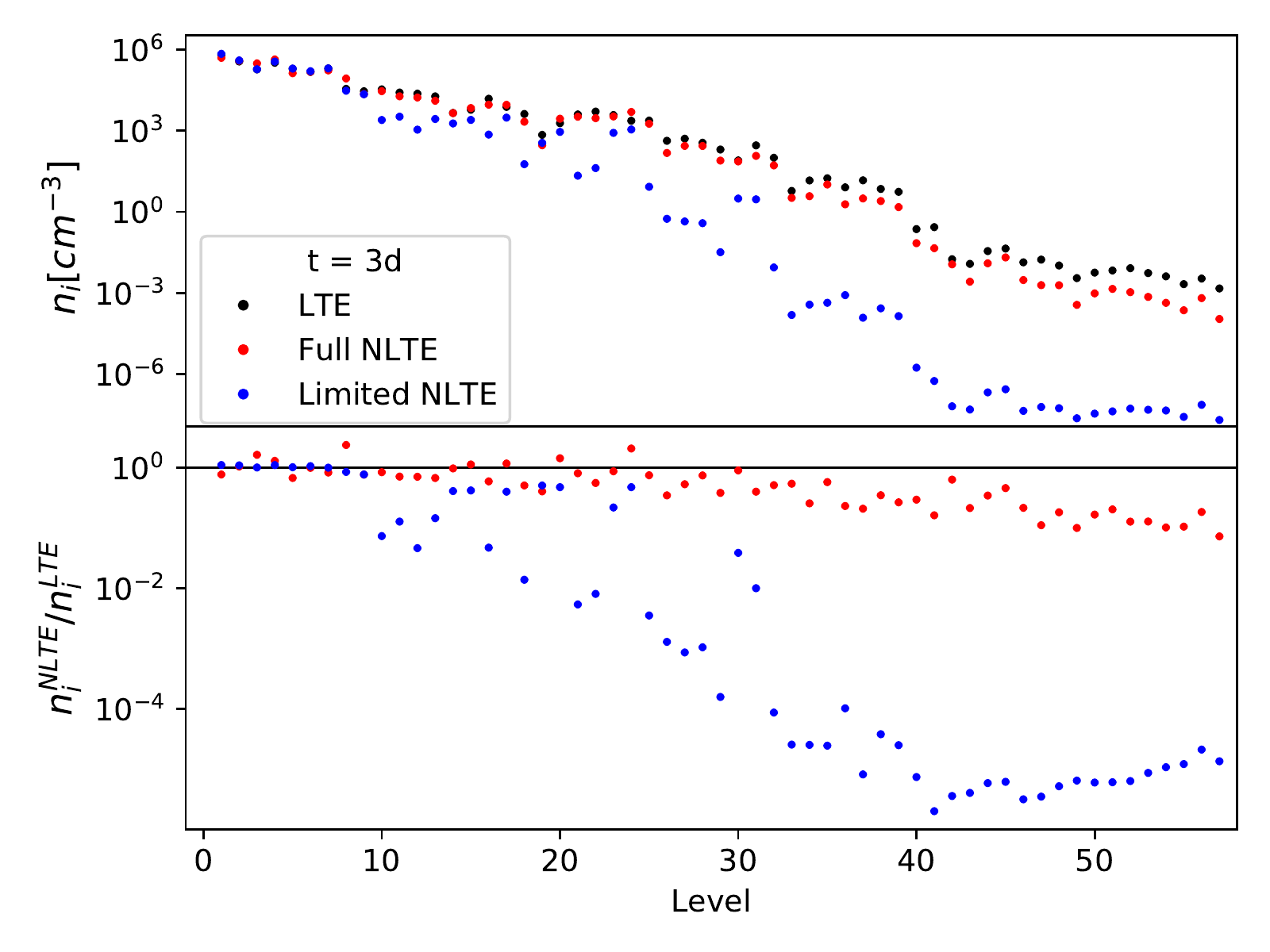}
    \includegraphics[trim={0.2cm 0.1cm 0.4cm 0.3cm},clip,width = 0.49\textwidth]{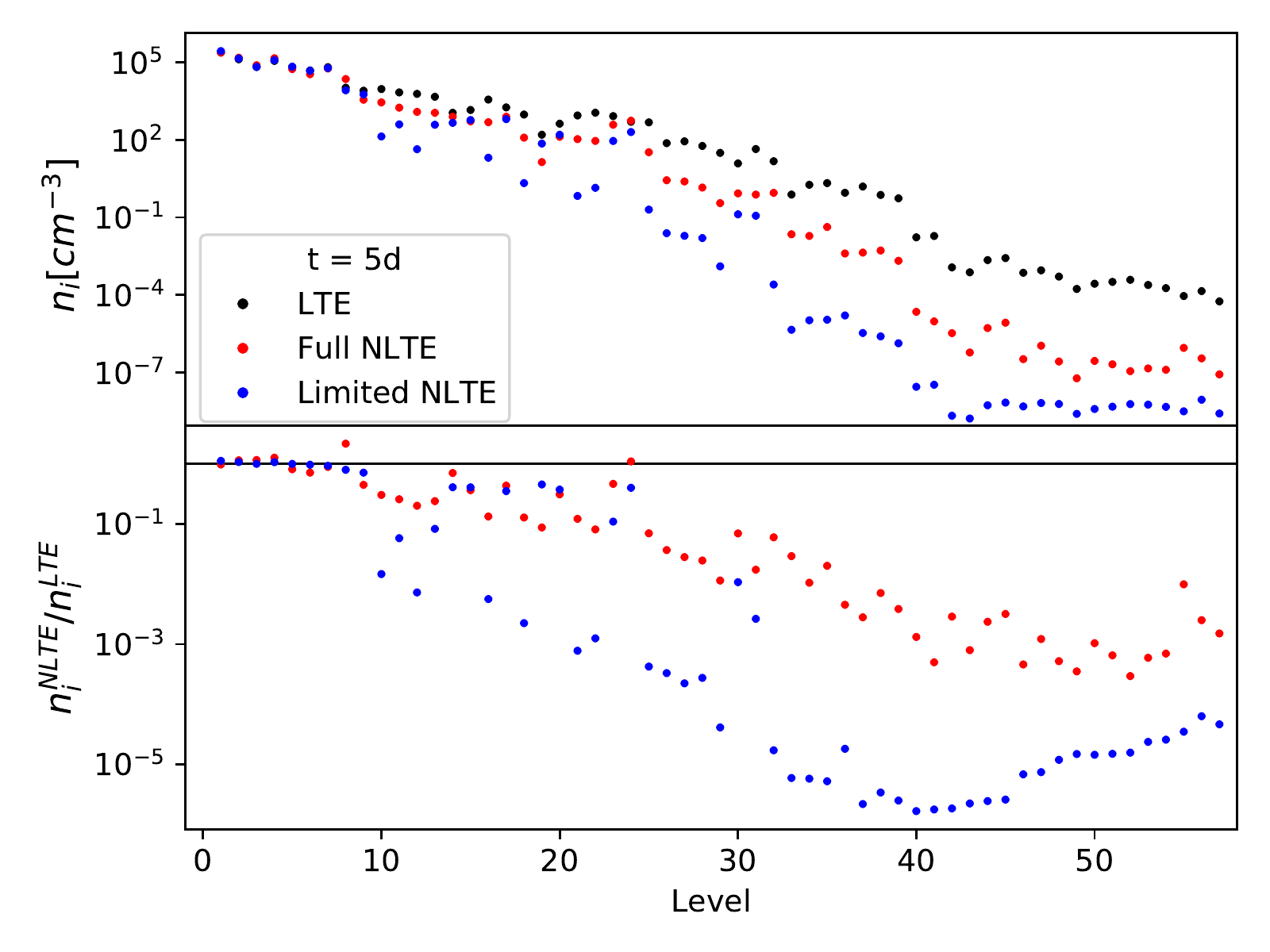} 
    \includegraphics[trim={0.2cm 0.1cm 0.4cm 0.3cm},clip,width = 0.49\textwidth]{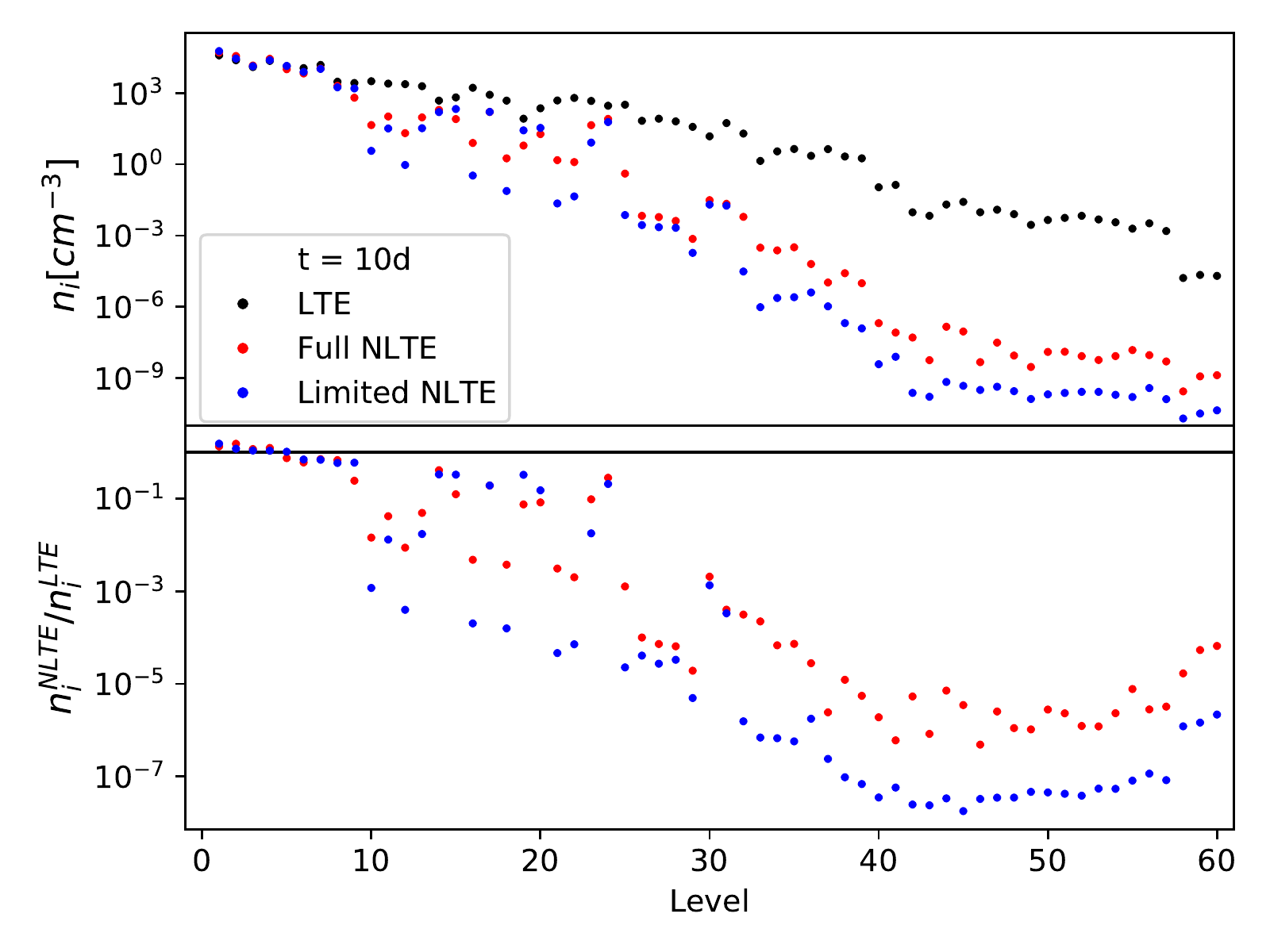} 
    \includegraphics[trim={0.2cm 0.1cm 0.4cm 0.3cm},clip,width = 0.49\textwidth]{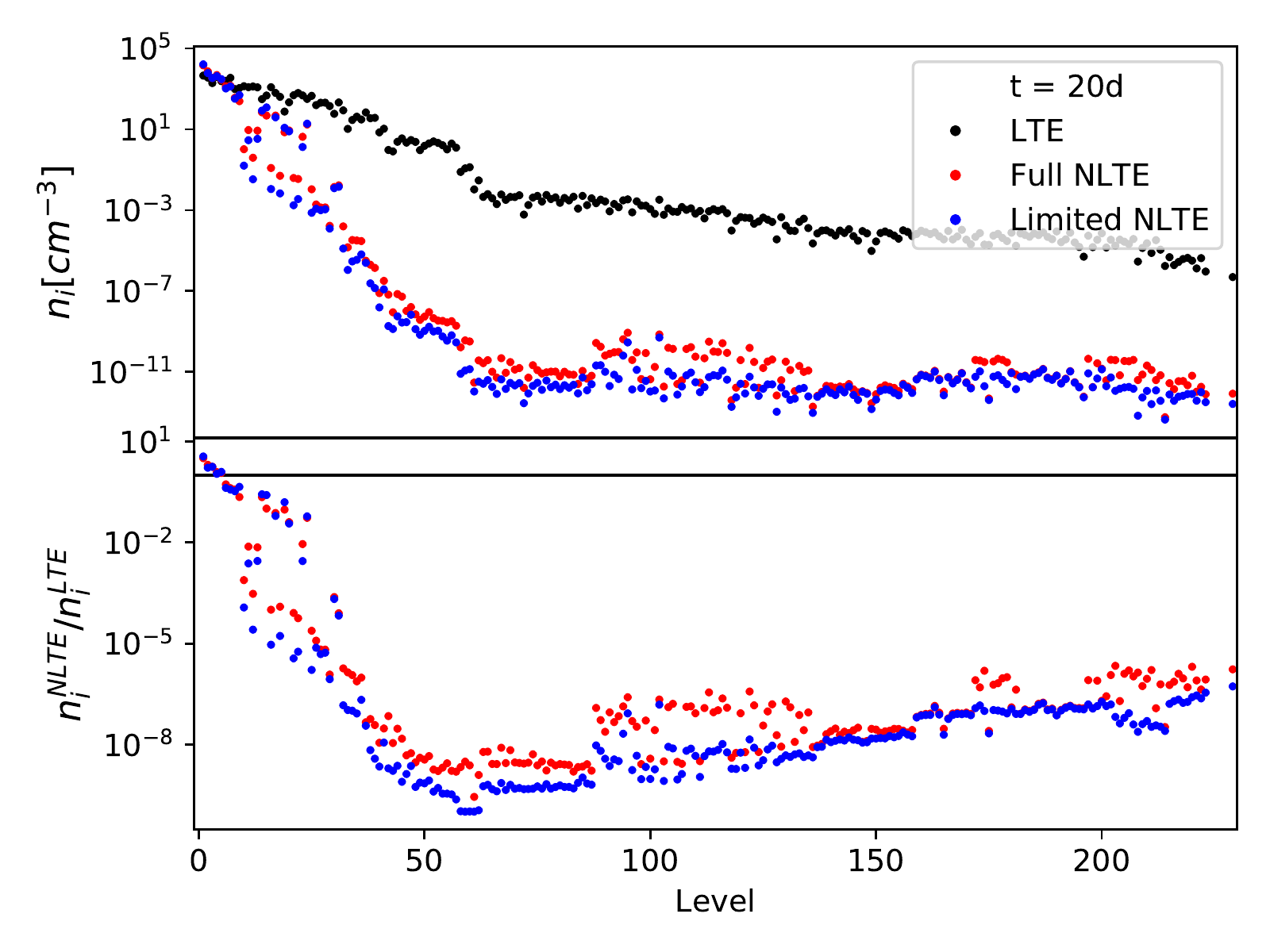} 
    \caption{Level populations of Ce III at 3, 5, 10, and 20 days after merger (top left to bottom right). States with an LTE level population smaller than the most populated state by a factor of $10^{10}$ are cut from these plots.}
    \label{fig:CeIII_levelpops}
\end{figure*}

\begin{figure*}
    \centering
    \includegraphics[trim={0.2cm 0.1cm 0.4cm 0.3cm},clip,width = 0.49\textwidth]{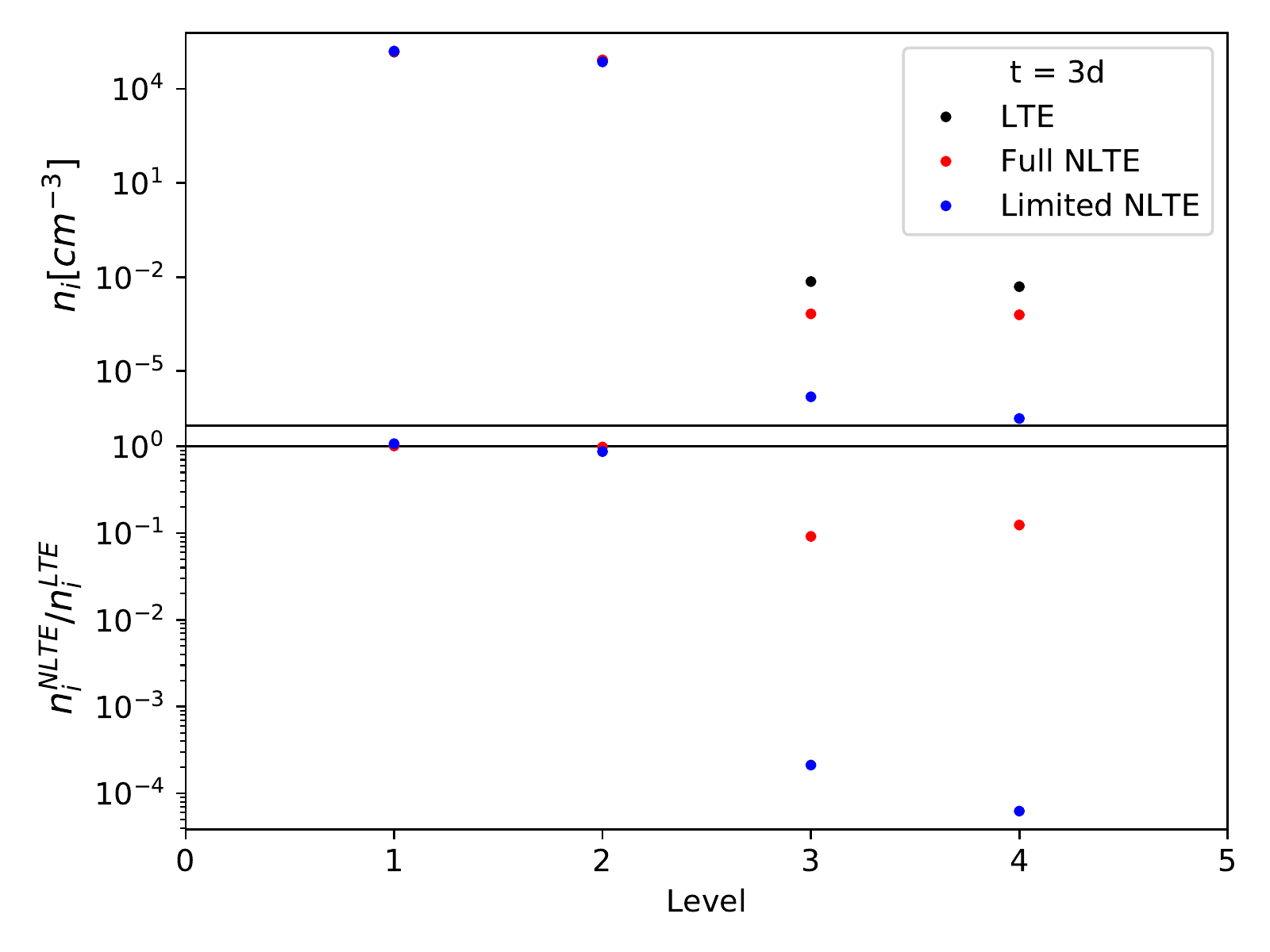}
    \includegraphics[trim={0.2cm 0.1cm 0.4cm 0.3cm},clip,width = 0.49\textwidth]{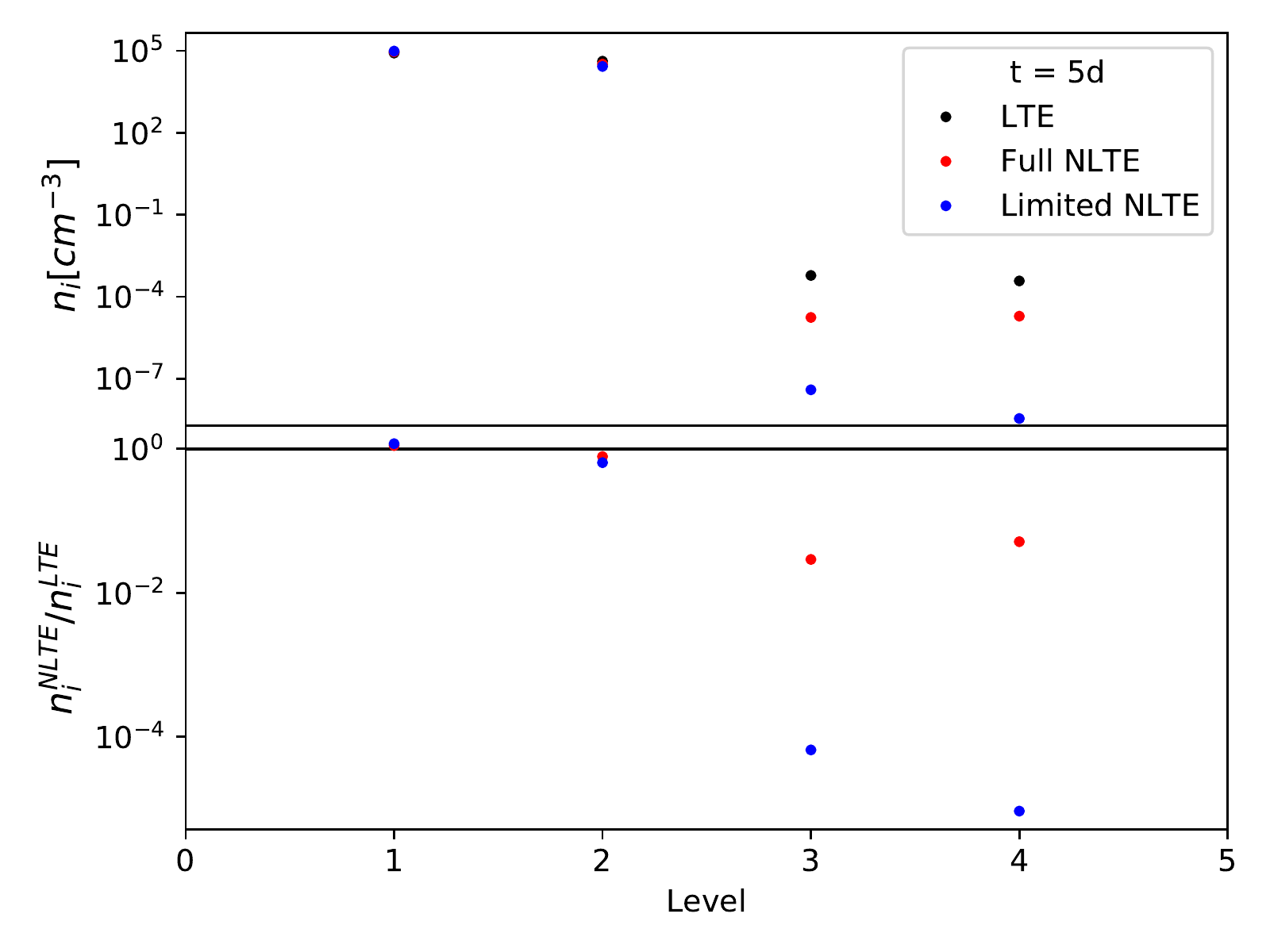} 
    \includegraphics[trim={0.2cm 0.1cm 0.4cm 0.3cm},clip,width = 0.49\textwidth]{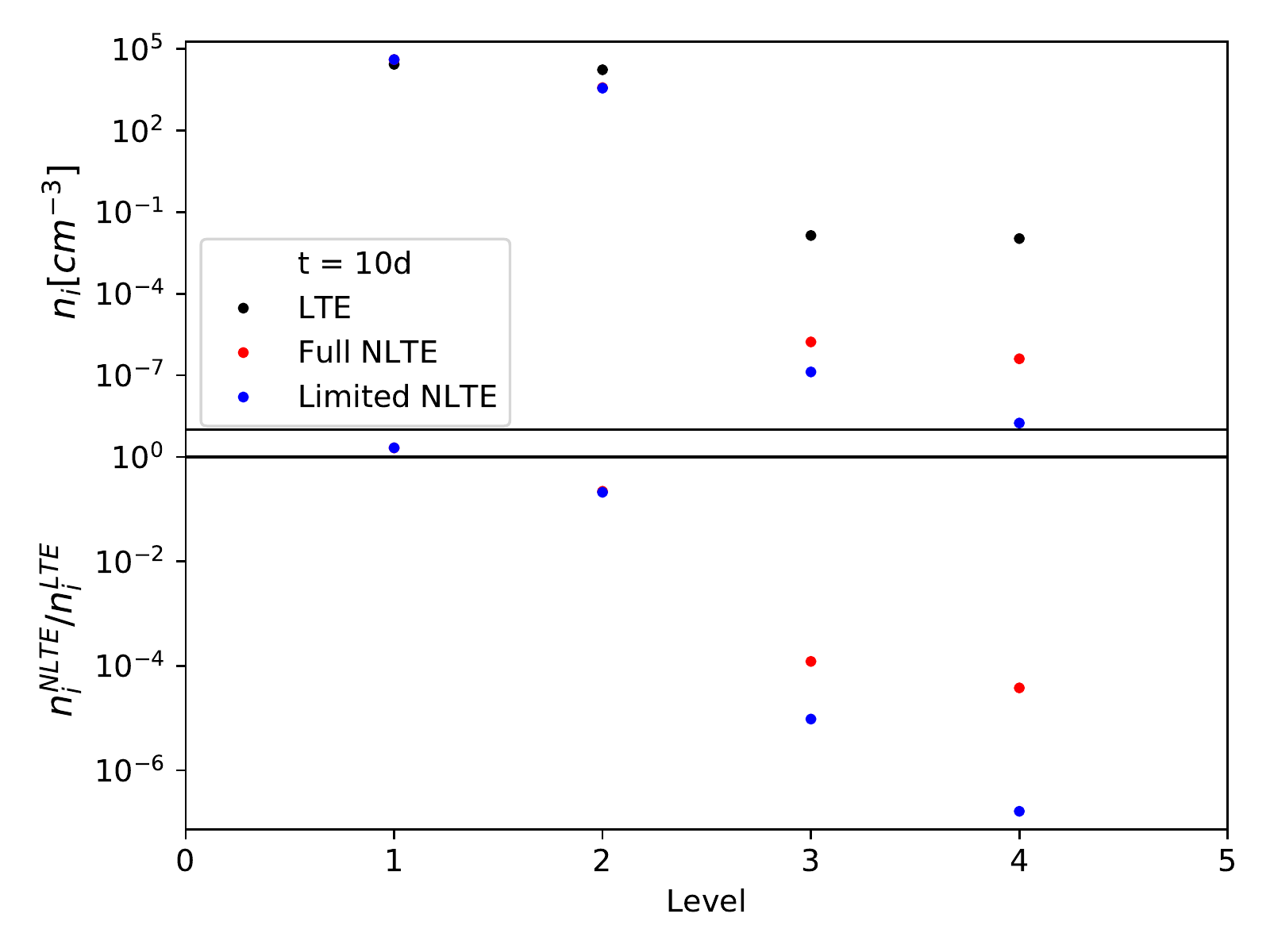} 
    \includegraphics[trim={0.2cm 0.1cm 0.4cm 0.3cm},clip,width = 0.49\textwidth]{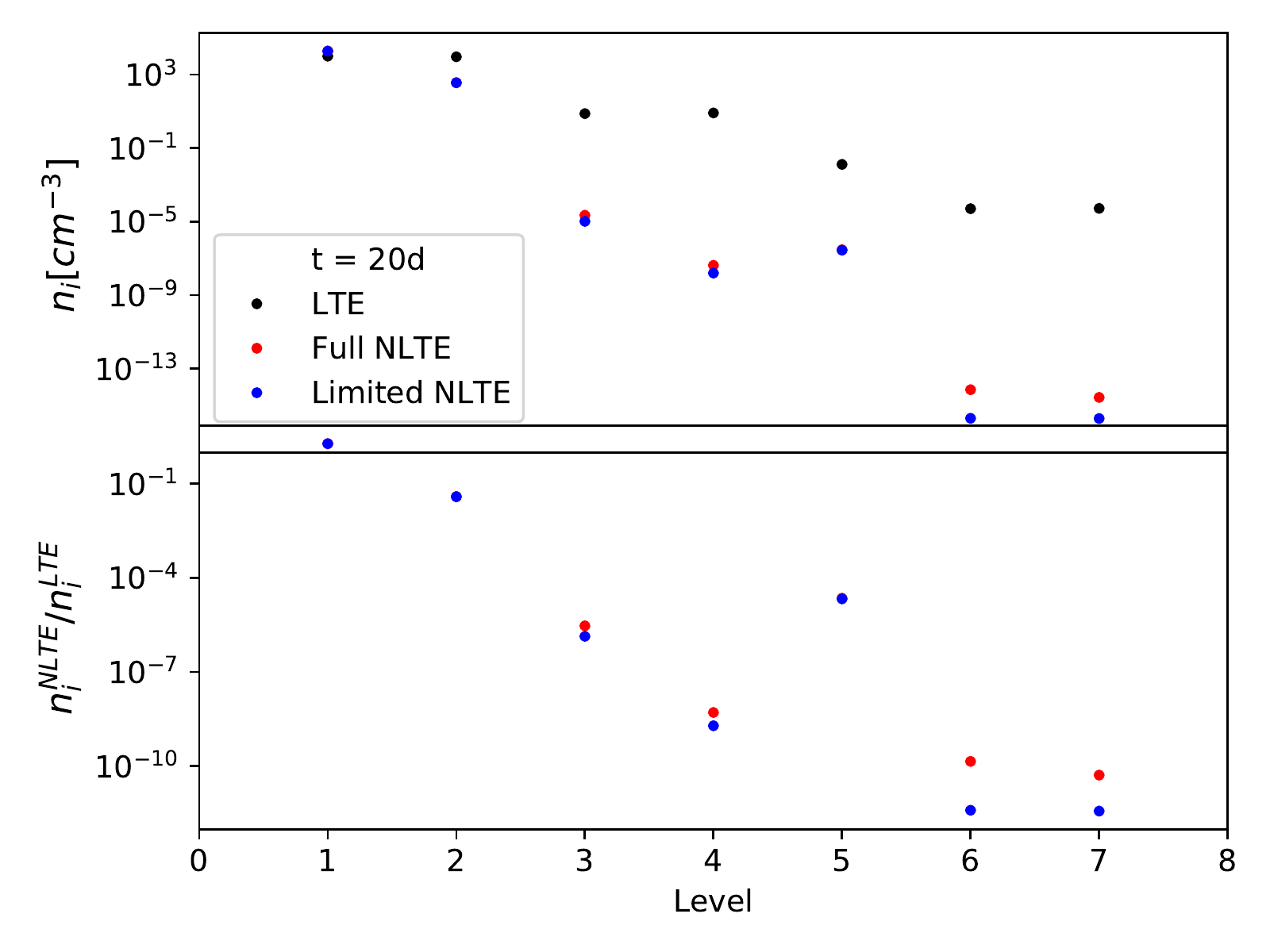} 
    \caption{Level populations of Ce IV at 3, 5, 10, and 20 days after merger (top left to bottom right). States with an LTE level population smaller than the most populated state by a factor of $10^{10}$ are cut from these plots.}
    \label{fig:CeIV_levelpops}
\end{figure*}

\begin{figure*}
    \centering
    \includegraphics[trim={0.2cm 0.1cm 0.4cm 0.3cm},clip,width = 0.49\textwidth]{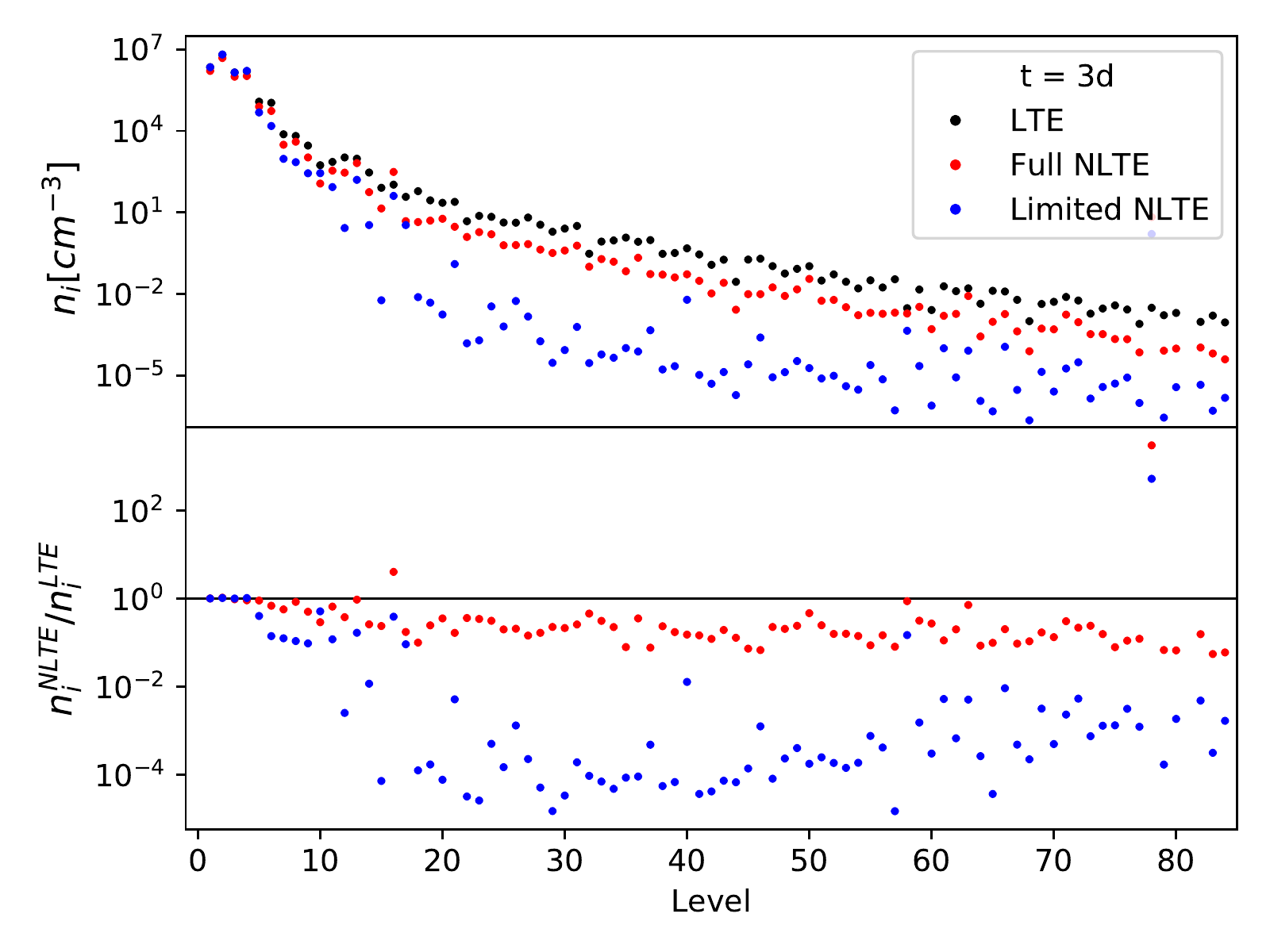}
    \includegraphics[trim={0.2cm 0.1cm 0.4cm 0.3cm},clip,width = 0.49\textwidth]{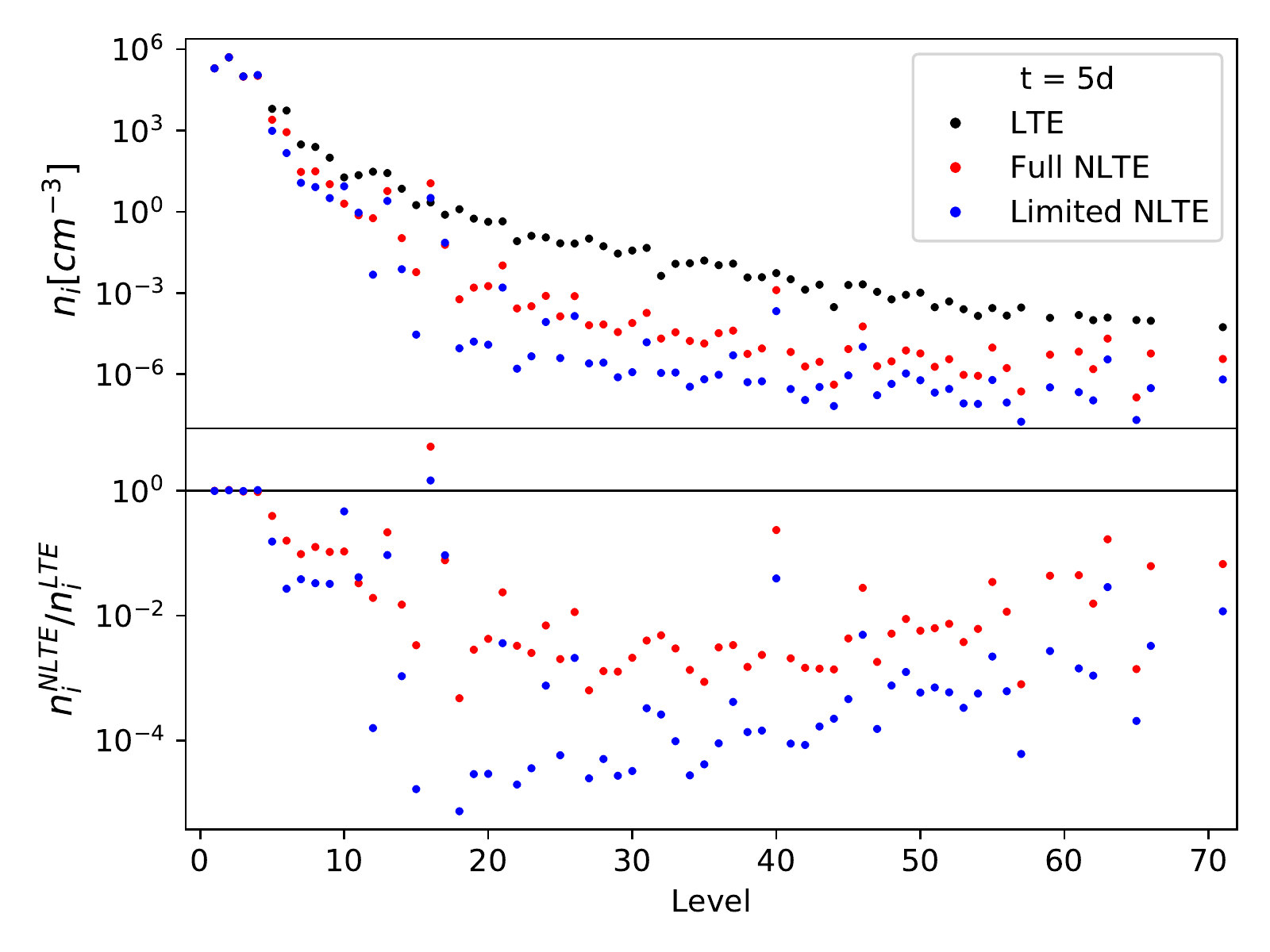} 
    \includegraphics[trim={0.2cm 0.1cm 0.4cm 0.3cm},clip,width = 0.49\textwidth]{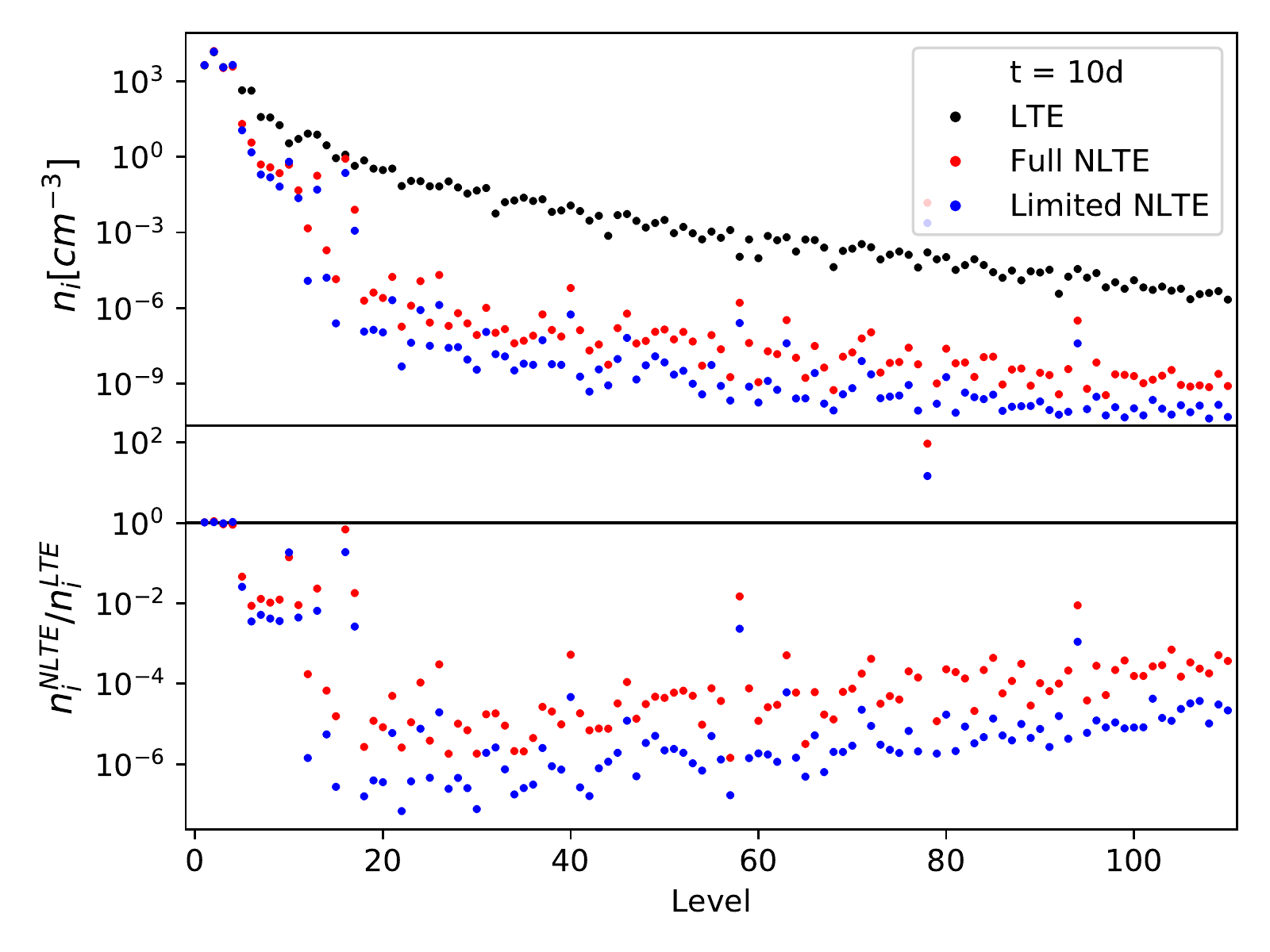} 
    \includegraphics[trim={0.2cm 0.1cm 0.4cm 0.3cm},clip,width = 0.49\textwidth]{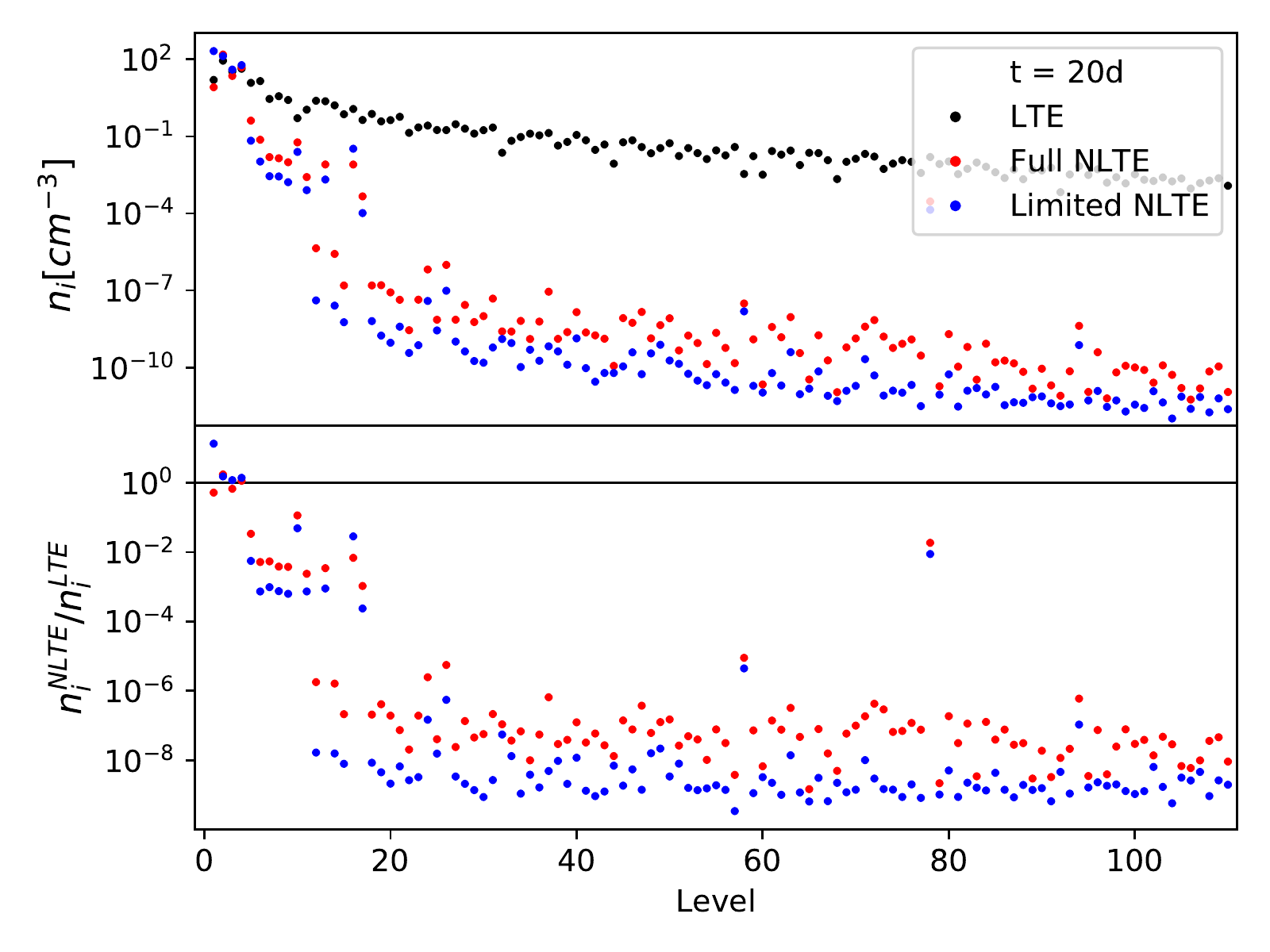} 
    \caption{Level populations of Pt I at 3, 5, 10, and 20 days after merger (top left to bottom right). States with an LTE level population smaller than the most populated state by a factor of $10^{10}$ are cut from these plots.}
    \label{fig:PtI_levelpops}
\end{figure*}

\begin{figure*}
    \centering
    \includegraphics[trim={0.2cm 0.1cm 0.4cm 0.3cm},clip,width = 0.49\textwidth]{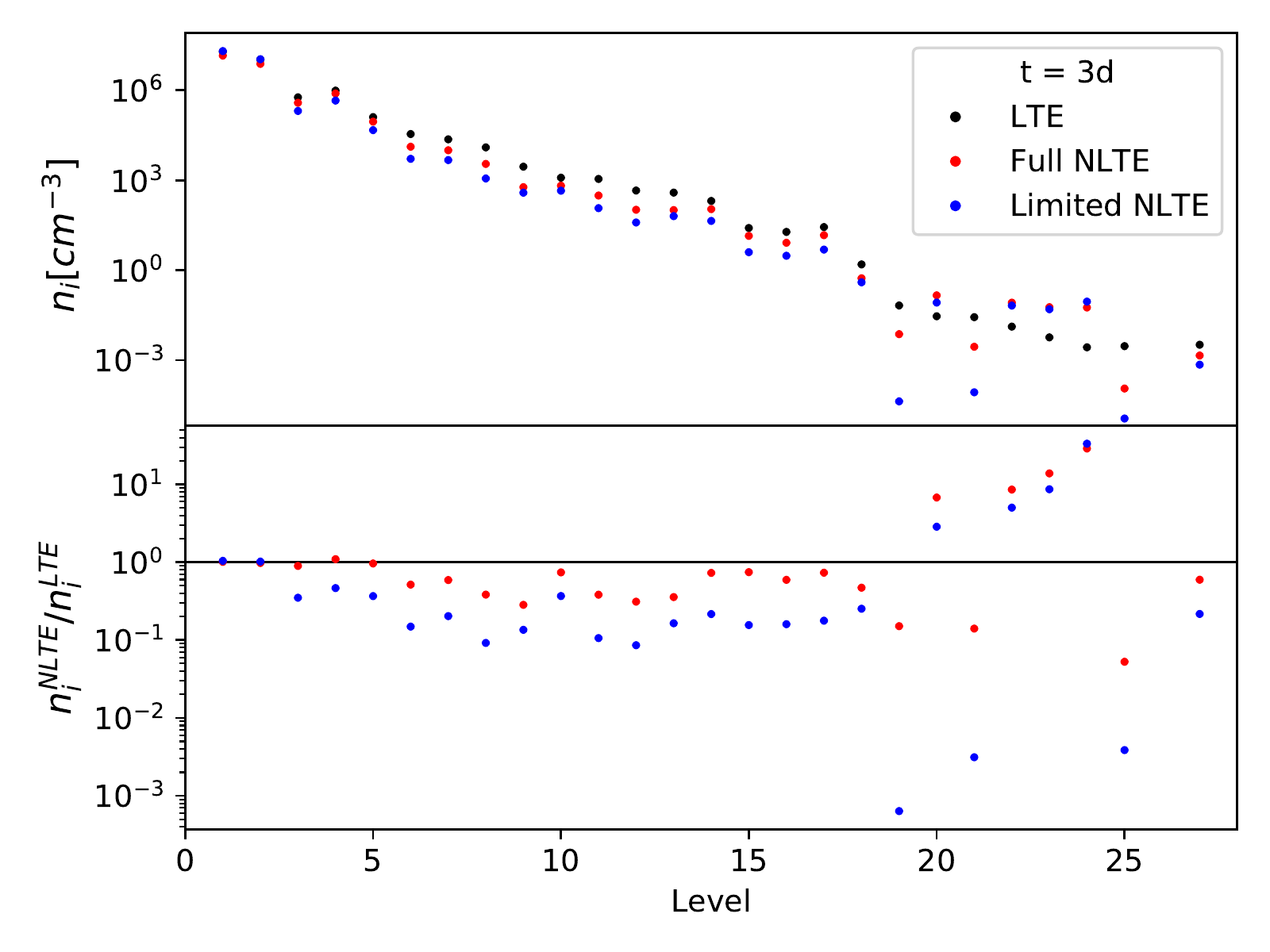}
    \includegraphics[trim={0.2cm 0.1cm 0.4cm 0.3cm},clip,width = 0.49\textwidth]{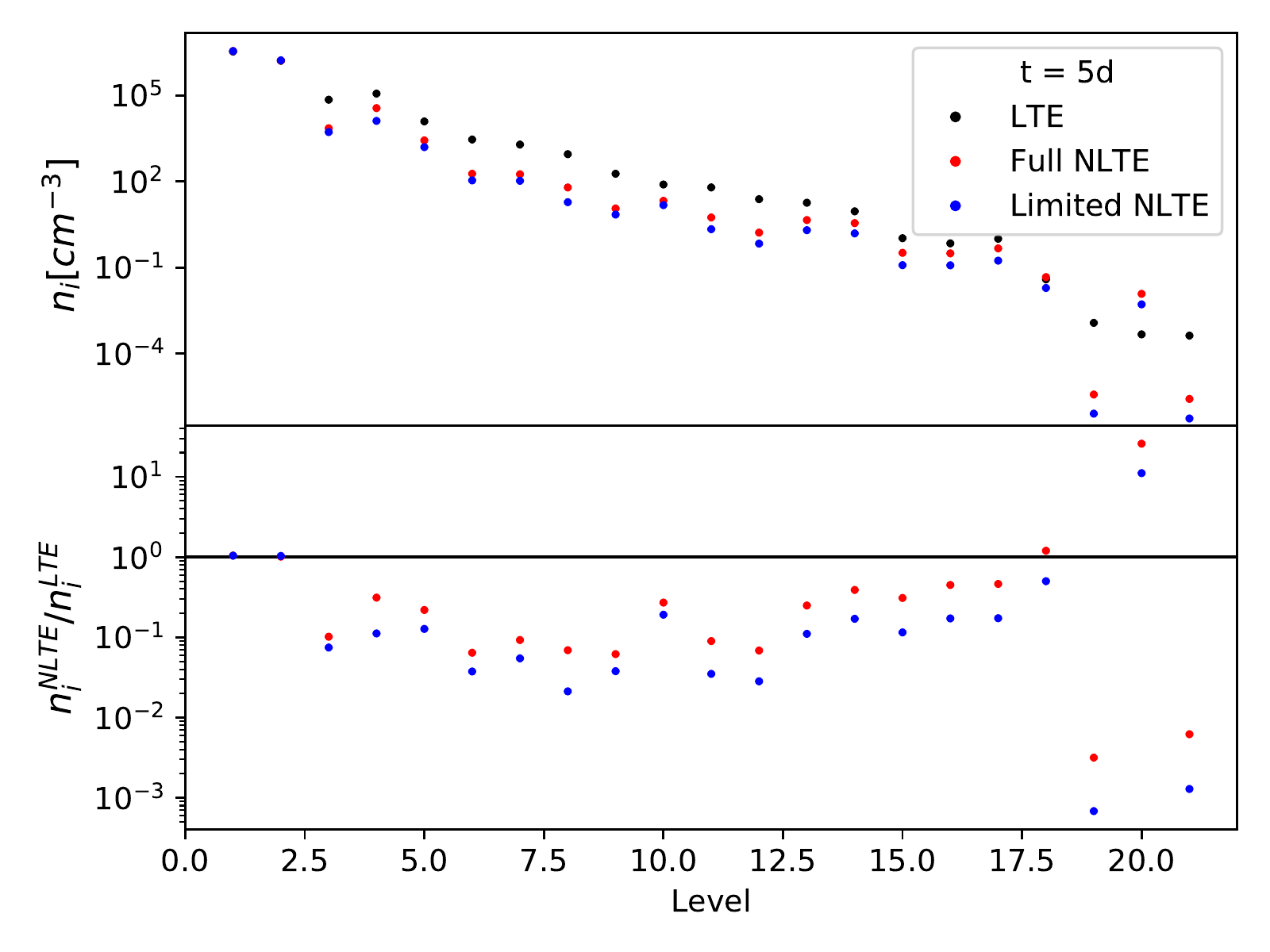} 
    \includegraphics[trim={0.2cm 0.1cm 0.4cm 0.3cm},clip,width = 0.49\textwidth]{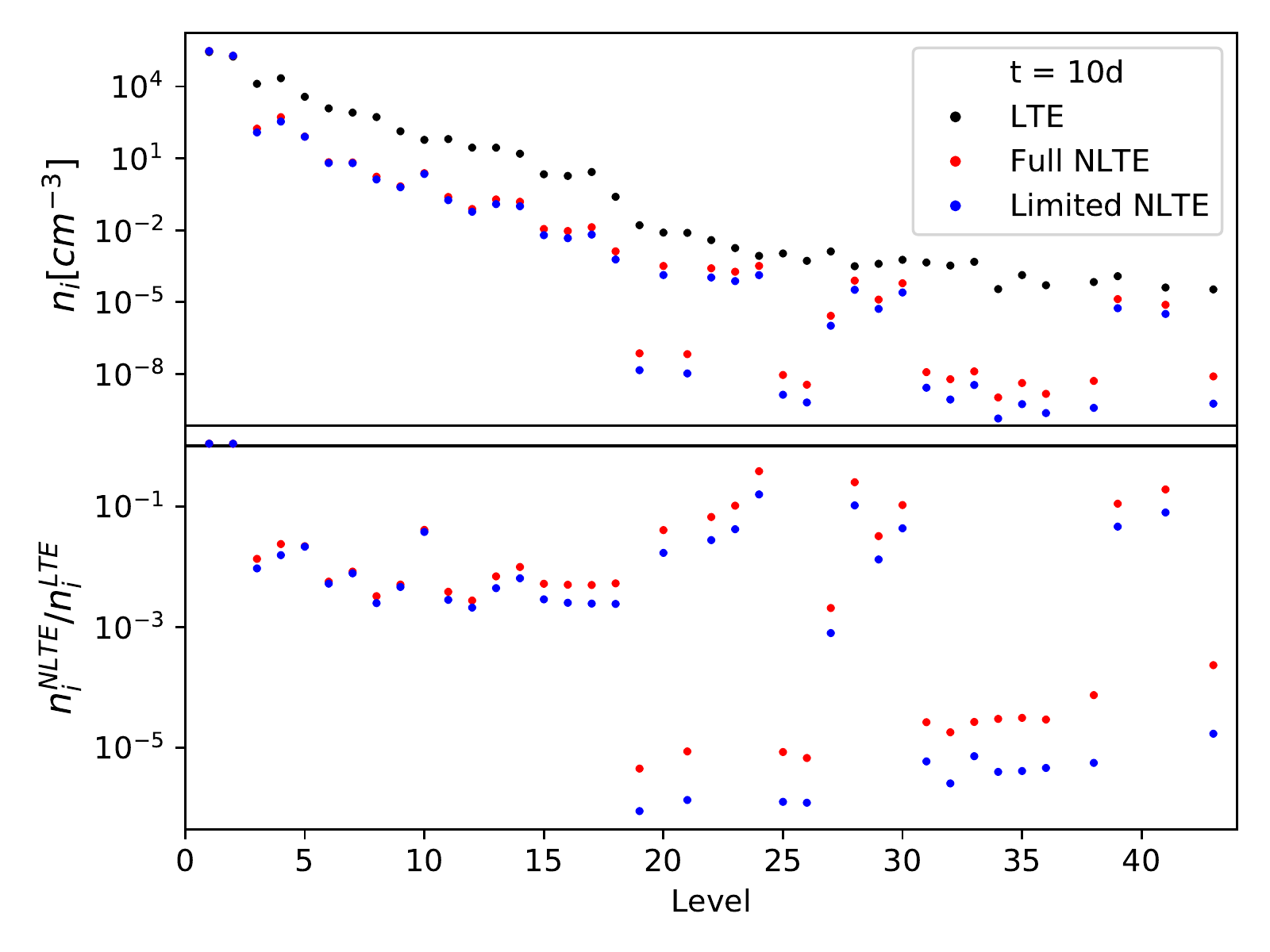} 
    \includegraphics[trim={0.2cm 0.1cm 0.4cm 0.3cm},clip,width = 0.49\textwidth]{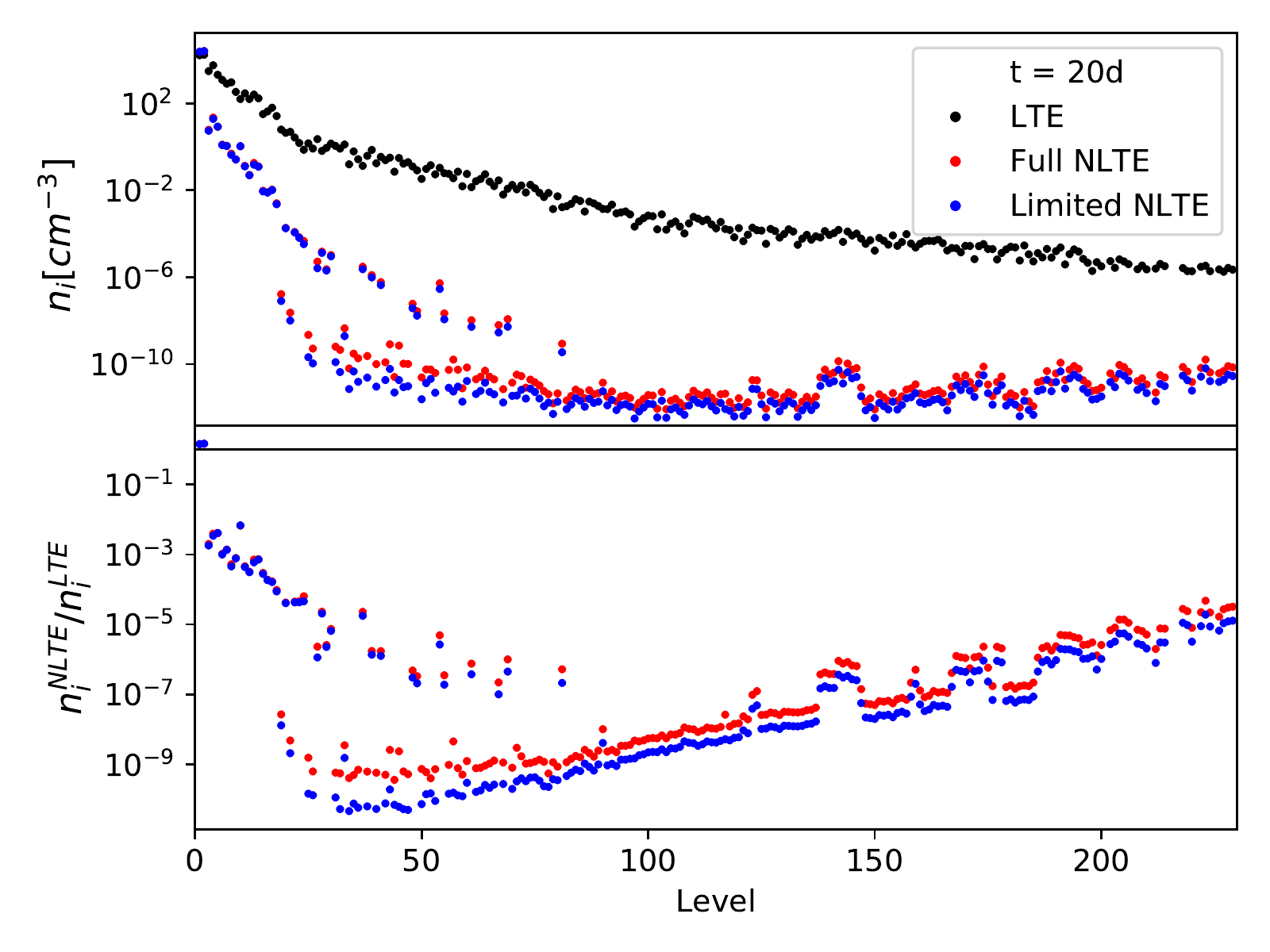} 
    \caption{Level populations of Pt II at 3, 5, 10, and 20 days after merger (top left to bottom right). States with an LTE level population smaller than the most populated state by a factor of $10^{10}$ are cut from these plots.}
    \label{fig:PtII_levelpops}
\end{figure*}

\begin{figure*}
    \centering
    \includegraphics[trim={0.2cm 0.1cm 0.4cm 0.3cm},clip,width = 0.49\textwidth]{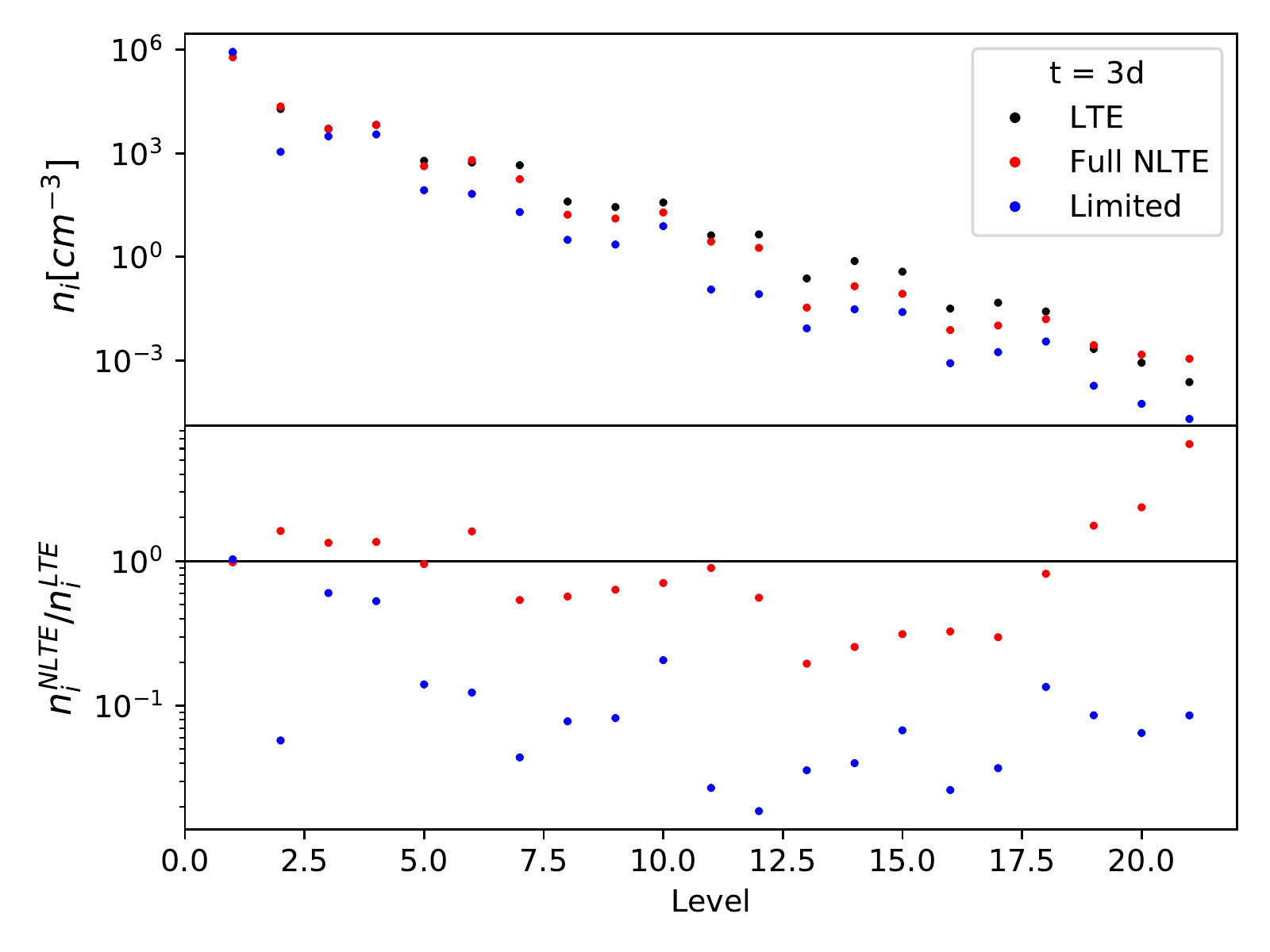}
    \includegraphics[trim={0.2cm 0.1cm 0.4cm 0.3cm},clip,width = 0.49\textwidth]{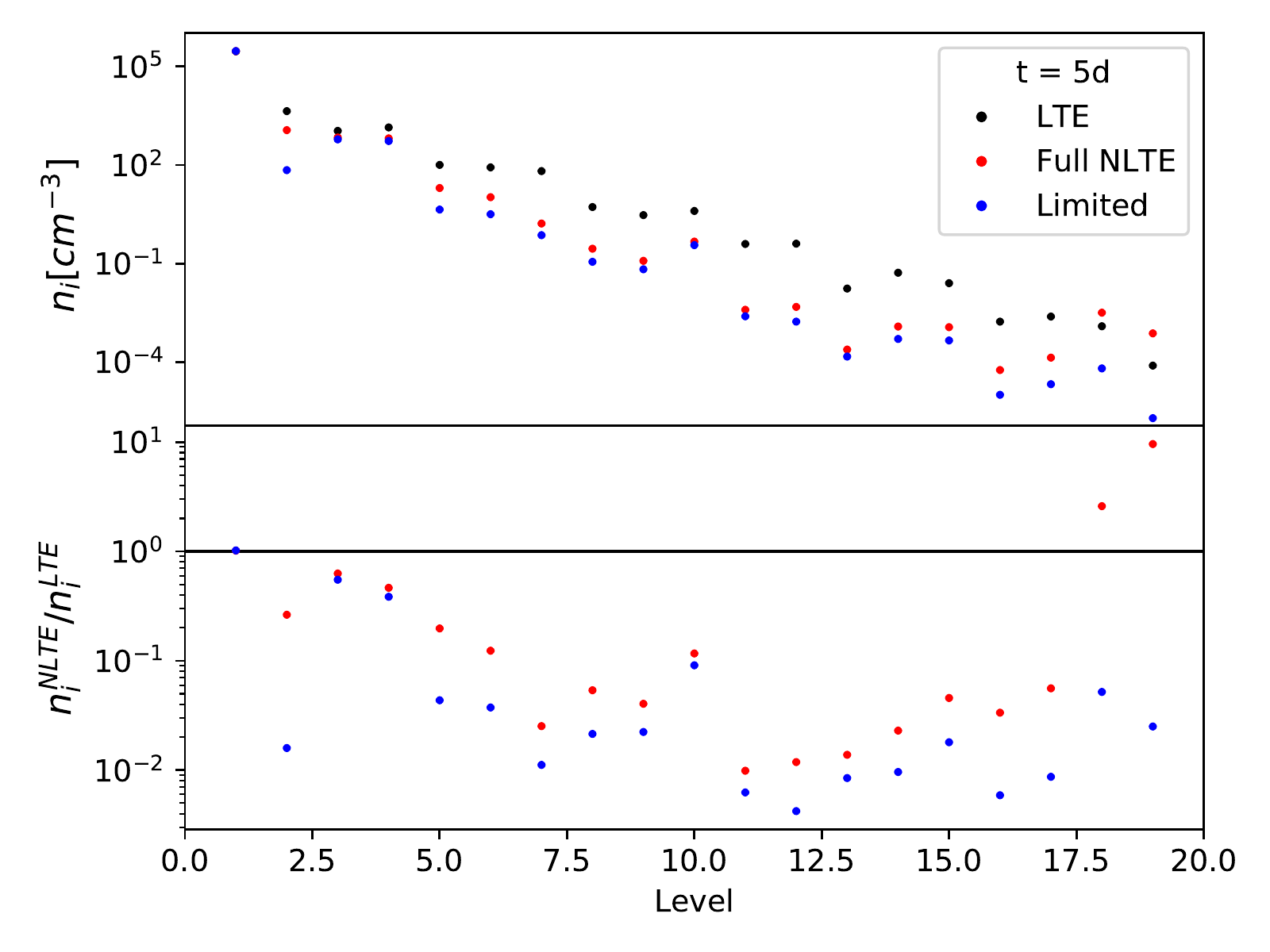} 
    \includegraphics[trim={0.2cm 0.1cm 0.4cm 0.3cm},clip,width = 0.49\textwidth]{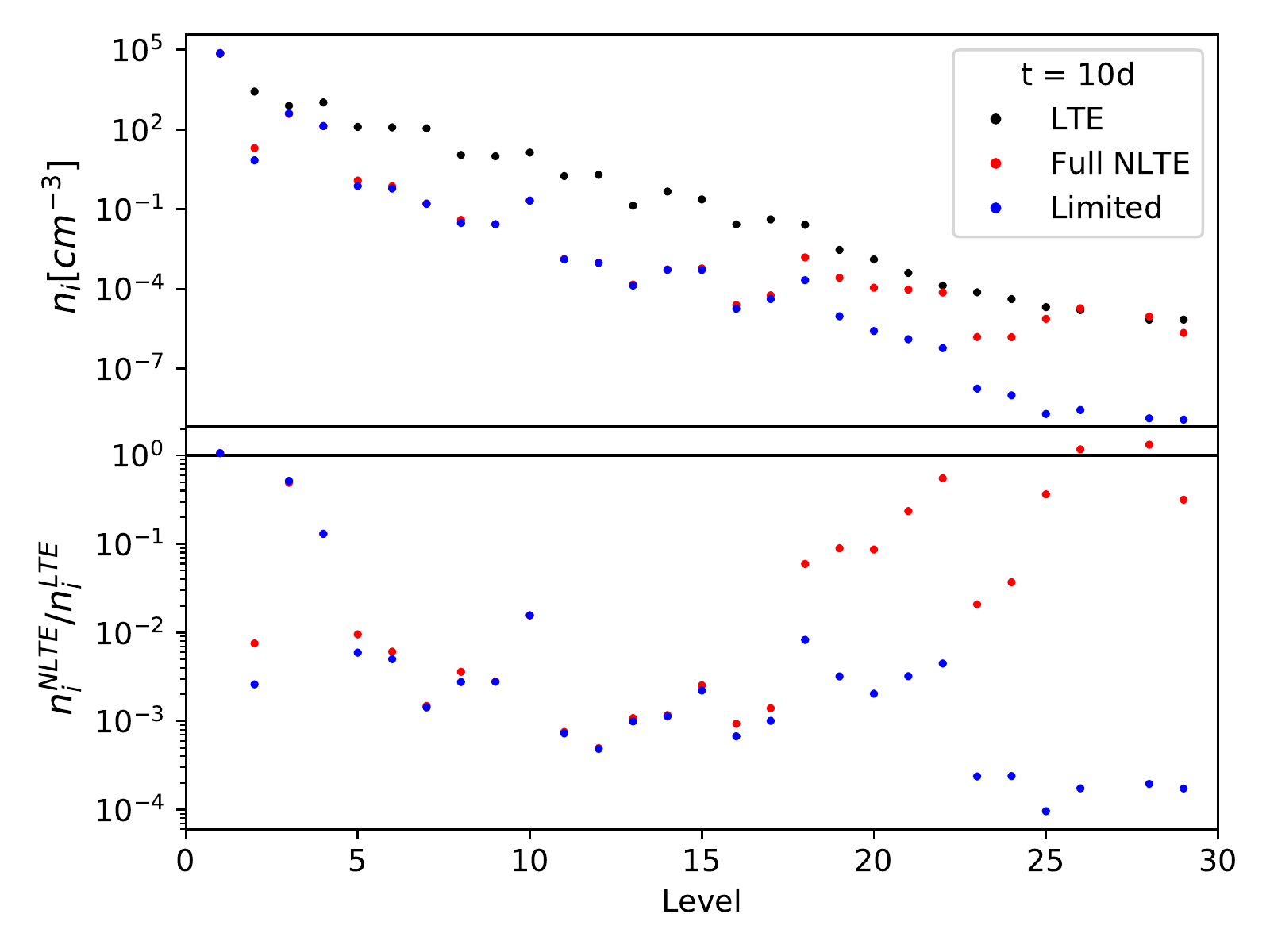} 
    \includegraphics[trim={0.2cm 0.1cm 0.4cm 0.3cm},clip,width = 0.49\textwidth]{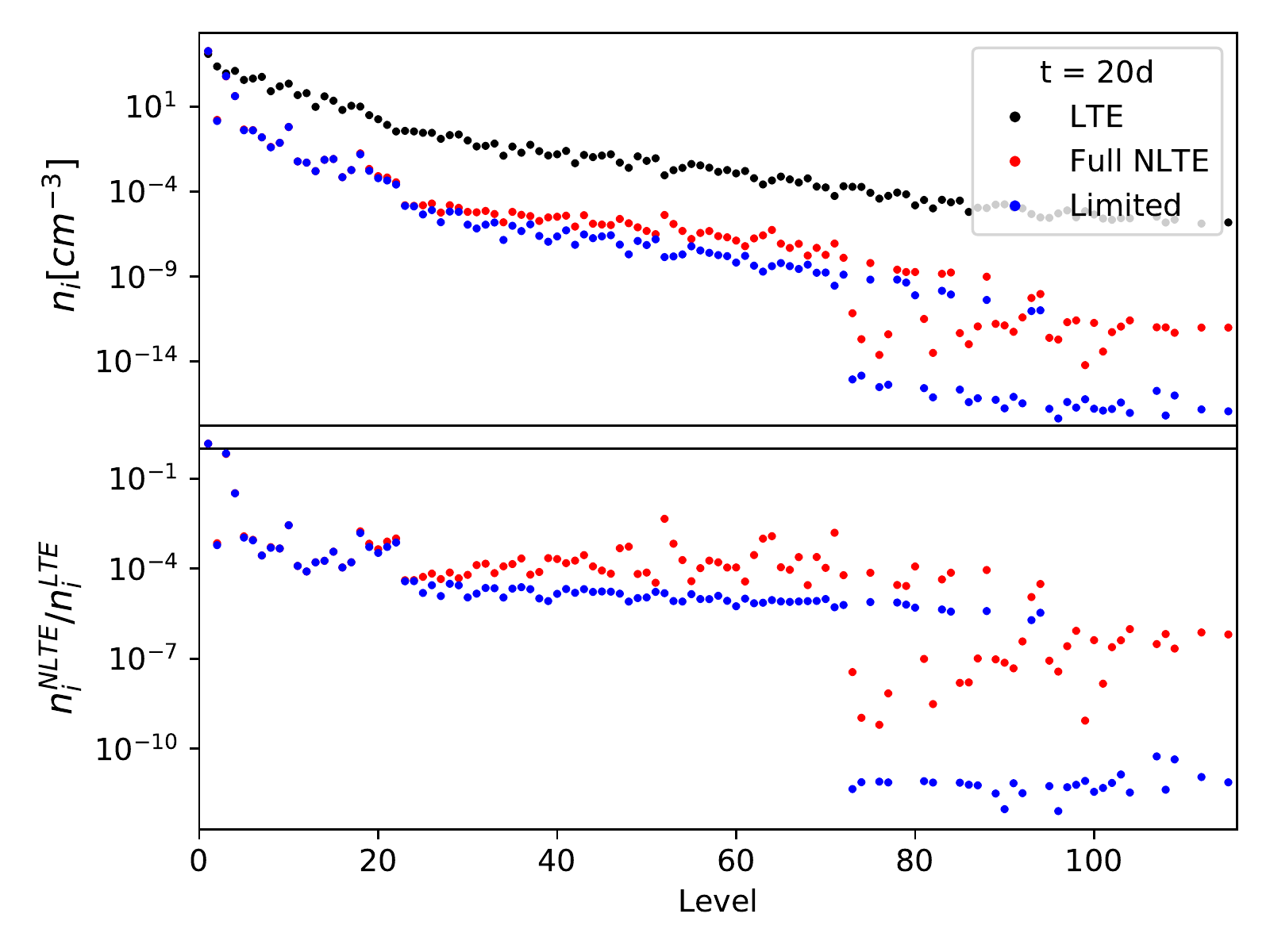} 
    \caption{Level populations of Pt IV at 3, 5, 10, and 20 days after merger (top left to bottom right). States with an LTE level population smaller than the most populated state by a factor of $10^{10}$ are cut from these plots.}
    \label{fig:PtIV_levelpops}
\end{figure*}

\begin{figure*}
    \centering
    \includegraphics[trim={0.2cm 0.1cm 0.4cm 0.3cm},clip,width = 0.49\textwidth]{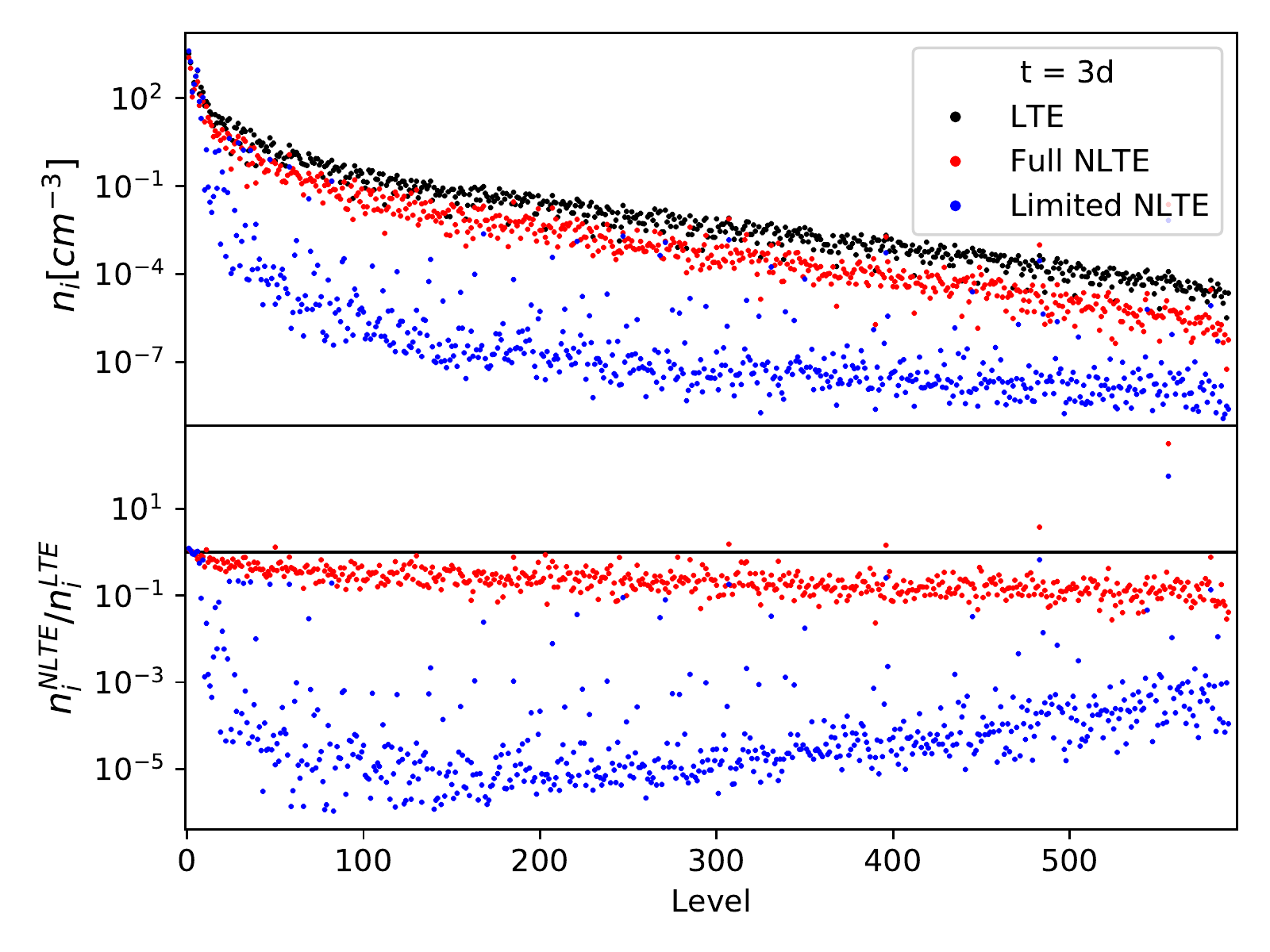}
    \includegraphics[trim={0.2cm 0.1cm 0.4cm 0.3cm},clip,width = 0.49\textwidth]{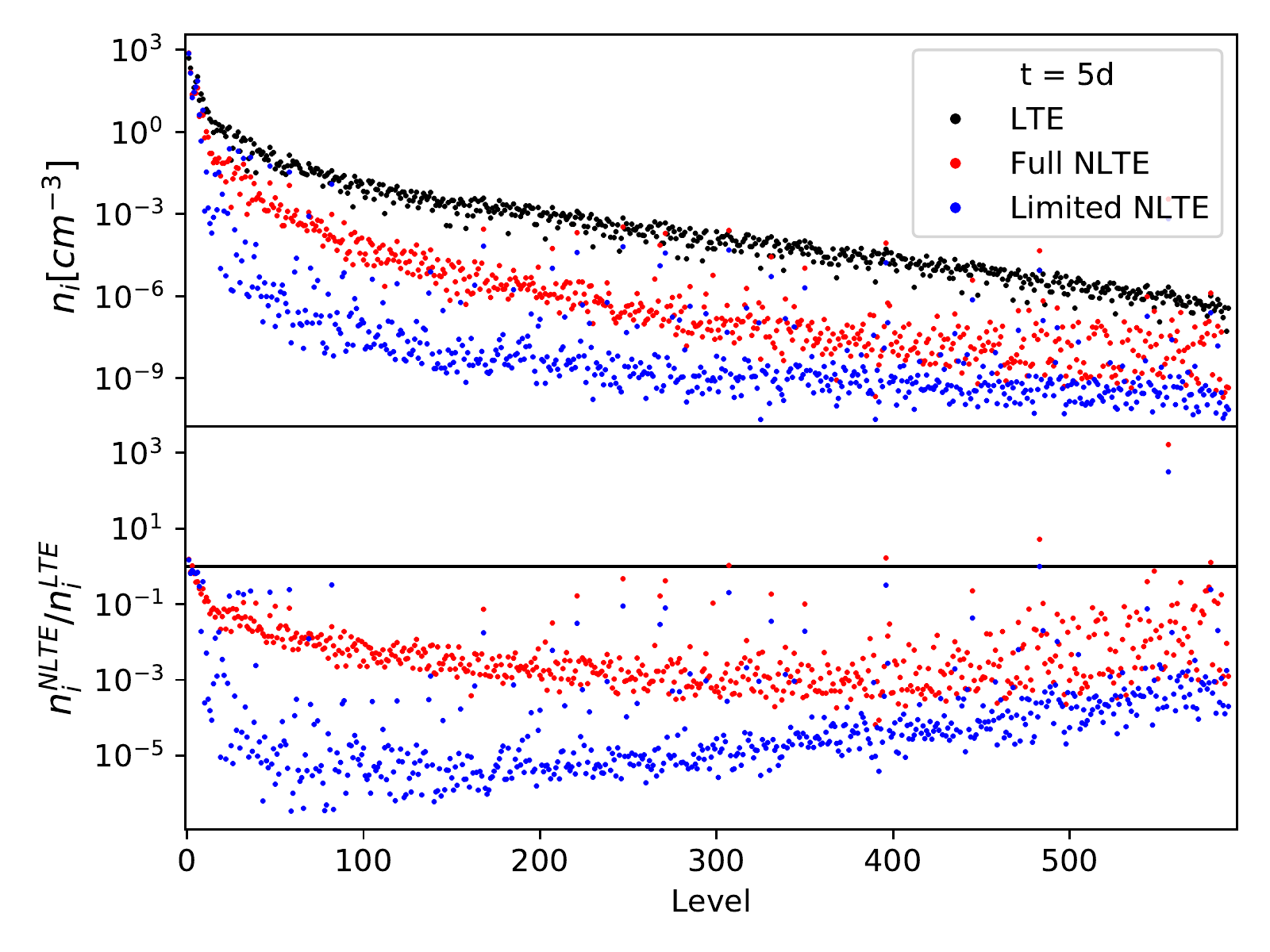} 
    \includegraphics[trim={0.2cm 0.1cm 0.4cm 0.3cm},clip,width = 0.49\textwidth]{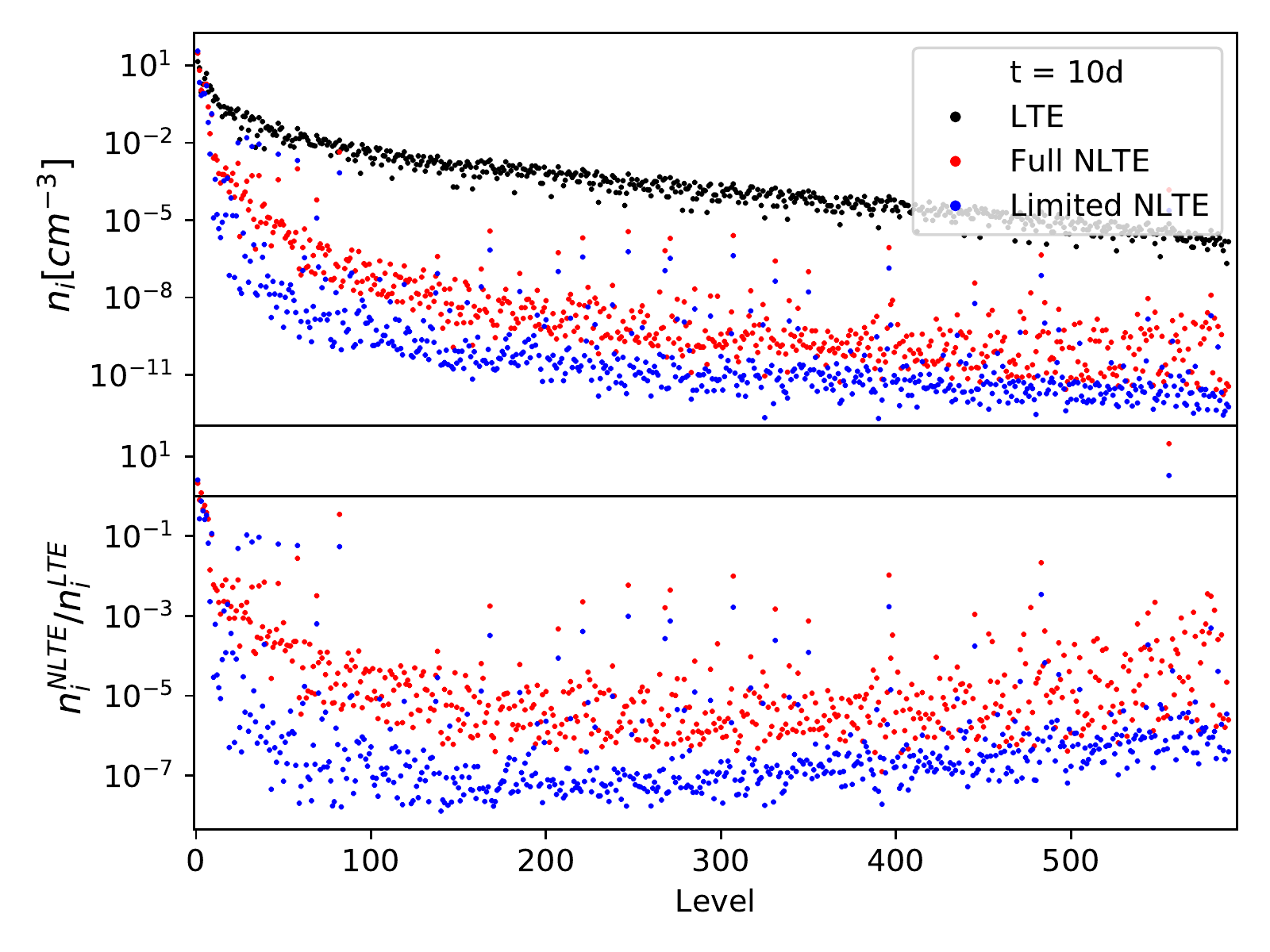} 
    \includegraphics[trim={0.2cm 0.1cm 0.4cm 0.3cm},clip,width = 0.49\textwidth]{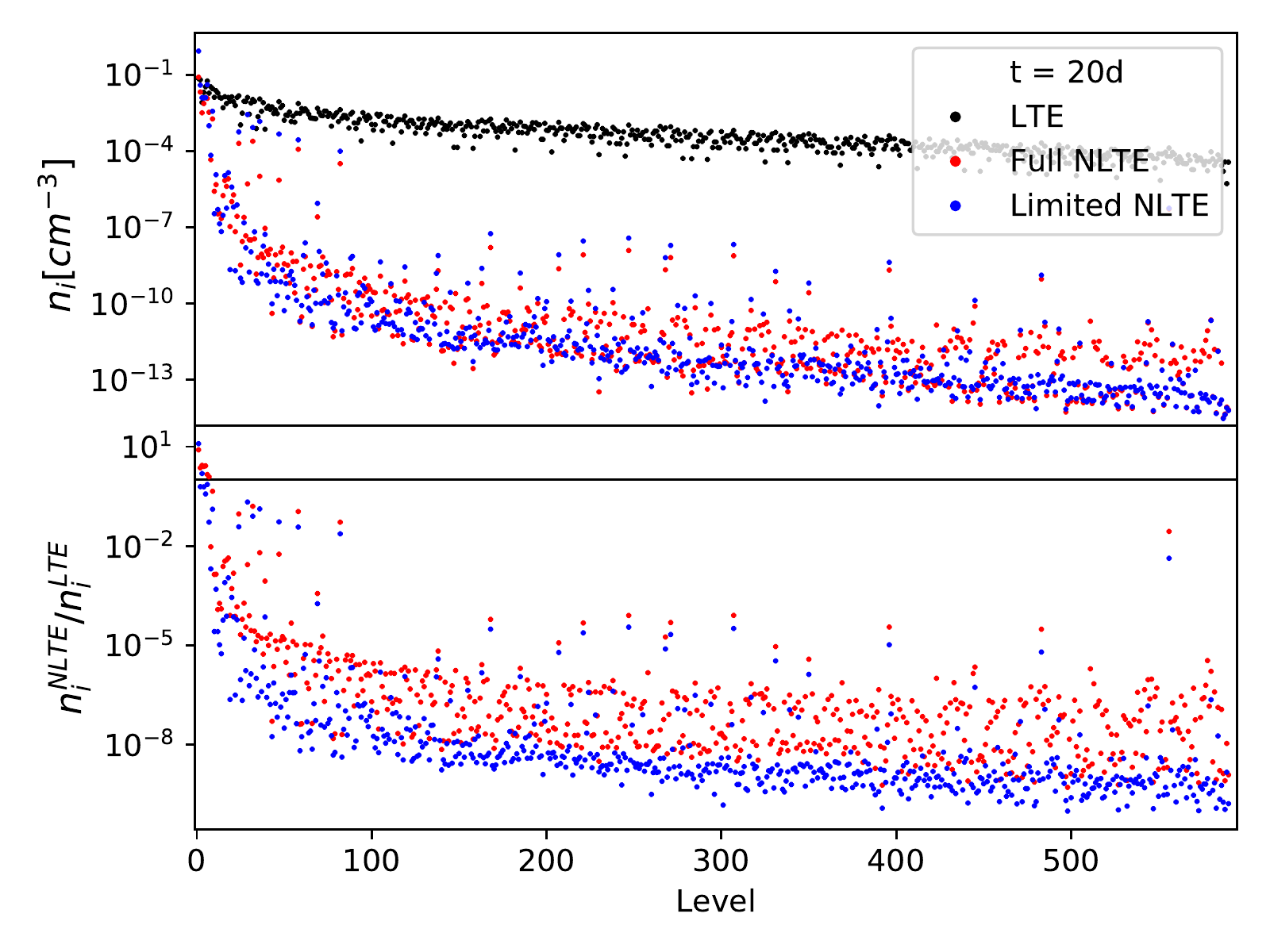} 
    \caption{Level populations of Th I at 3, 5, 10, and 20 days after merger (top left to bottom right). States with an LTE level population smaller than the most populated state by a factor of $10^{10}$ are cut from these plots.}
    \label{fig:ThI_levelpops}
\end{figure*}

\begin{figure*}
    \centering
    \includegraphics[trim={0.2cm 0.1cm 0.4cm 0.3cm},clip,width = 0.49\textwidth]{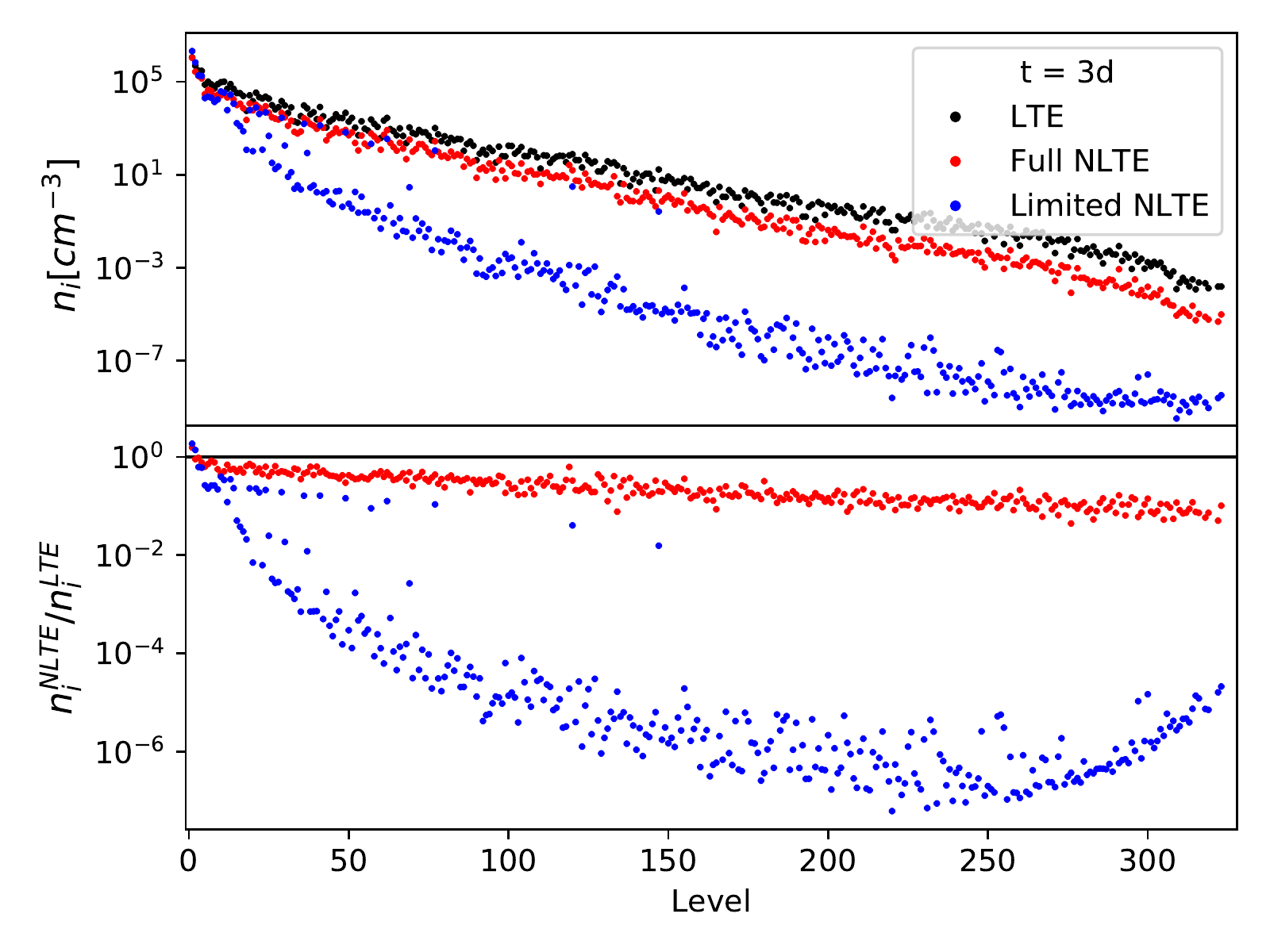}
    \includegraphics[trim={0.2cm 0.1cm 0.4cm 0.3cm},clip,width = 0.49\textwidth]{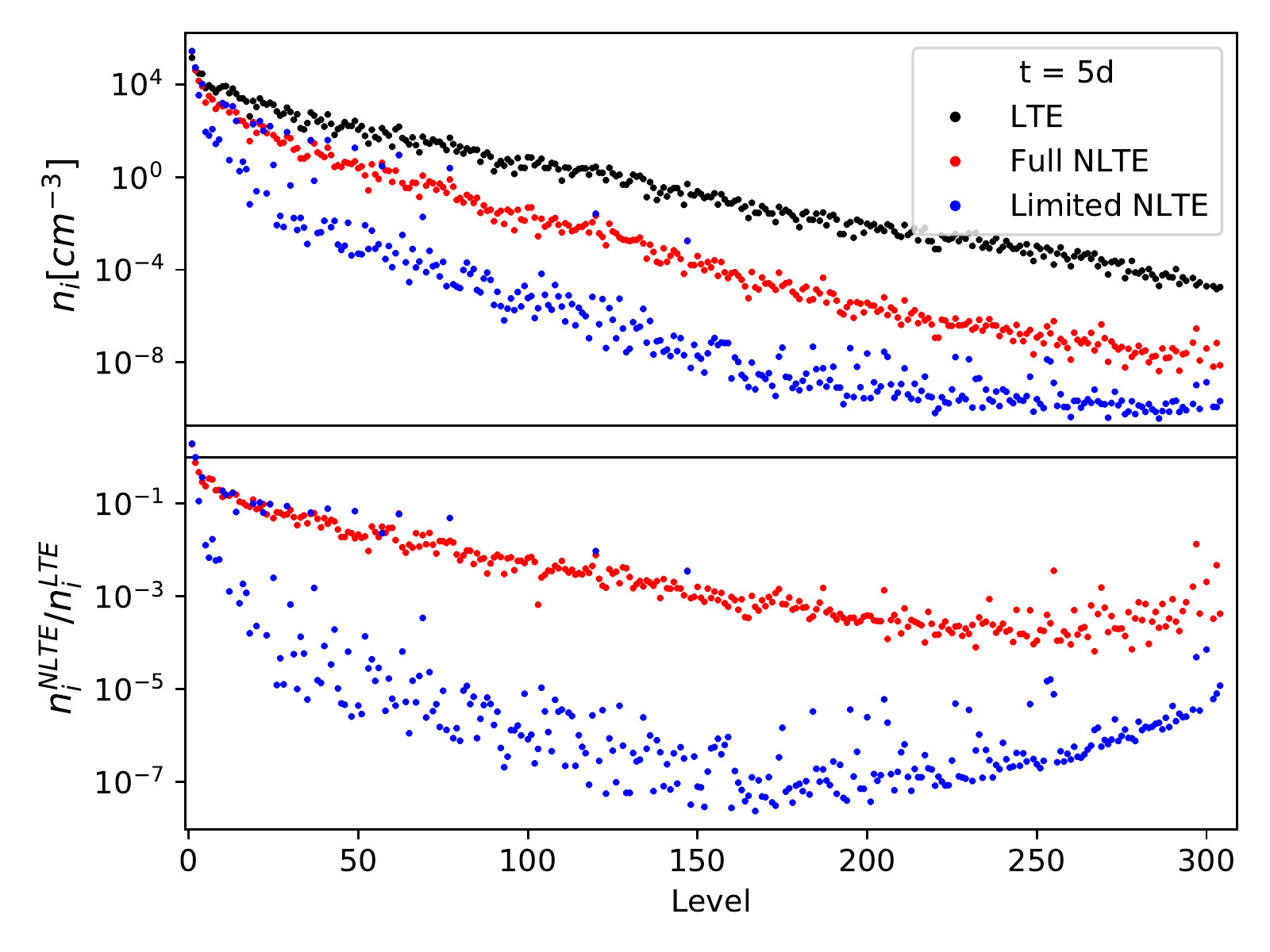} 
    \includegraphics[trim={0.2cm 0.1cm 0.4cm 0.3cm},clip,width = 0.49\textwidth]{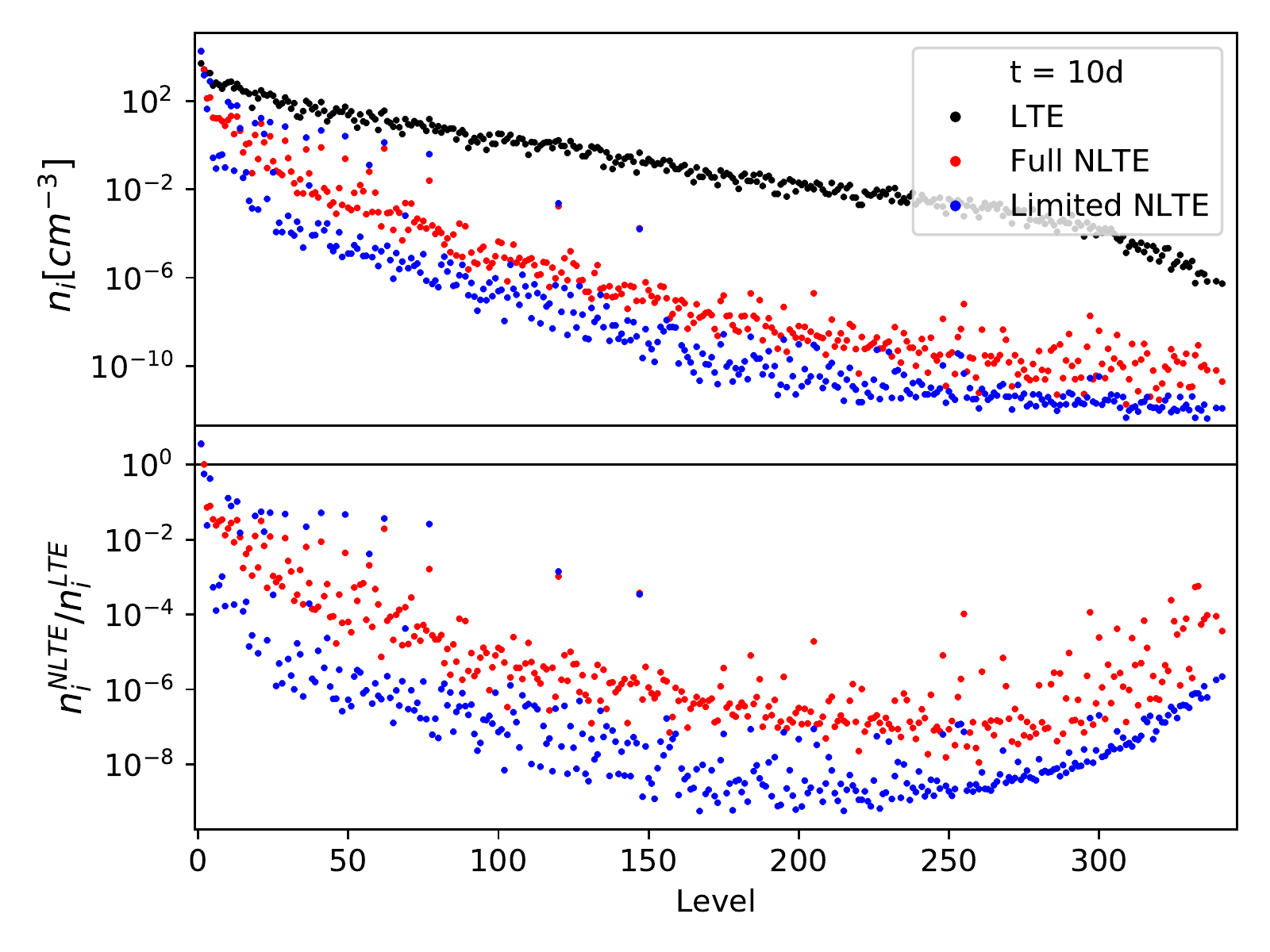} 
    \includegraphics[trim={0.2cm 0.1cm 0.4cm 0.3cm},clip,width = 0.49\textwidth]{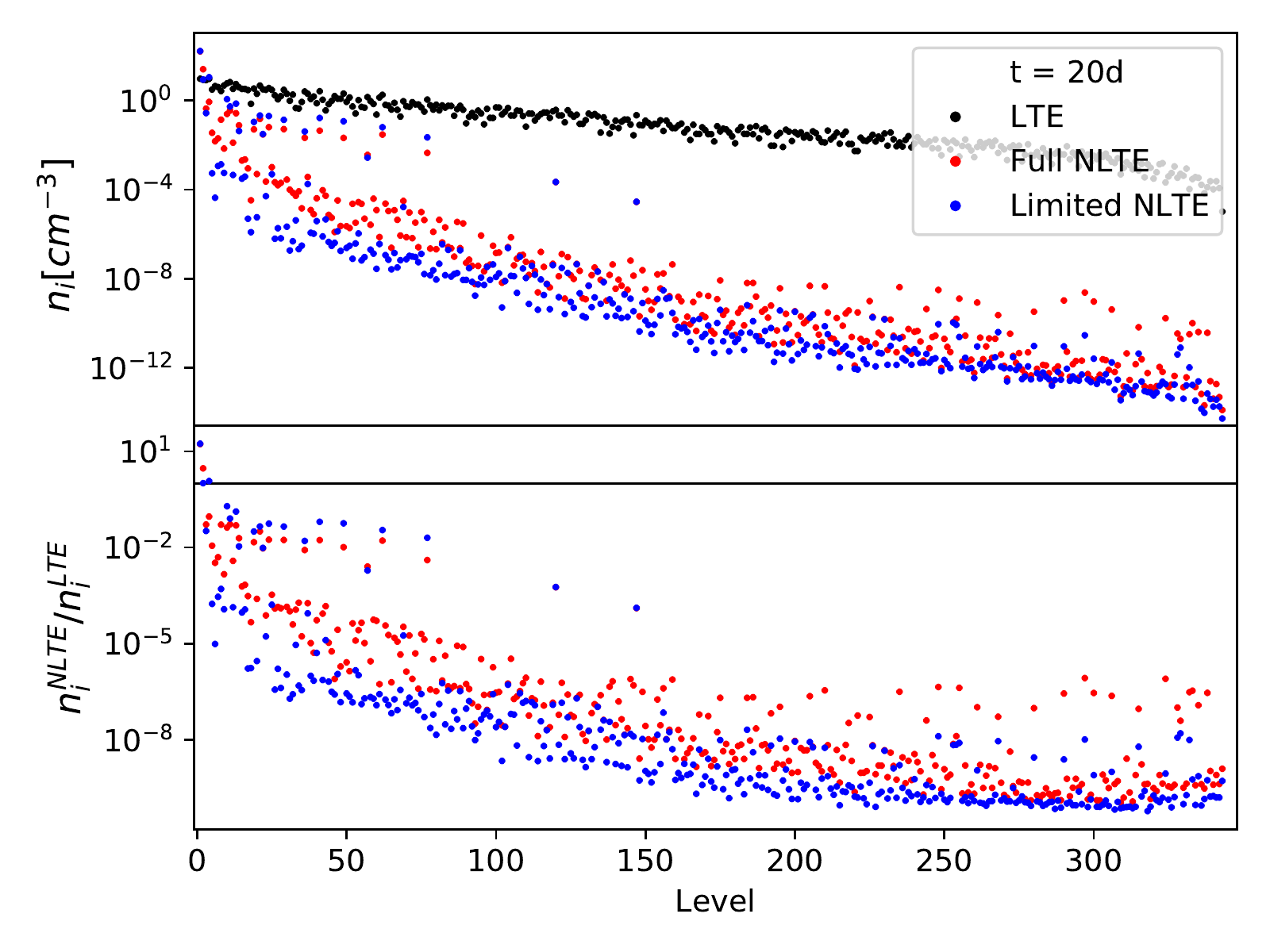} 
    \caption{Level populations of Th II at 3, 5, 10, and 20 days after merger (top left to bottom right). States with an LTE level population smaller than the most populated state by a factor of $10^{10}$ are cut from these plots.}
    \label{fig:ThII_levelpops}
\end{figure*}

\begin{figure*}
    \centering
    \includegraphics[trim={0.2cm 0.1cm 0.4cm 0.3cm},clip,width = 0.49\textwidth]{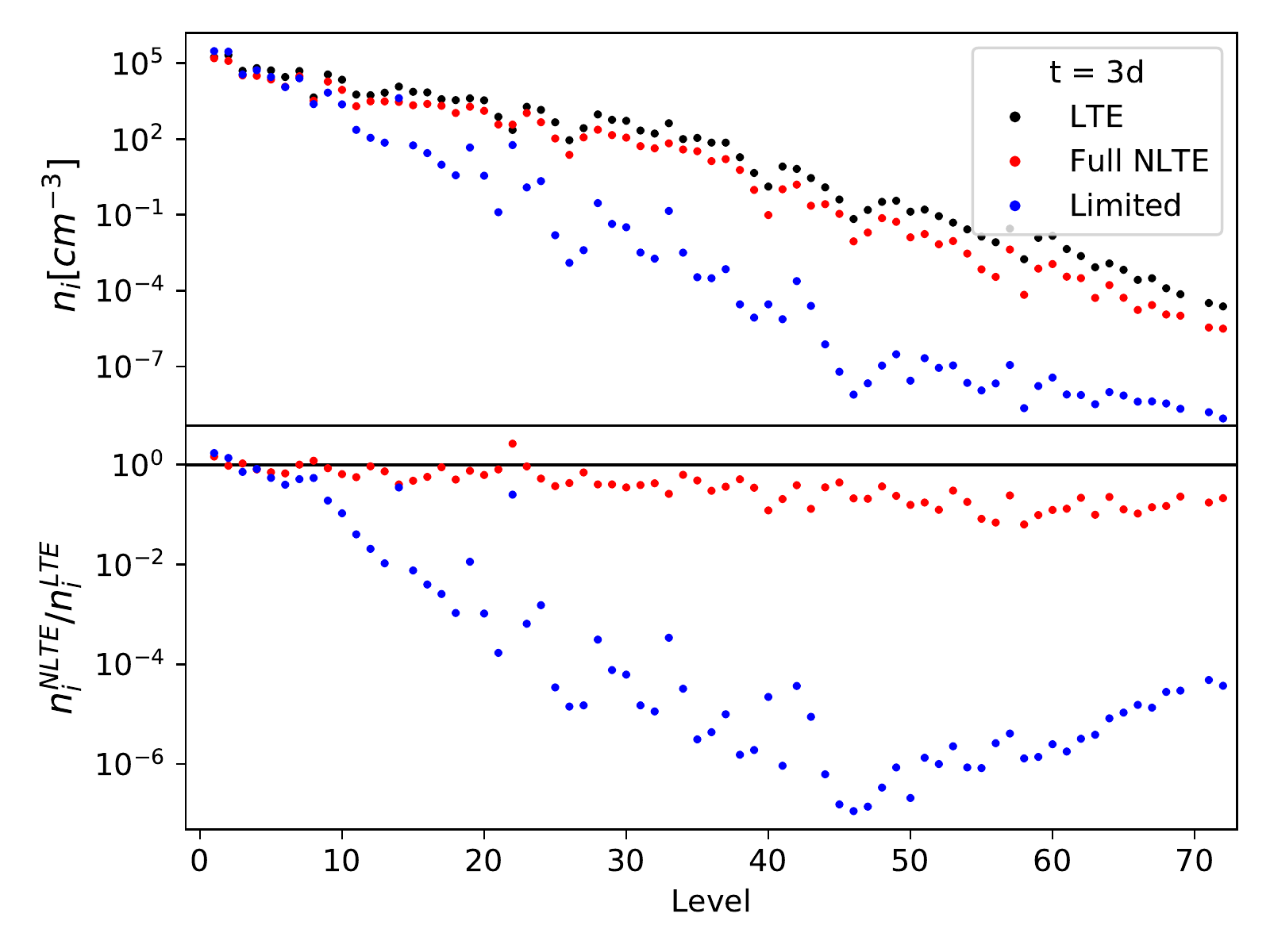}
    \includegraphics[trim={0.2cm 0.1cm 0.4cm 0.3cm},clip,width = 0.49\textwidth]{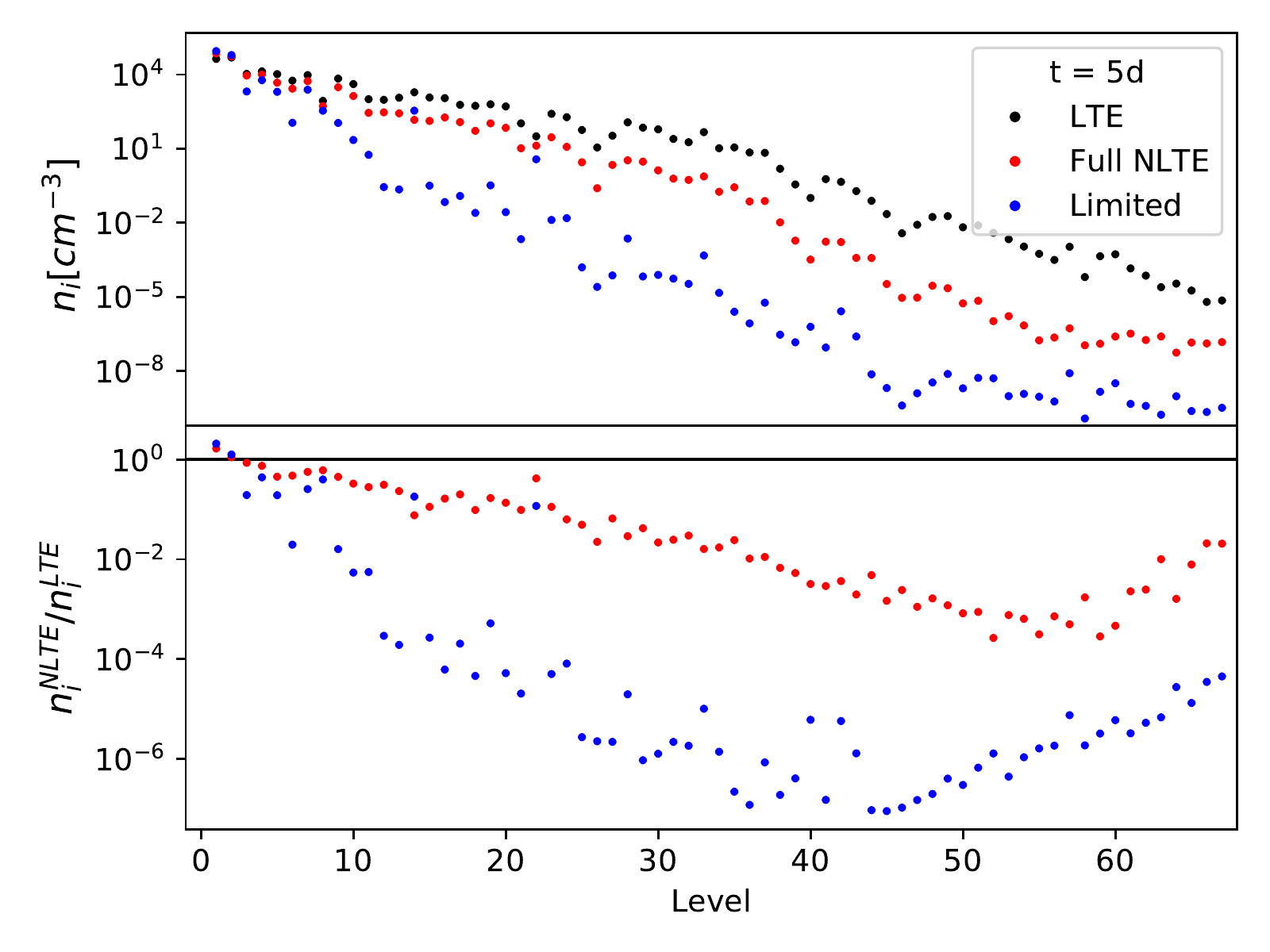} 
    \includegraphics[trim={0.2cm 0.1cm 0.4cm 0.3cm},clip,width = 0.49\textwidth]{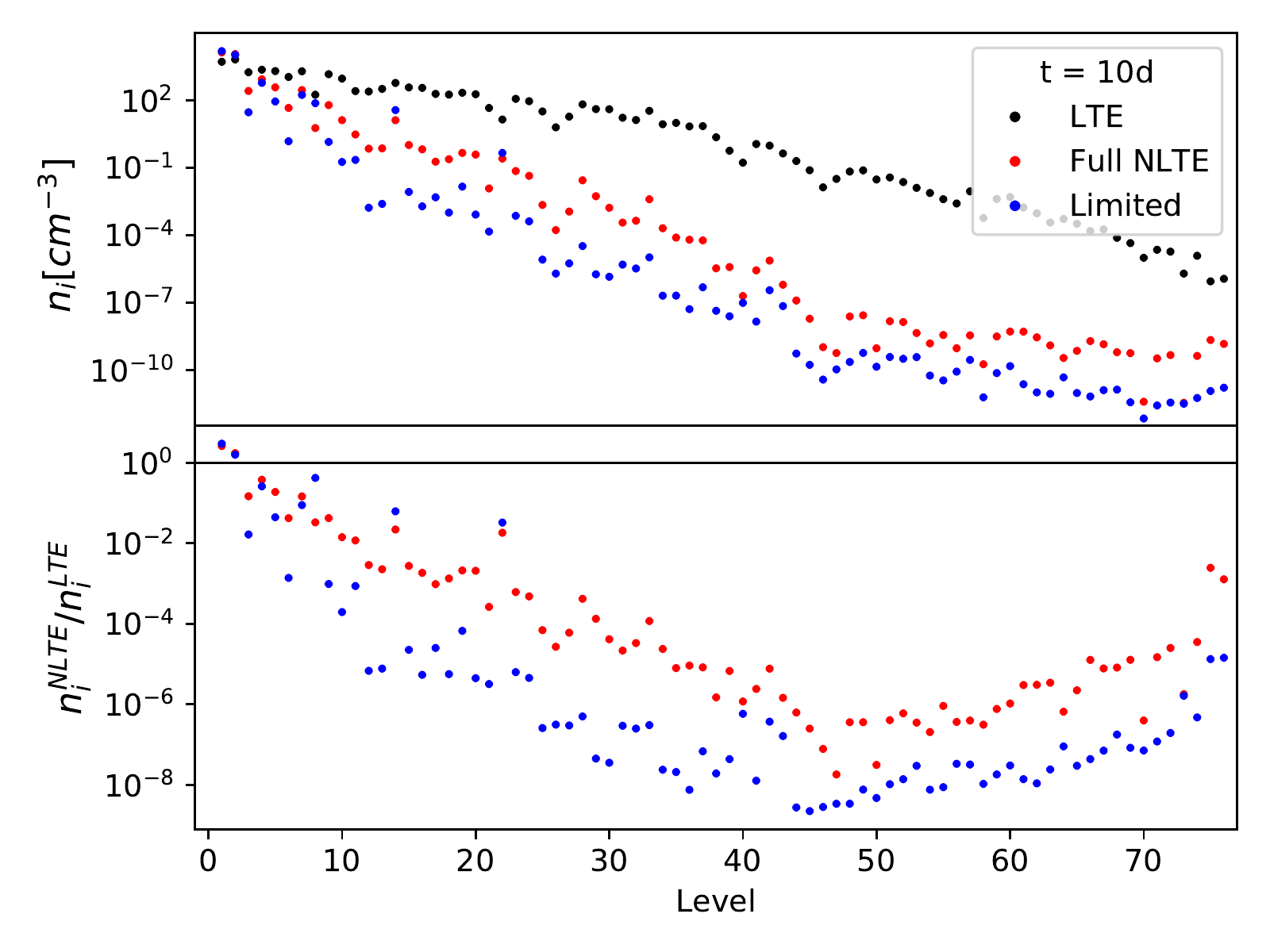} 
    \includegraphics[trim={0.2cm 0.1cm 0.4cm 0.3cm},clip,width = 0.49\textwidth]{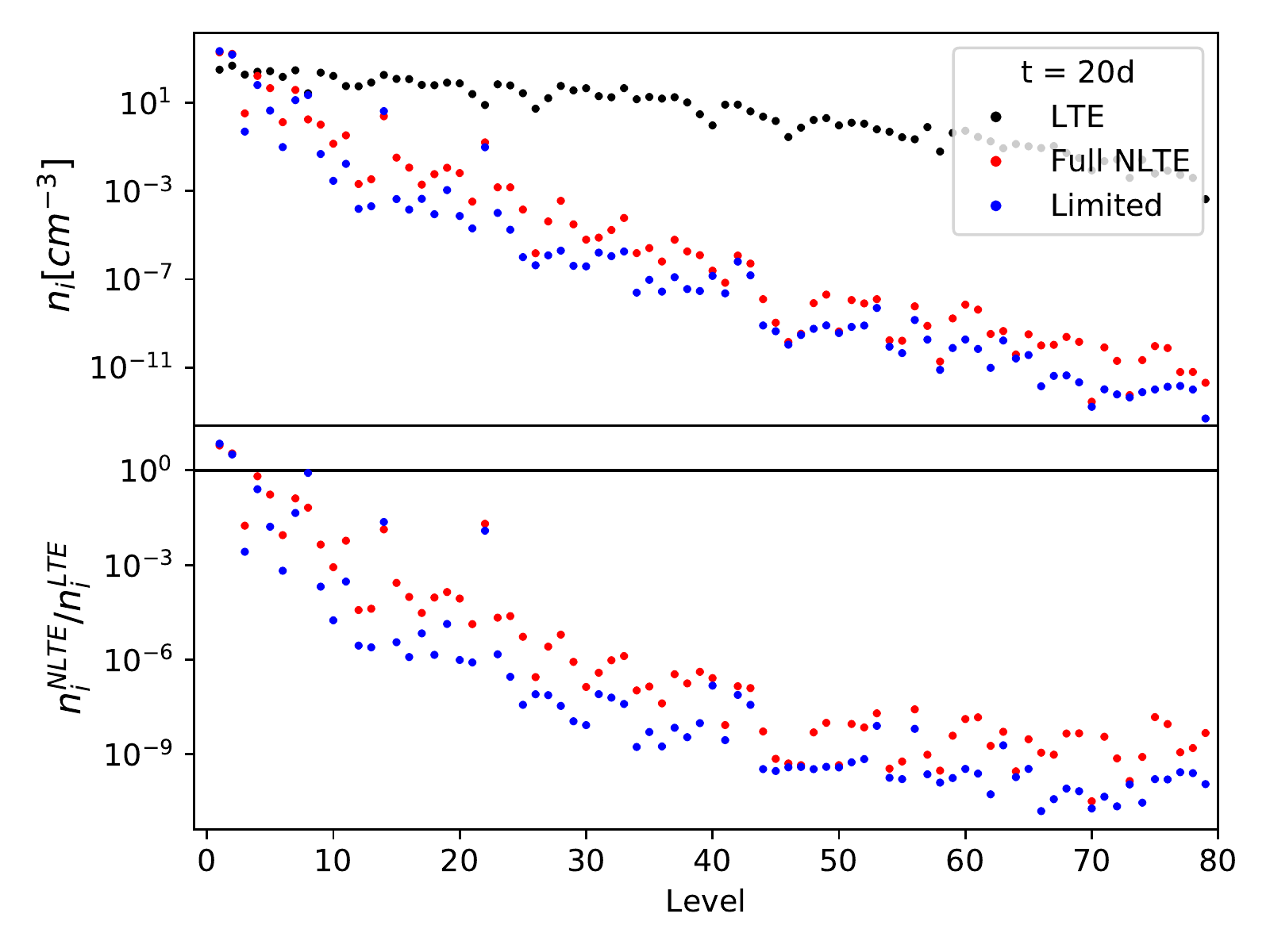} 
    \caption{Level populations of Th III at 3, 5, 10, and 20 days after merger (top left to bottom right). States with an LTE level population smaller than the most populated state by a factor of $10^{10}$ are cut from these plots.}
    \label{fig:ThIII_levelpops}
\end{figure*}

\begin{figure*}
    \centering
    \includegraphics[trim={0.2cm 0.1cm 0.4cm 0.3cm},clip,width = 0.49\textwidth]{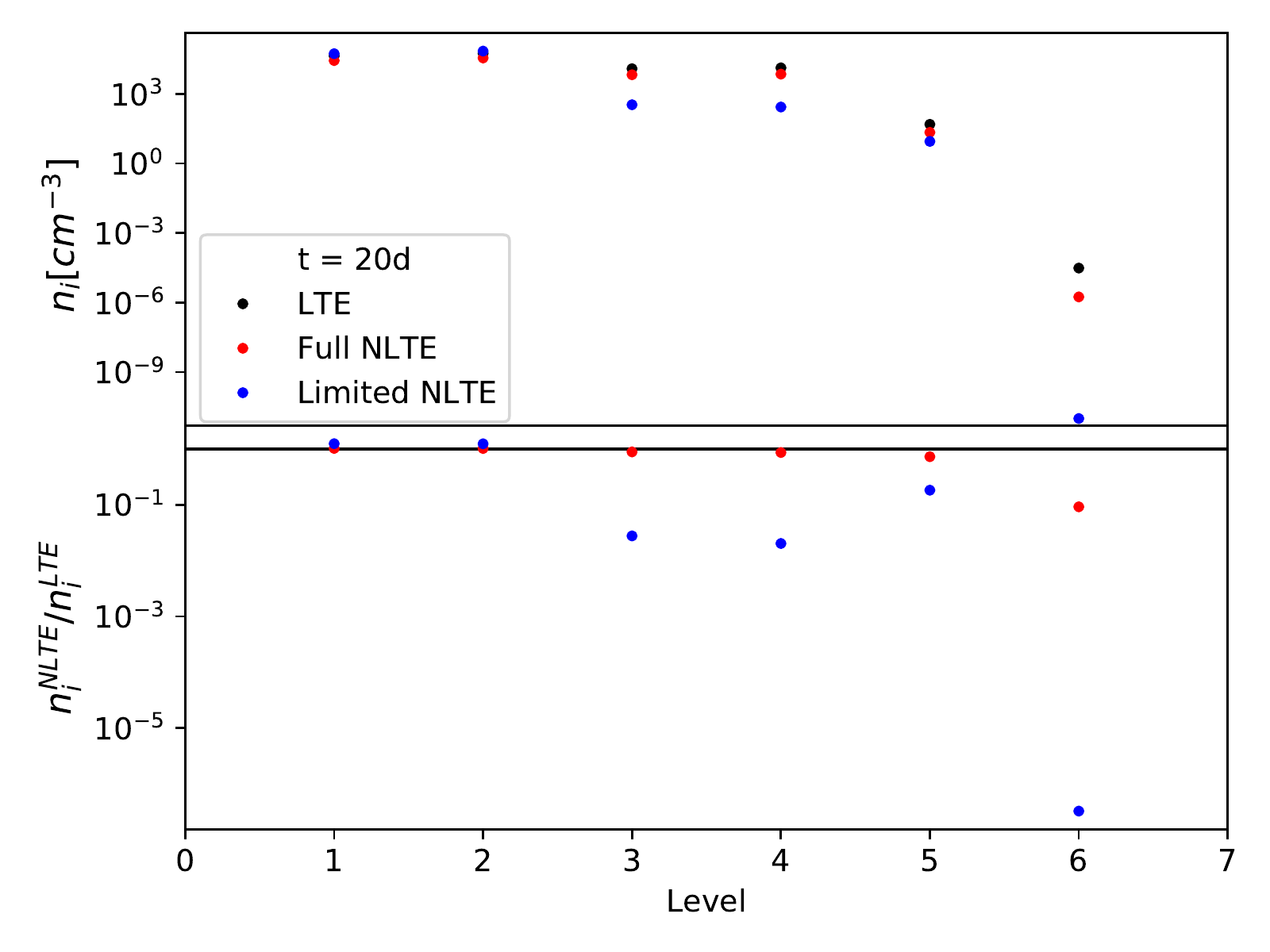}
    \includegraphics[trim={0.2cm 0.1cm 0.4cm 0.3cm},clip,width = 0.49\textwidth]{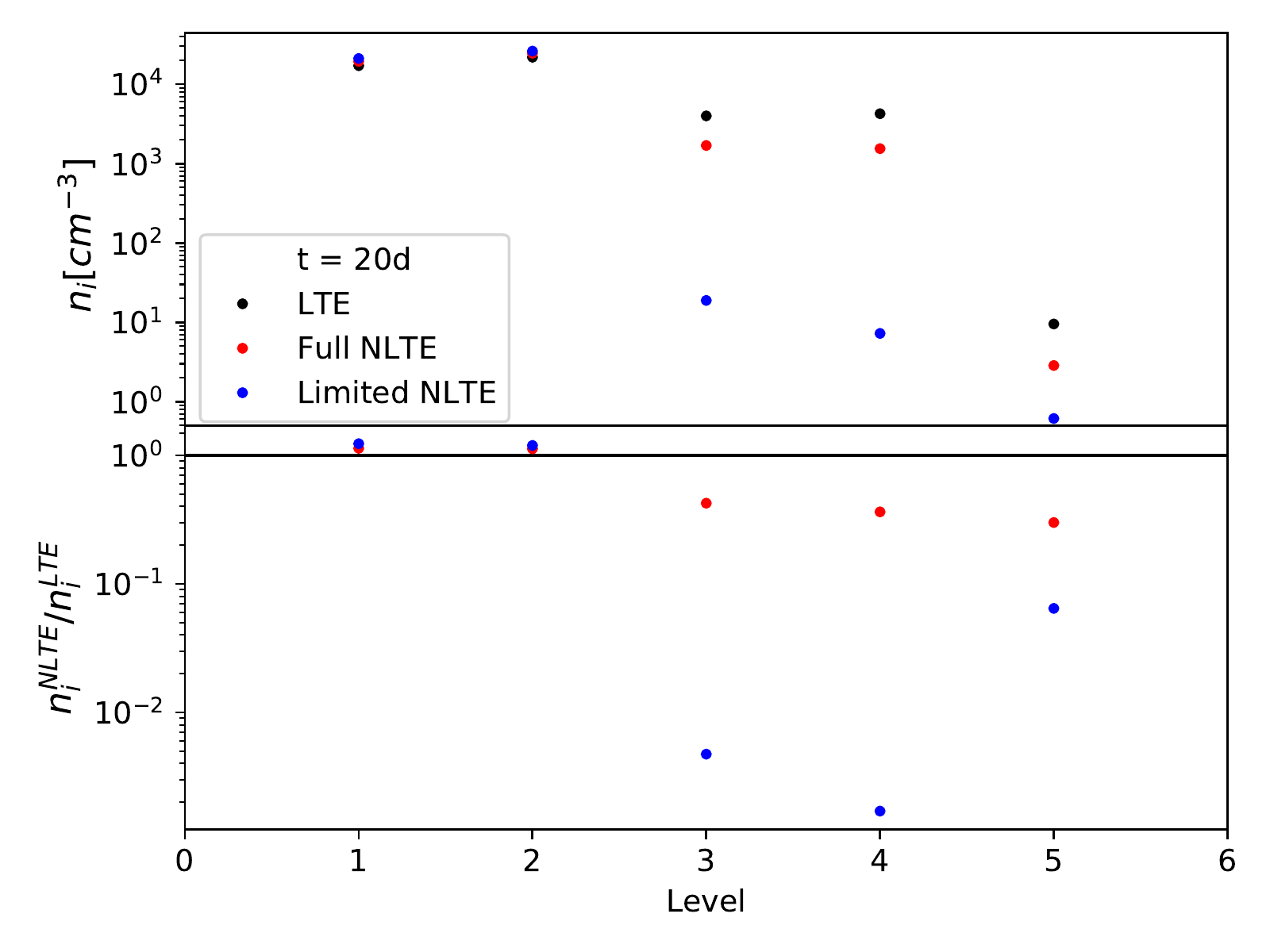} 
    \includegraphics[trim={0.2cm 0.1cm 0.4cm 0.3cm},clip,width = 0.49\textwidth]{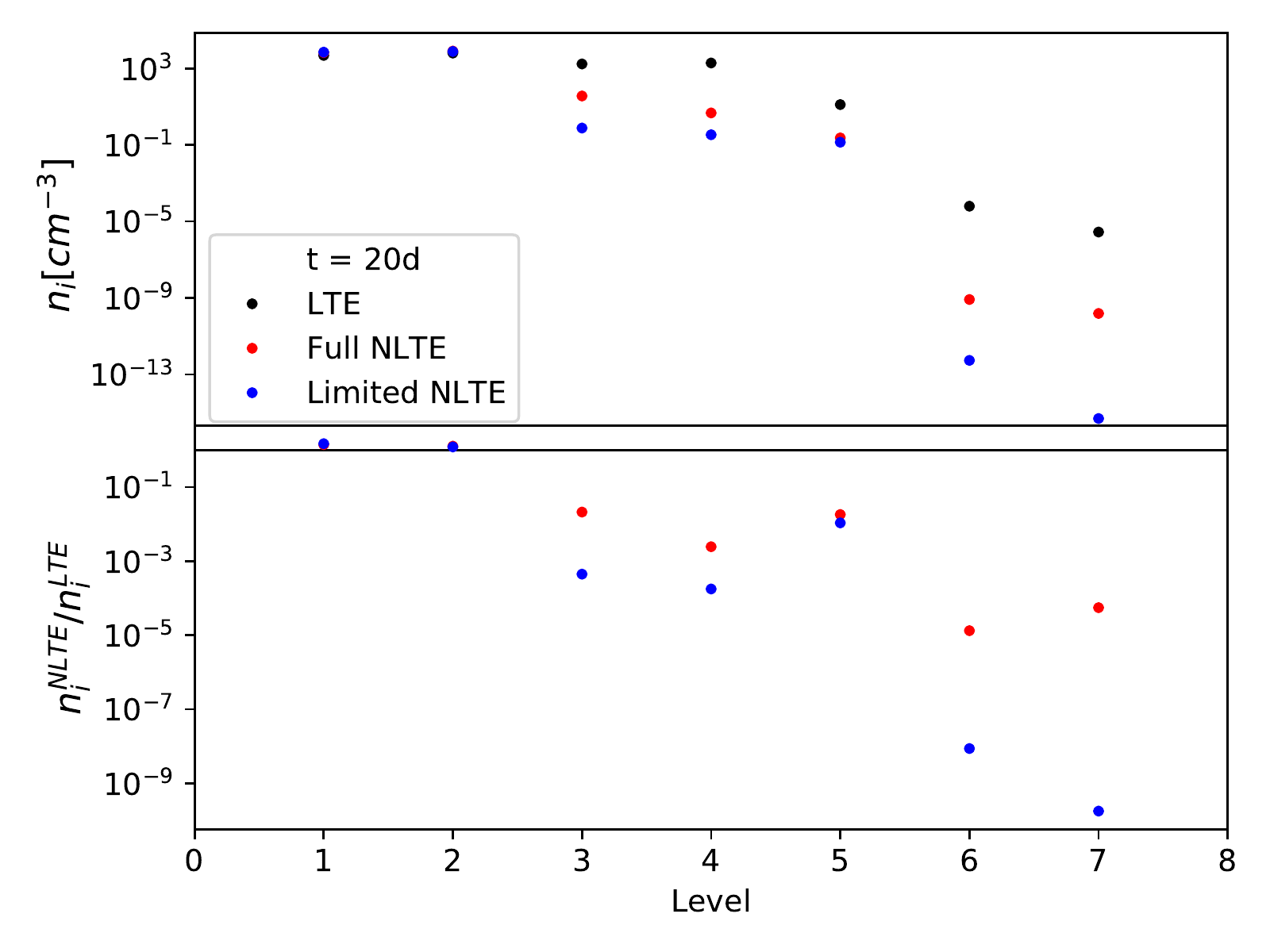} 
    \includegraphics[trim={0.2cm 0.1cm 0.4cm 0.3cm},clip,width = 0.49\textwidth]{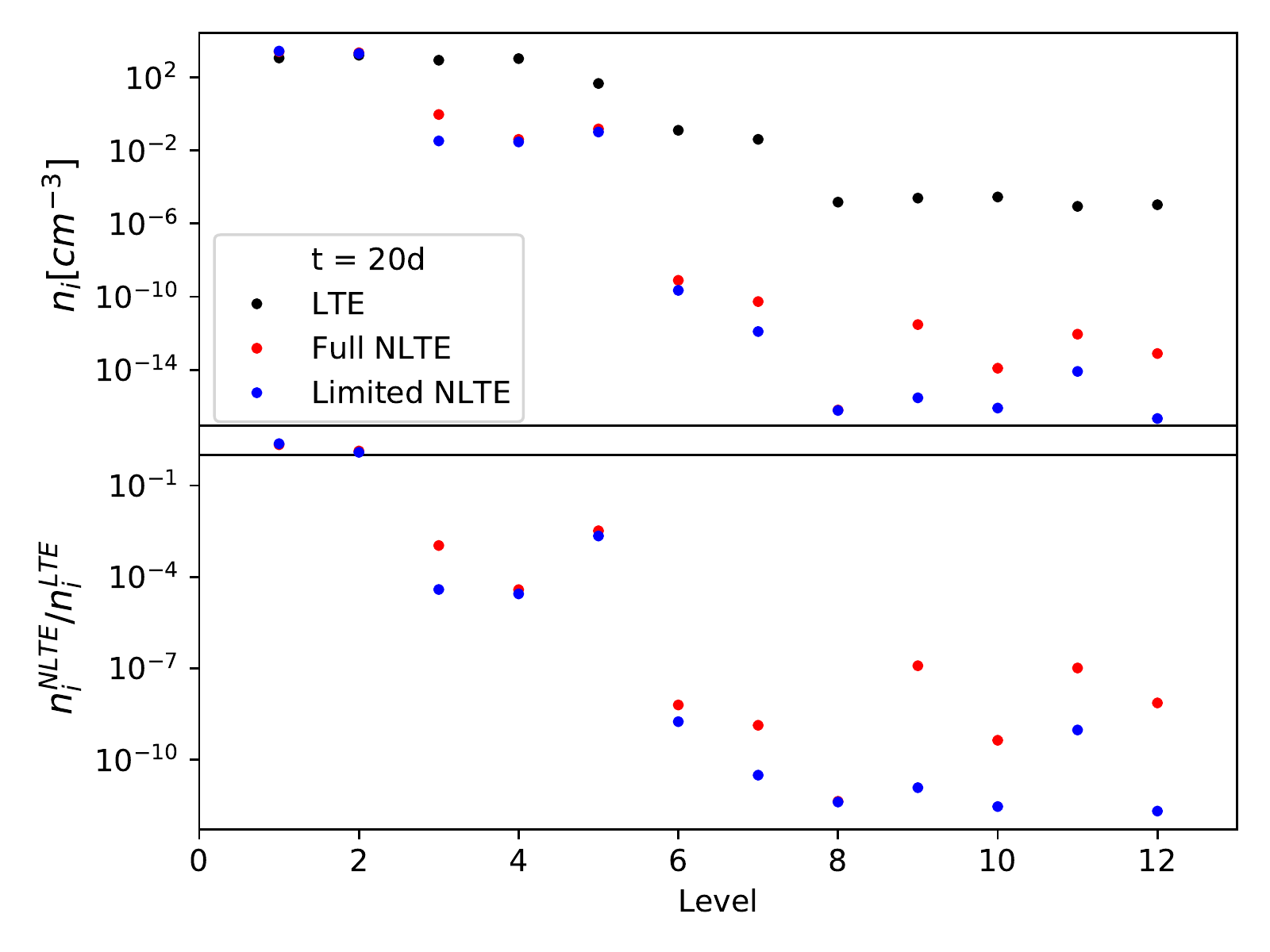} 
    \caption{Level populations of Th IV at 3, 5, 10, and 20 days after merger (top left to bottom right). States with an LTE level population smaller than the most populated state by a factor of $10^{10}$ are cut from these plots.}
    \label{fig:ThIV_levelpops}
\end{figure*}

In this appendix we present the level plots from the other ions in a fashion similar to Figures \ref{fig:CeII_levelpops} and \ref{fig:PtIII_levelpops}, that is epoch evolving from 3 to 20 days after merger, top left to bottom right. Te I - Te IV are presented in Figures \ref{fig:TeI_levelpops} - \ref{fig:TeIV_levelpops}, Ce I - Ce IV in Figures \ref{fig:CeI_levelpops} - \ref{fig:CeIV_levelpops} (omitting Ce II), Pt I - Pt IV in Figures \ref{fig:PtI_levelpops} - \ref{fig:PtIV_levelpops} (omitting Pt III), and Th I - Th IV in Figures \ref{fig:ThI_levelpops} - \ref{fig:ThIV_levelpops}.

\section{Elemental Opacity Plots}
\label{app:opacity}

In this appendix, we present the opacities of individual ions calculated using NLTE ionisation and Boltzmann level populations (same as in Figure \ref{fig:opacities_individual}) at the other epochs: 3, 10 and 20 days after merger.

\begin{figure*}
    \centering
    \includegraphics[trim={0.cm 0.cm 0.4cm 0.4cm},clip,width = 0.49\textwidth]{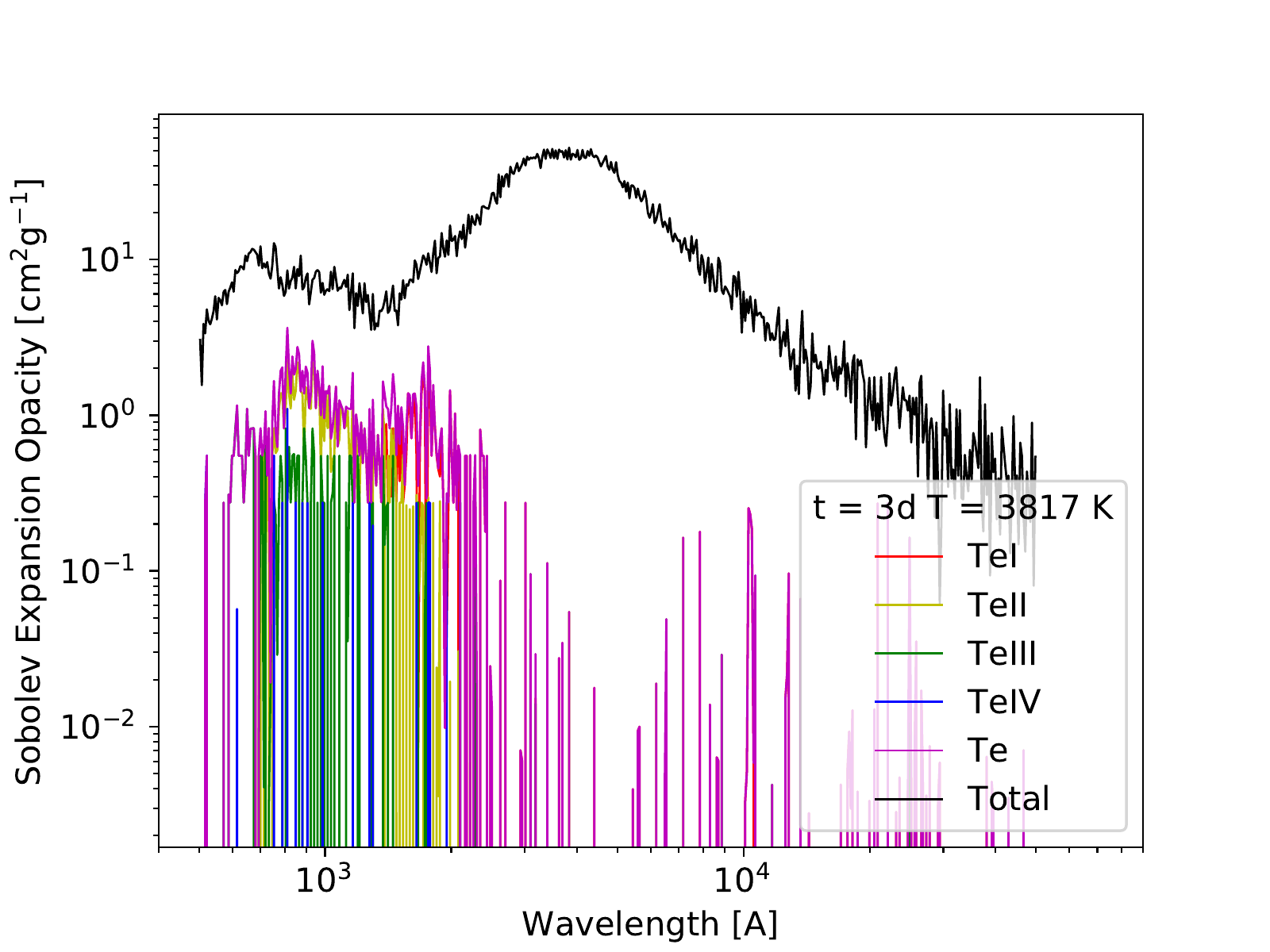}
    \includegraphics[trim={0.cm 0.cm 0.4cm 0.4cm},clip,width = 0.49\textwidth]{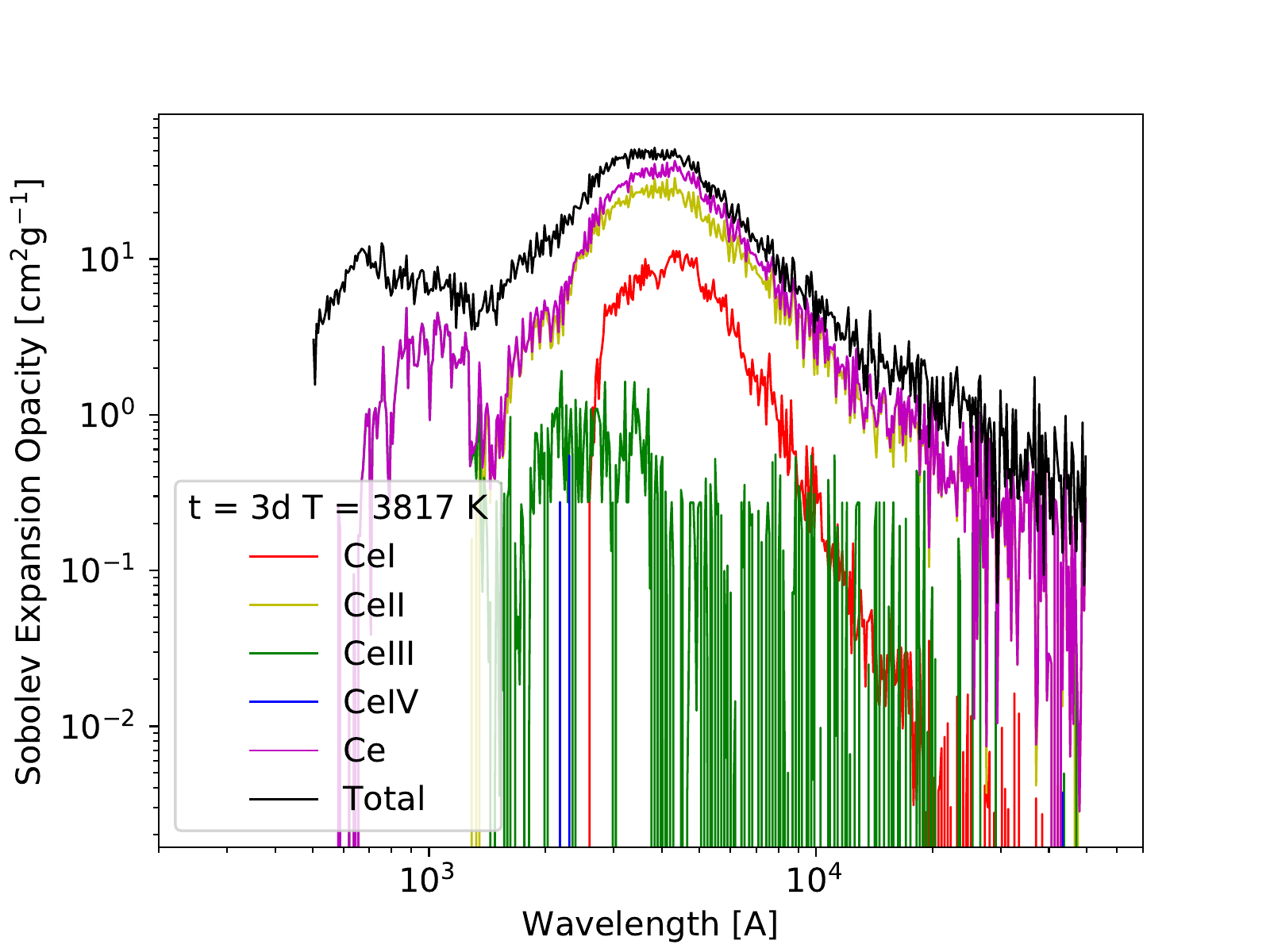}
    \includegraphics[trim={0.cm 0.cm 0.4cm 0.4cm},clip,width = 0.49\textwidth]{opacities/fig6a.pdf}
    \includegraphics[trim={0.cm 0.cm 0.4cm 0.4cm},clip,width = 0.49\textwidth]{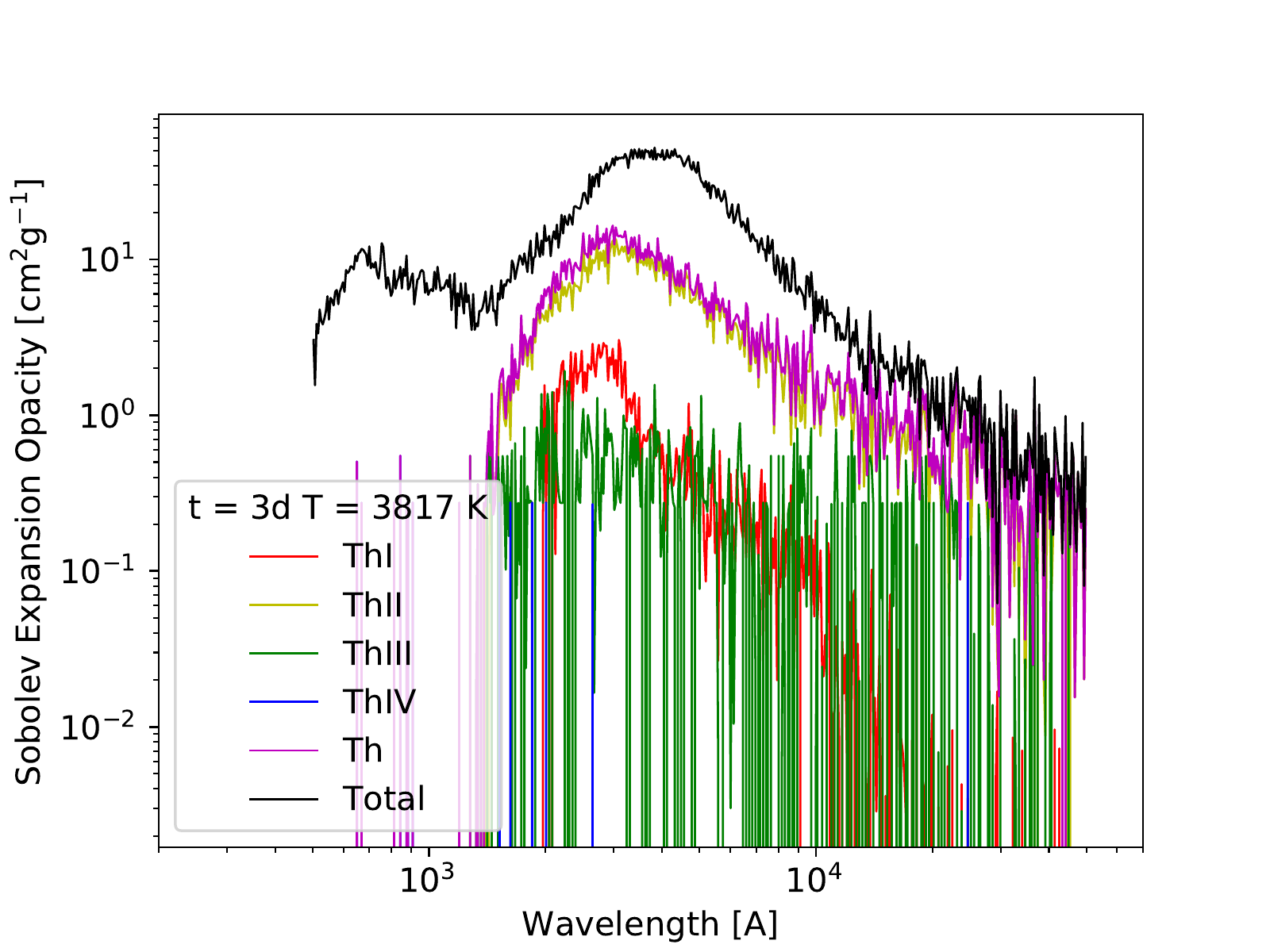}
    \caption{Expansion opacities calculated from the Boltzmann level populations for each individual ion, and summed up for each element, at 3 days after merger. The total opacity of the limited LTE model is also shown for reference.}
    \label{fig:opacities_3d}
\end{figure*}

\begin{figure*}
    \centering
    \includegraphics[trim={0.cm 0.cm 0.4cm 0.4cm},clip,width = 0.49\textwidth]{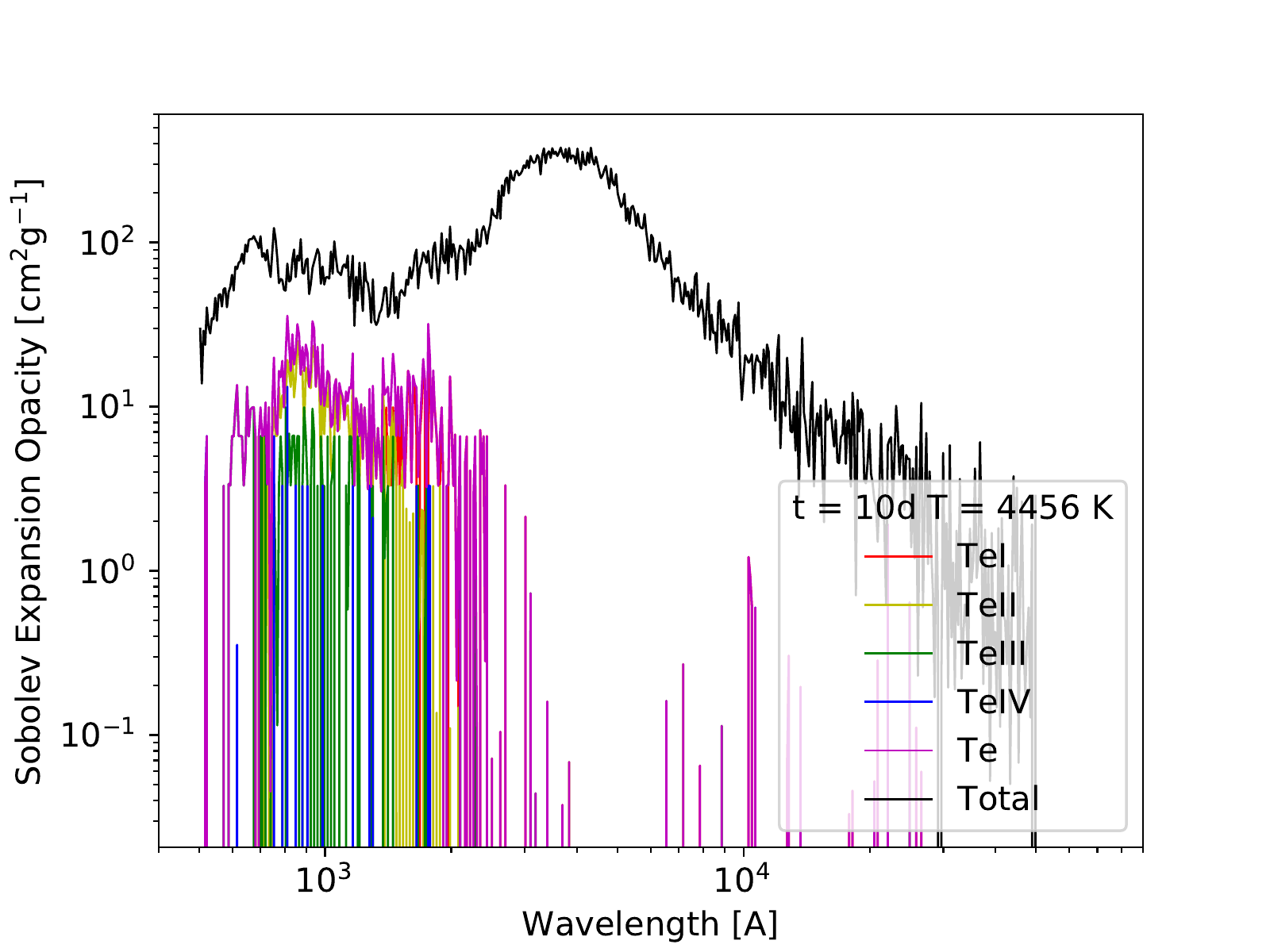} 
    \includegraphics[trim={0.cm 0.cm 0.4cm 0.4cm},clip,width = 0.49\textwidth]{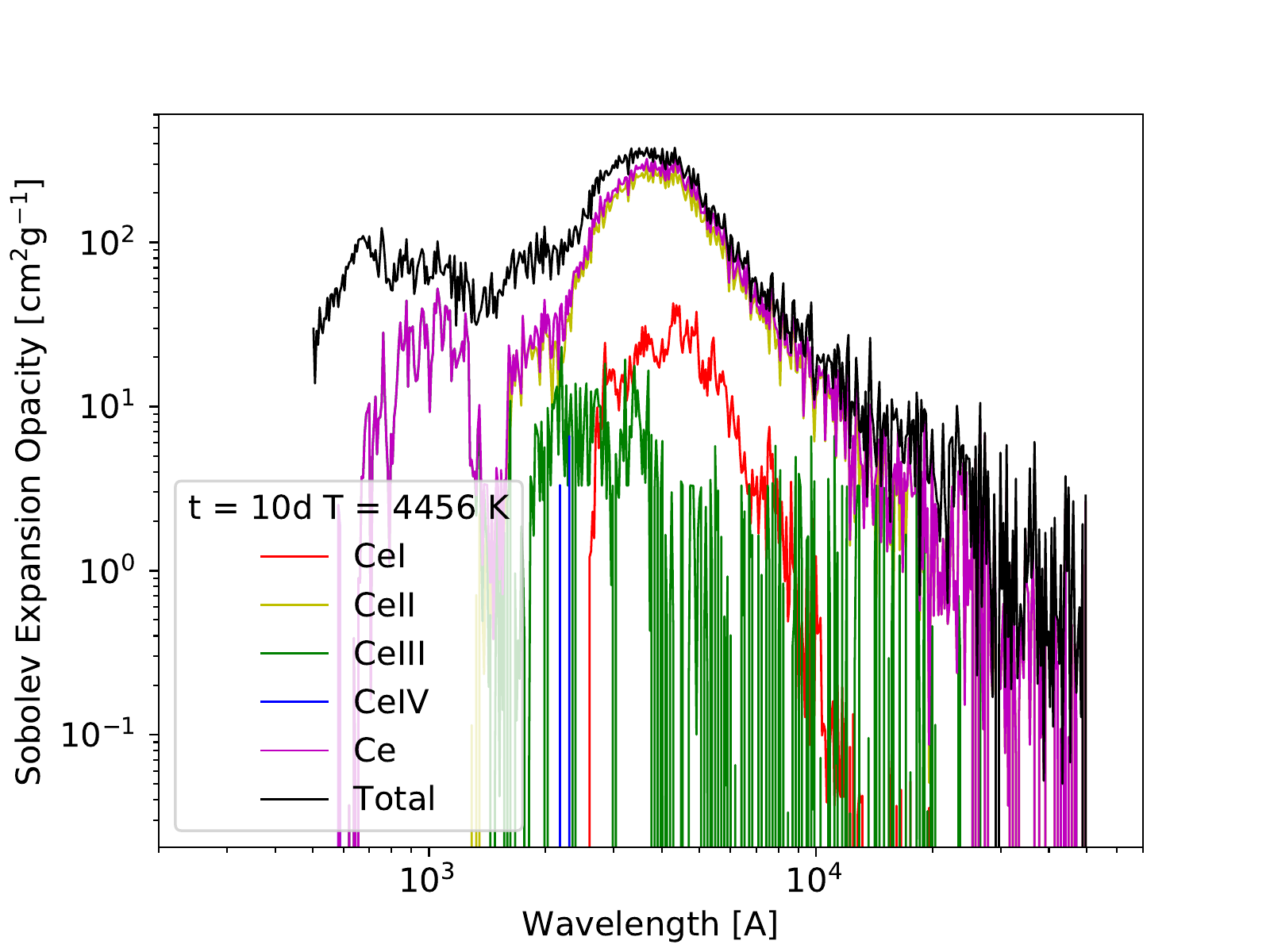}
    \includegraphics[trim={0.cm 0.cm 0.4cm 0.4cm},clip,width = 0.49\textwidth]{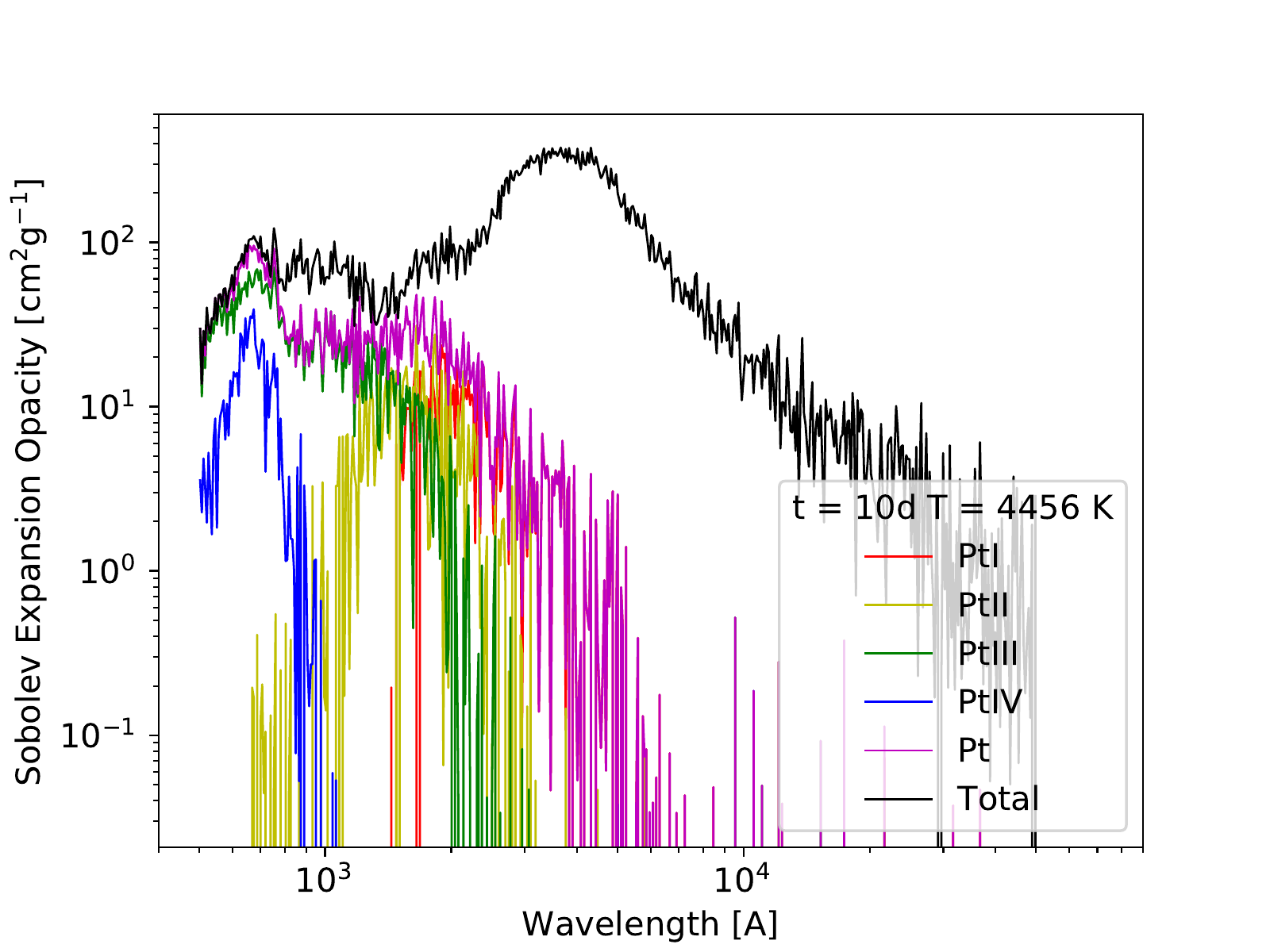} 
    \includegraphics[trim={0.cm 0.cm 0.4cm 0.4cm},clip,width = 0.49\textwidth]{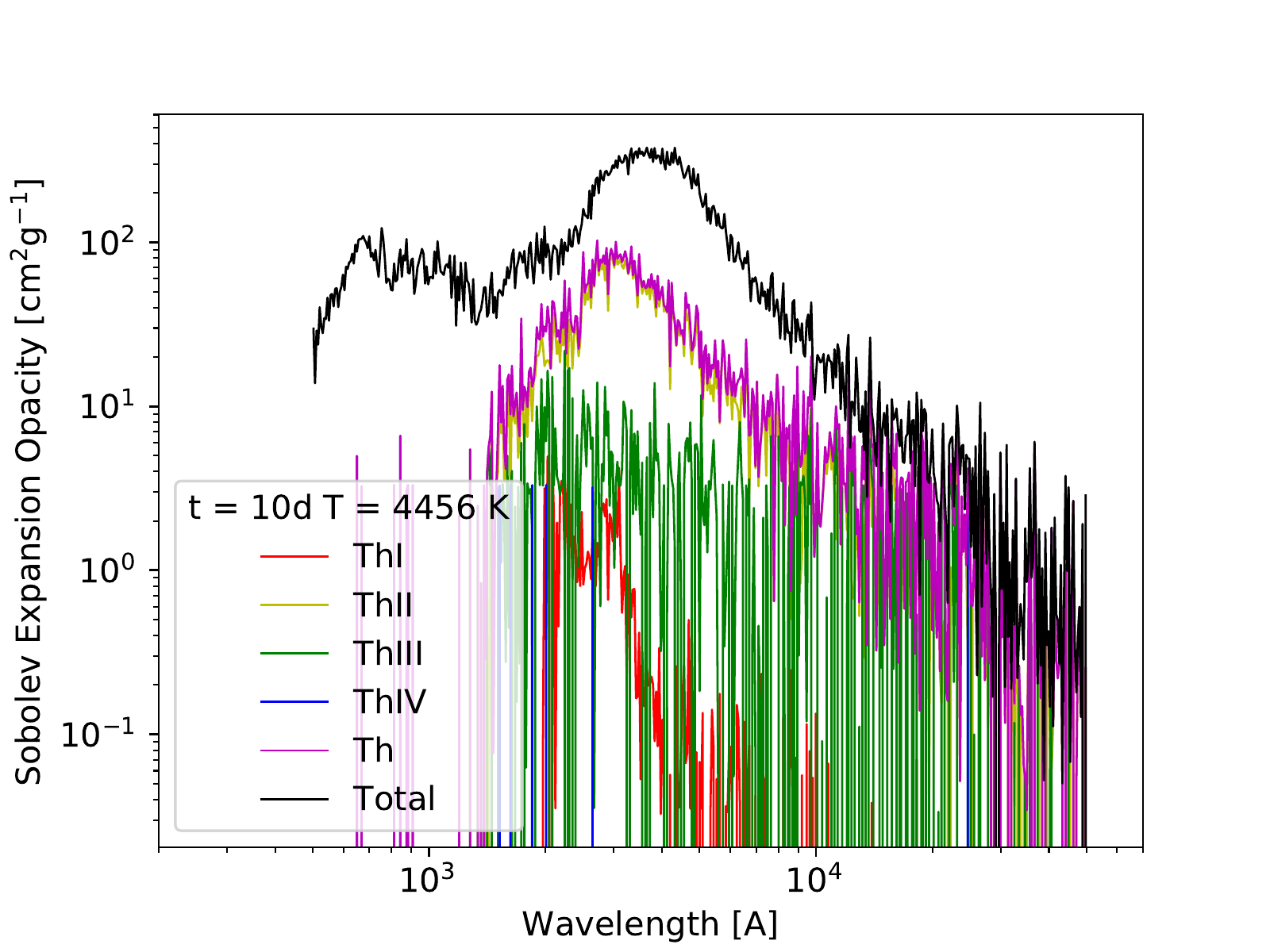}
    \caption{Expansion opacities calculated from the Boltzmann level populations for each individual ion, and summed up for each element, at 10 days after merger. The total opacity of the limited LTE model is also shown for reference.}
    \label{fig:opacities_10d}
\end{figure*}

\begin{figure*}
    \centering
    \includegraphics[trim={0.cm 0.cm 0.4cm 0.4cm},clip,width = 0.49\textwidth]{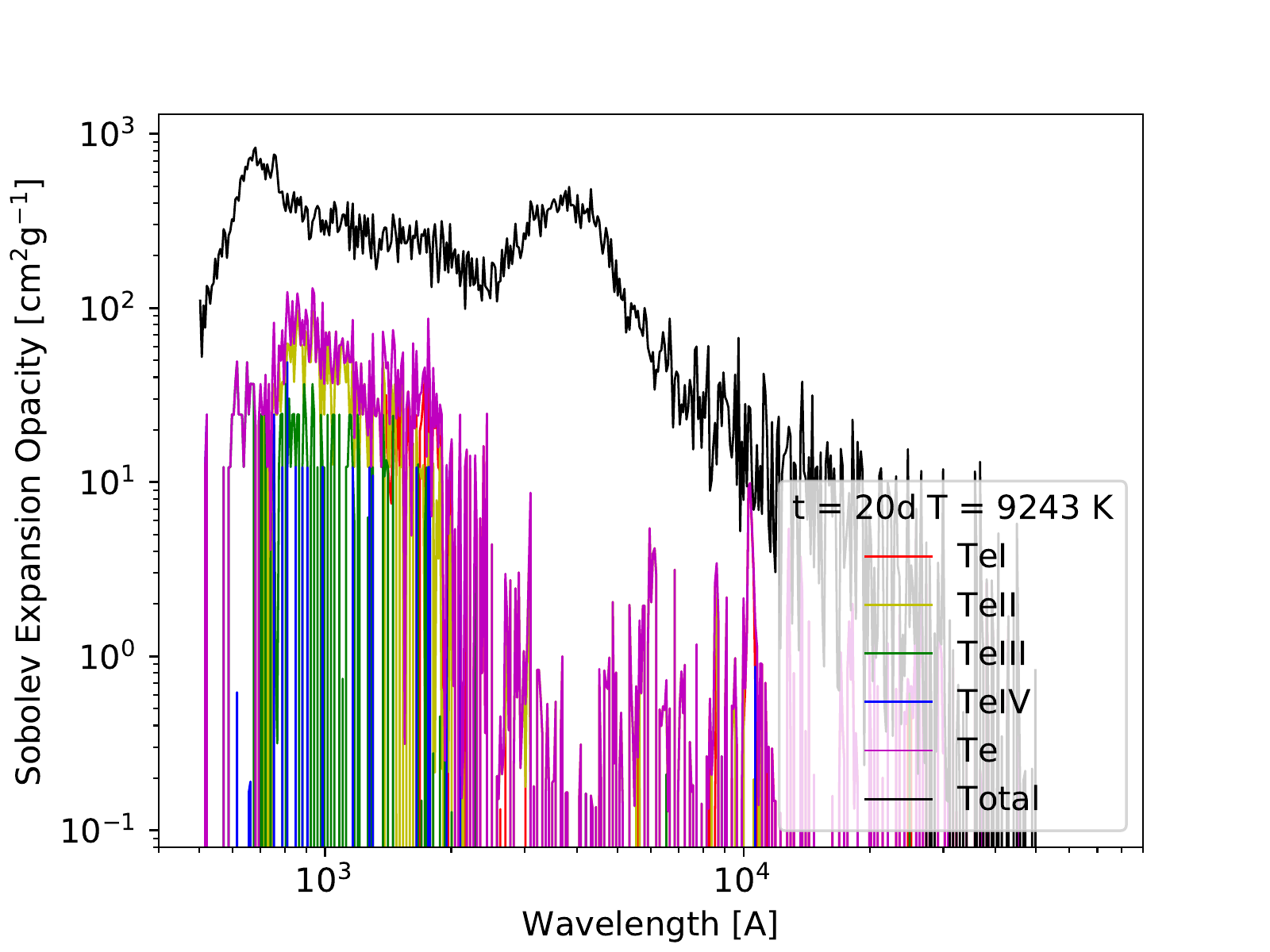} 
    \includegraphics[trim={0.cm 0.cm 0.4cm 0.4cm},clip,width = 0.49\textwidth]{opacities/fig6c.pdf}
    \includegraphics[trim={0.cm 0.cm 0.4cm 0.4cm},clip,width = 0.49\textwidth]{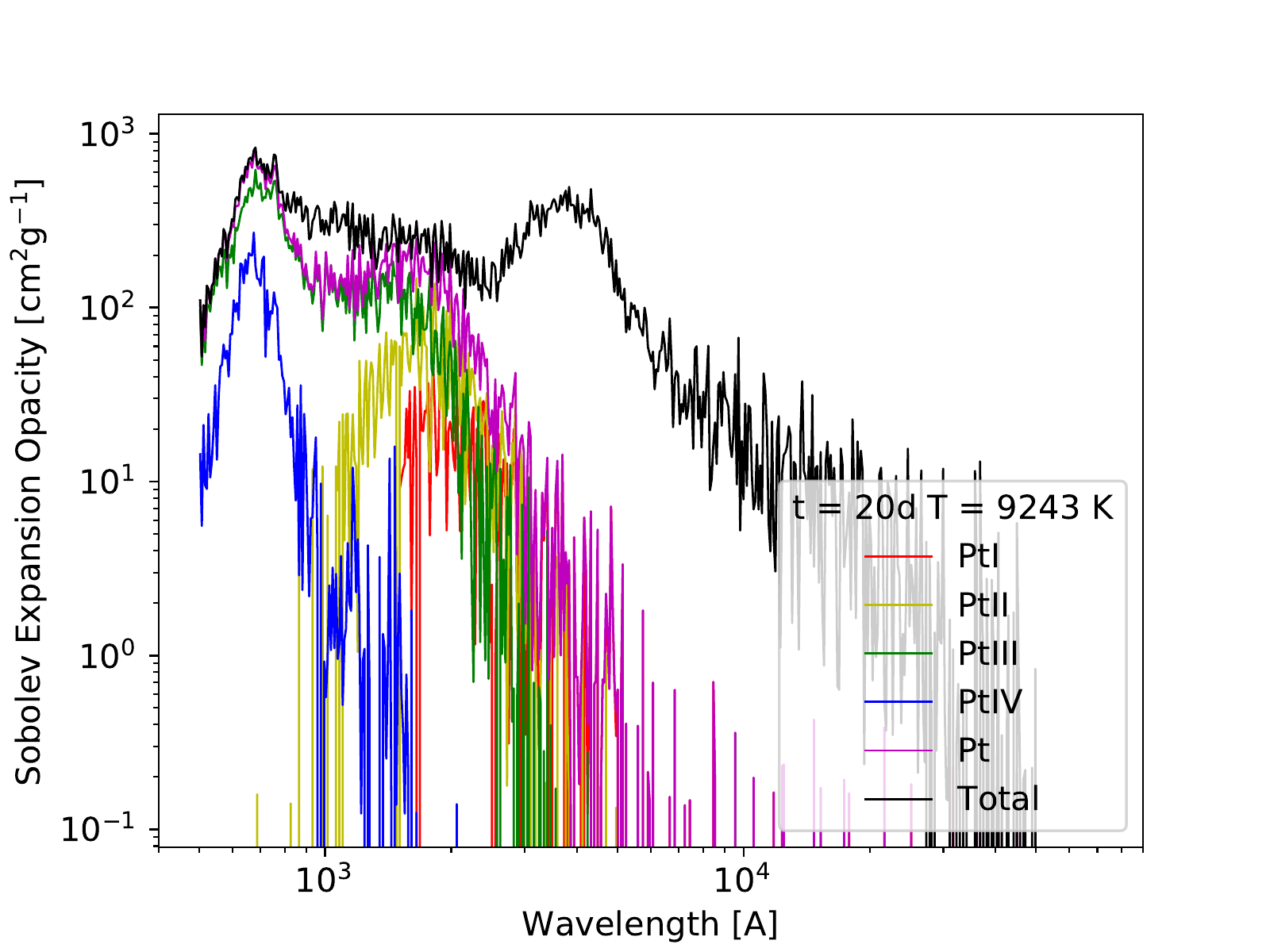} 
    \includegraphics[trim={0.cm 0.cm 0.4cm 0.4cm},clip,width = 0.49\textwidth]{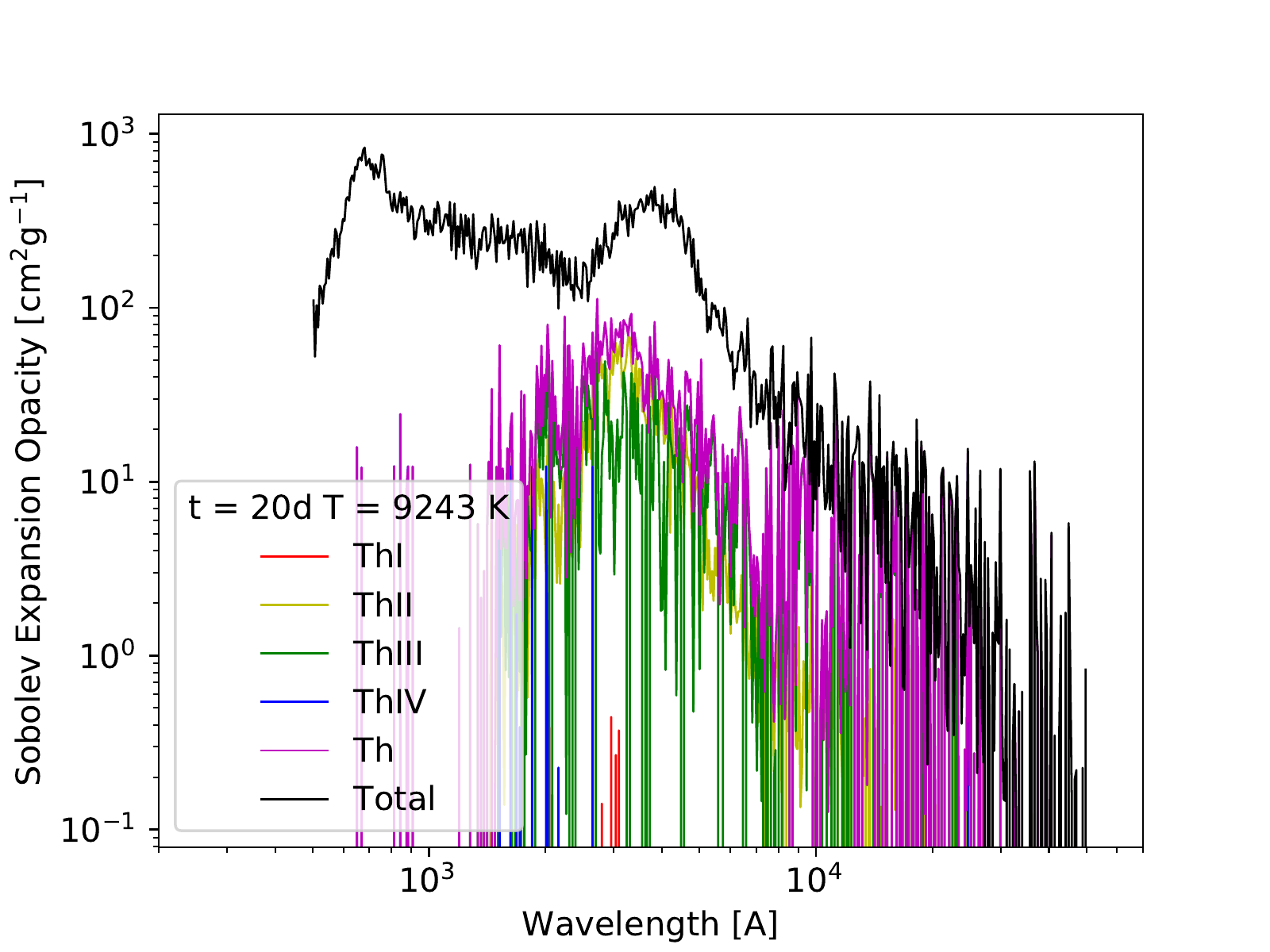}
    \caption{Expansion opacities calculated from the Boltzmann level populations for each individual ion, and summed up for each element, at 20 days after merger. The total opacity of the limited LTE model is also shown for reference.}
    \label{fig:opacities_20d}
\end{figure*}


\bsp	
\label{lastpage}
\end{document}